\documentclass[journal ]{new-aiaa}
\usepackage[utf8]{inputenc}
\usepackage{textcomp}

\usepackage[final]{changes} 

\usepackage{amsmath}
\usepackage[version=4]{mhchem}
\usepackage{siunitx}
\usepackage{longtable,tabularx}
\DeclareMathAlphabet\mathsfbi{OT1}{phv}{b}{it}

\usepackage{epstopdf, epsfig}
\usepackage{lscape}
\usepackage{mathtools}
\usepackage{xfrac}
\usepackage{pgfplots} 
\pgfplotsset{compat=newest}
\pgfplotsset{plot coordinates/math parser=false}
\newlength\figureheight
\newlength\figurewidth 
\usepackage{tikz,tikz-3dplot} 
\usetikzlibrary{decorations.pathreplacing,angles,quotes}
\usepackage{xifthen}
\usetikzlibrary{arrows}
\newcommand{\defeq}{\vcentcolon=}

\title{Simulating Airplane Aerodynamics with Body Forces: Actuator Line Method for Non-planar Wings}

\author{Vitor G. Kleine\footnote{PhD student, Department of Engineering Mechanics, FLOW.}~\footnote{Assistant Professor, Instituto Tecnol\'{o}gico de Aeron\'{a}utica, Pra\c{c}a Marechal Eduardo Gomes, 50, Vila das Ac\'{a}cias, 12228-900, S\~{a}o Jos\'{e} dos Campos - SP, Brazil.}, Ardeshir Hanifi\footnote{Researcher, Department of Engineering Mechanics, FLOW.} and Dan S. Henningson\footnote{Professor, Department of Engineering Mechanics, FLOW.}}
\affil{KTH Royal Institute of Technology, Stockholm, SE-10044, Sweden}

\begin{document}

\maketitle

\begin{abstract}
Two configurations typical of fixed-wing aircraft are simulated with the actuator line method (ALM): a wing with winglets and a T-tail. The ALM is extensively used in rotor simulations to model the blades by body forces, which are calculated from airfoil data and the relative flow velocity. This method has not been used to simulate airplane aerodynamics, despite its advantage of allowing coarser grids. This may be credited to the failure of the uncorrected ALM to accurately predict forces near the tip of wings, even for simple configurations. The recently-proposed vortex-based smearing correction shows improved results, suggesting those limitations are part of the past. For the non-planar configurations studied in this work, differences between the ALM with the original smearing correction and a non-linear lifting line method (LL) are observed near the intersection of surfaces, because the circulation generated in the numerical simulation differs from the calculated corrected circulation. A vorticity magnitude correction is proposed, which improves the agreement between ALM and LL. This second-order correction resolves the ambiguity in the velocity used to define the lift force. The good results indicate that the improved ALM can be used for airplane aerodynamics, with an accuracy similar to the LL.
\end{abstract}

\section*{Nomenclature}

{\renewcommand\arraystretch{1.0}
\noindent\begin{longtable*}{@{}l @{\quad=\quad} l@{}}

$\mathsfbi{A_y}$, $\mathsfbi{A_z}$ & $N \times N$ matrix of influence coefficients \\
ALM                     & actuator line method \\
ALM$^1$                 & actuator line method without vorticity magnitude correction \\
ALM$^{VMC}$             & actuator line method with vorticity magnitude correction \\
$b_{y}$,$b_{z}$         & measure of the sensitivity of the circulation to a change in the velocities in $y$ and $z$ directions \\
$c$                     & chord of the airfoil \\
$C_d$,$C_l$             & local airfoil drag and lift coefficients, define based on the local velocity \\
$C_{l U_0}^c$          & lift coefficient defined based on the corrected force and the velocity at infinity $C_{l U_0}^c=2F_l^c/(\rho U_0^2 c)$ \\
CFD                     & computational fluid dynamics \\
$\mathbf{f}$            & body force, in three dimensions \\
$\mathbf{F}_{2D}$       & force per spanwise unit length \\
$F_d$,$F_l$             & drag and lift forces per spanwise unit length \\
LL                      & non-linear lifting line method \\
$L_X,L_Y,L_Z$           & size of numerical domain\\
$N$                     & number of control points \\
$p$                     & pressure \\
$R$                     & span \\
$Re$                    & Reynolds number $Re=U_{0} R / \nu$ \\
$s$                     & spanwise distance to the center of the wing \\
$\mathbf{u}$            & flow velocity, in three dimensions $\mathbf{u}=(u_x,u_y,u_z)$ \\
$u_{r}$                 & magnitude of the relative velocity between the flow and the airfoil, $u_{r}=\sqrt{u_y^2 + u_z^2}$ \\
$\mathsfbi{u_y}$,$\mathsfbi{u_z}$            & $N \times 1$ vector of flow velocity values \\
$\mathbf{u}^c$          & corrected velocity \\
$\mathbf{u}^m$          & missing velocity \\
$\mathbf{u}^s$          & velocity sampled from the numerical simulation \\
$\mathbf{u}^{v}$        & velocity induced by vortex of Gaussian core \\
$\mathbf{u}^{vi}$       & velocity induced by ideal vortex \\
$U_0$                   & undisturbed flow velocity, $U_0=1$ in the $Z$-direction \\
$\mathsfbi{U_y}$,$\mathsfbi{U_z}$            & $N \times 1$ vector of undisturbed velocity values \\
$(x,y,z)$               & local reference system defined by a cross-section of the wing \\
$(X,Y,Z)$               & global reference system \\
$\alpha$                & effective angle of attack \\
$\alpha_g$              & geometric angle of attack \\
$\beta$                 & normalized difference of circulations \\
$\boldsymbol{\Gamma}$   & vector of circulation values \\
$\Gamma$                & circulation \\
$\delta()$              & Dirac delta function \\
$\Delta x$              & average grid spacing \\
$\varepsilon$           & smearing parameter, length parameter of the Gaussian function \\
$\eta()$,$\eta_3()$     & sne and three-dimensional Gaussian functions \\
$\nu$                   & kinematic viscosity \\
$\rho$                  & density \\
$\boldsymbol{\omega}$   & vorticity, in three dimensions $\boldsymbol{\omega}=(\omega_X,\omega_Y,\omega_Z)$ \\
$\defeq$                & equal to by definition \\

\multicolumn{2}{@{}l}{Subscripts and Superscripts}\\
$()^c$ & quantity corrected by the smearing correction \\
$()^{ll}$ & quantity related to lifting line method \\
$()_{ref}$ & reference quantities, corresponding to two-dimensional flow with $U_0$ and $\alpha_g$ \\
$()^s$ & quantity related to numerical simulation \\

\end{longtable*}}
\section{Introduction}
The actuator line method (ALM) was originally developed by \citet{sorensen2002numerical} to represent blades of wind turbines in numerical simulations of the Navier-Stokes equations, focused on simulations of wind turbine wakes. In this method, the blade is modeled by body forces distributed along a line. The body forces are calculated using the local relative flow velocity and tabulated airfoil data~\citep{sorensen2002numerical}.

Since then, the ALM has become one of the most used methods for modeling wind turbine blades in numerical simulations. The reviews by \citet{sorensen2011aerodynamic,sanderse2011review} and \citet{breton2017survey} can provide an overview of how this method relates to other numerical methods used in wind energy aerodynamics. A few of the early examples of applications of this method for the study of horizontal-axis wind turbine wakes are summarized in~\citep{sorensen2015simulation}. It has also been used in other applications, such as propellers~\citep{jones2018influence,schollenberger2018boundary,stokkermans2019validation}, helicopter and rotorcraft blades~\citep{forsythe2015coupled,buhler2018actuator,merabet2022hovering,zhang2022toward}, tidal turbines~\citep{churchfield2013large,baba2017validation,apsley2018actuator}, vertical-axis turbines~\citep{shamsoddin2014large,mendoza2019near} and kite-based power systems~\citep{gaunaa2020engineering,fredriksson2021modelling}.

The main advantage of the ALM is that it allows coarser resolution in the regions around the blade, because there is no need to capture the boundary layer or even the geometry of the airfoil in order to accurately calculate the forces, greatly reducing the grid size. This advantage is not unique to rotating blades, it can be a way to include wings in Navier-Stokes (or Reynolds-averaged Navier–Stokes) simulations, for low and middle-fidelity simulations, in diverse applications. Nevertheless, so far, this method has not been applied to wings and tails of airplanes. Translating wings are routinely simulated with the ALM. However, the motivation for these simulations of non-rotating wings is, generally, to validate the method, in the process of applying the method to rotating blades~\citep{shives2013mesh,jha2013accuracy,schluntz2015actuator,jha2018actuator,martinez2019filtered,meyer2019vortex,dag2020new,kleine2022non}. Thus, usually, the simulations are limited to simple geometries such as elliptic~\citep{shives2013mesh,jha2013accuracy,schluntz2015actuator,jha2018actuator,martinez2019filtered,meyer2019vortex} or constant chord planar wings~\citep{jha2013accuracy,dag2020new,martinez2019filtered,meyer2019vortex,kleine2022non}.

One possible explanation for the lack of adoption of this technique for airplane aerodynamics is that, until recently, the ALM misrepresented the forces near the tip of the wings~\citep{jha2013accuracy,dag2020new}. Hence, a tip correction is usually employed in rotor simulations. Since most tip corrections are tuned for rotating blades, the forces on translating wings did not agree with theoretical predictions.

However, this drawback does not seem to be an issue anymore. \citet{dag2020new} (originally in~\citep{dag2017combined}) proposed a vortex-based correction that is tuning-free and makes the results of the ALM similar to that of a numerical lifting line. \citet{dag2020new} observed that the bound vortex created by the actuator line possesses a Gaussian core, similar to a Lamb-Oseen vortex~\citep{oseen1911uber,lamb1932hydrodynamics,saffman1992vortex}. The authors then conjectured that this pattern would be extended to the vortex sheet and proposed a correction to the velocity that accounted for the difference in the velocity induced by a Lamb-Oseen vortex and an ideal vortex (vortex filament with infinitesimal vortex core). This correction was later proved to be based on the mathematical and physical properties of the model by \citet{martinez2019filtered}. They recognized that the Gaussian distribution of vorticity for the bound vortex had been proved by \citet{forsythe2015coupled} and proved that the shed vorticity would also have a similar distribution of vorticity. \citet{martinez2019filtered}(originally in \citep{martinez2017large}) also proposed a similar correction for translating wings, calling it ``subfilter-scale velocity correction'' or ``filtered actuator line model''. \citet{meyer2019vortex} showed that the results of an ALM without correction agree with the results of a vortex method with a Gaussian core, while the results of the ALM with a vortex-based smearing correction agree with the results of a vortex method with ideal vortices.

The vortex-based smearing corrections of~\citet{dag2020new} and~\citet{meyer2019vortex} relied on iterative methods. Recently, \citet{kleine2022non} proposed a non-iterative method, based on the linearization of the equations, that replaces the iterative method with the solution of a small linear system, inspired by the linear lifting line method. Additional contributions of that work include a more accurate calculation of the difference of the velocity induced by a smeared vortex segment, based on analytical integration of the equations of the velocity induced by a vortex with Gaussian core, and the implementation of a free-vortex wake method, that maintains the generality of the method when compared to the implementation with a prescribed vortex sheet.

All the mentioned smearing corrections mentioned differ slightly on implementation choices, but the corrections are based on the same general idea. For the details on the implementations, the readers are referred to each original work, but~\citep{kleine2022non} also provides some comparison between previous works~\citep{dag2020new,martinez2019filtered,meyer2019vortex}. In the present study and our previous work~\citep{kleine2022non} we focus on the accuracy and generality of the correction, but if computational cost is favored, \citet{meyer2020brief} presents other implementation choices that reduce the computational time for rotating blades.

In the present work, we use the term ``vortex-based smearing correction'' or simply ``smearing correction'', a nomenclature popularized by the works of \citet{meyer2019vortex,meyer2019wake,meyer2020brief}. The works of \citet{dag2020new} and \citet{meyer2019vortex} compared the results of translating wings to the results of the lifting line method, showing good agreement. Nevertheless, \citet{kleine2022non} improved on that agreement by constructing a lifting line method and an actuator line method that are compatible with each other. The results of the ALM with vortex-based smearing correction were observed to be equal to the results of a non-linear lifting line method within the limits of a first-order method, for a planar wing with constant chord. This means that, for an induced velocity in the order of $10^{-2}$ of the freestream velocity, the differences were observed to be in the order of $10^{-4}$. Hence, for practical purposes, the ALM with smearing correction can be considered equivalent to a lifting line method that can be integrated into a Navier-Stokes solver.

In summary, the ALM enables the integration of lifting lines with a CFD (Computational Fluid Dynamics) solver with viscous effects, which is not allowed by the classical lifting line, allowing the study of the interaction of lifting lines with more complex flows. Also, it generally does not suffer from the instabilities of traditional free-vortex wake methods. Nevertheless, the main advantage is that it allows grids that are coarser than traditional CFD methods that model the blades as wall boundary conditions. Similar to immersed boundary methods, it allows the use of simple grids for complex geometries or moving lifting surfaces, with the advantage of allowing a coarser discretization than immersed boundary methods. Thus, the ALM can be considered a viable alternative for low and middle-fidelity simulations of wings.

Due to the recent advancements and the remarkable agreement with the lifting line method, there are reasons to believe that the ALM with smearing correction achieved a level of accuracy sufficient to be used to simulate lifting surfaces of fixed-wing aircraft. This is the main concept explored in this work. Some lifting surface configurations typical of fixed-wing aircraft are simulated with the ALM and compared to a non-linear iterative lifting line method. We focus on two complex geometries commonplace in airplanes: a wing with winglets and a T-tail configuration. These configurations have not been studied in the past, due to not being common in rotating blades and the known limitations of the ALM before the conception of the smearing correction.

Initially, for these non-planar wings, our results using the ALM with smearing correction presented higher-than-expected errors. Hence, a second-order correction is proposed, which is necessary for geometries such as wings with winglets and T-tail configurations. This second-order correction is also based on the same ideas that led to the design of the vortex-based smearing correction: the vorticity created by smeared body forces. As explained in section \ref{sec:alm}, this correction is applied to the body force to correct the magnitude of the vorticity generated in the numerical simulation. 

Regarding the traditional application of the ALM, to wind turbines and rotors in general, the vorticity magnitude correction would also be useful in simulations of rotating blades with winglets or other non-planar wingtip devices. These configurations have been drawing more attention for their possible increase in efficiency and performance~\citep{hansen2018winglet,farhan2019numerical,khaled2019investigation} of a single turbine or as a mechanism to change the behavior of the near or far wake~\citep{muhle2020experimental,schroder2021experiments,schroder2022experimental}, which could improve the power production of wind farms or reduce vortex-induced vibrations and noise. 

The vorticity magnitude correction is presented in section \ref{sec:alm}, together with a brief introduction to the actuator line method with smearing correction. The numerical method used in the simulations is described in section \ref{sec:numericalmethods} and the results are shown in section \ref{sec:results}. Finally, section \ref{sec:conclusions} contains the main conclusions of the work.

\section{The actuator line method} \label{sec:alm}

In this section, a short introduction to the actuator line method (ALM) with vortex-based smearing correction is provided. For a more detailed description of the method used in this work and its theoretical background, the reader is referred to~\citep{kleine2022non,kleine2022onstability}.

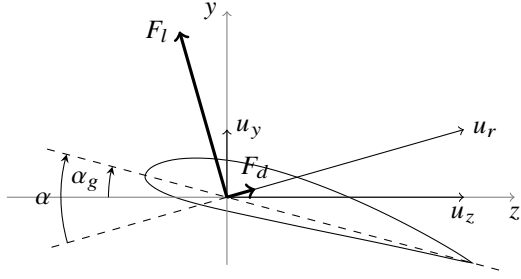
\begin{figure}
  \centering
  \begin{tikzpicture}
  \def\c{4.5}
  \def\r{0.35*\c}
  \draw[scale=\c,rotate=-15] plot file{naca4415.dat} -- cycle;
  \draw[rotate=-15,dashed] (-0.55*\c,0)--(0.85*\c,0);
  \draw[->,gray] (-0.65*\c,0)--(0.85*\c,0) node[below,black]{$z$};
  \draw[->,gray] (0,-0.2*\c)--(0,0.55*\c) node[left,black]{$y$};
  \draw[->] (0,0)--(0.7*\c,0) node[below]{$u_z$};
  \draw[->] (0,0)--(0,0.2*\c) node[right]{$u_y$};
  \draw[->] (0,0)--(0.7*\c,0.2*\c) node[right]{$u_{r}$};
  \draw[dashed] (0,0)--(-0.7*0.75*\c,-0.2*0.75*\c);
  \draw[->,very thick] (0,0)--(0.7*0.12*\c,0.2*0.12*\c) node[above]{$F_{d}$};
  \draw[->,very thick] (0,0)--(-0.2*0.7*\c,0.7*0.7*\c) node[left]{$F_{l}$};
  \draw[-stealth] (0,0) +(180:\r) arc(180:165:\r) node[left,midway]{$\alpha_g$};
  \draw[-stealth] (0,0) +(195.945:1.4*\r) arc(195.945:165:1.4*\r) node[left,midway]{$\alpha$};
\end{tikzpicture}
  \caption{Local reference system defined by a cross-section of the wing.}
  \label{fig:2dref}
\end{figure}

The incompressible Navier-Stokes equation written in primitive variables (pressure $p$ and velocity $\mathbf{u}$) are~\citep{mikkelsen2003actuator,troldborg2009actuator}:
\begin{equation}
  \frac{\partial \mathbf{u}}{\partial t} + \mathbf{u} \cdot \nabla \mathbf{u} = -\frac{1}{\rho} \nabla p + \nu \nabla^2 \mathbf{u} + \mathbf{f}
\end{equation}
where $\rho$ and $\nu$ are the density and kinematic viscosity of the fluid, respectively. The body force term, $\mathbf{f}$, in the case of the actuator line method, is used to model the turbine~\citep{sorensen2002numerical,mikkelsen2003actuator,troldborg2009actuator}. The body forces are based on the two-dimensional force per spanwise unit length, $\mathbf{F}_{2D}$, given by
\begin{equation}
  \mathbf{F}_{2D} = (F_l,F_d) = \left( \frac{1}{2}\rho \, u_{r}^2 \, c \, C_l,\frac{1}{2}\rho \, u_{r}^2 \, c \, C_d \right)
\end{equation}
where $F_l$ and $F_d$ are the lift and drag forces (lift is perpendicular to the relative velocity and drag is parallel to the relative velocity, see figure \ref{fig:2dref}), calculated from the relative velocity $u_{r}$, the local chord $c$ and the two-dimensional lift and drag coefficients, $C_l$ and $C_d$, which are obtained from the airfoil data at the local Reynolds number and angle of attack $\alpha$ (calculated using the local relative velocity).

In the ALM, to avoid numerical problems related to singularities, the forces need to be distributed smoothly on several mesh points. The forces are usually projected into the grid by the convolution with a regularization kernel which is usually a three-dimensional Gaussian kernel ($\eta_3$) with the same constant Gaussian width, $\varepsilon$, in the three directions
\begin{equation}
  \label{eq:eta3d}
  \eta_3(x,y,z) \defeq \frac{1}{\pi^{3/2}\varepsilon^3} \exp{\left(-\frac{x^2+y^2+z^2}{\varepsilon^2}\right)} .
\end{equation}
Non-uniform and anisotropic kernels have also been proposed~\citep{mikkelsen2003actuator,shives2013mesh,martinez2017optimal,churchfield2017advanced,jha2018actuator,cormier2021evaluation}, but these are not investigated in the present work. The body force term, $\mathbf{f}$, is then calculated as
\begin{equation}
  \mathbf{f} = \left( \frac{1}{\rho} \mathbf{F}_{2D} \delta(y)\delta(z) \right) * \eta_3
\end{equation}
where $*$ indicates the convolution of the two functions and $\delta$ is the Dirac delta function. This can be written as
\begin{equation}
  \label{eq:bodyforce}
  \mathbf{f} = \left( \frac{1}{\rho} \mathbf{F}_{2D} * \eta \right)(x) \, \eta(y) \, \eta(z) 
\end{equation}
where $x$ is the local spanwise direction and $\eta$ is the one-dimensional Gaussian function
\begin{equation}
  \eta(x) \defeq \frac{1}{\pi^{1/2}\varepsilon} \exp{\left(-\frac{x^2}{\varepsilon^2}\right)} .
\end{equation}
If the force $\mathbf{F}_{2D}$ is considered constant in at each spanwise section, an analytical formula for the convolution is available, which is provided in~\citep{kleine2022non}.

However, the smearing of the forces by the convolution operation makes the vorticity generated in the numerical simulation to also be spread. As a consequence, the velocity induced by this smeared vorticity would present errors, that would be carried into the force calculations. Historically, ad-hoc or semi-empirical tip corrections were used to correct the forces or velocities. Nevertheless, the recently derived vortex-based smearing correction~\citep{dag2020new,martinez2019filtered,meyer2019vortex,kleine2022non} is a correction that is based on the mathematical and physical properties of the simulation that corrects the velocity used to calculate the forces.

In the smearing correction, a corrected velocity $\mathbf{u}^{c}$ is calculated from the velocity sampled from the numerical simulations $\mathbf{u}^{s}$ as
\begin{equation}
  \mathbf{u}^{c} = \mathbf{u}^{s} + \mathbf{u}^{m}
\end{equation}
where $\mathbf{u}^{m}$, termed ``missing velocity'', is a velocity defined as the difference of the velocity induced by ideal vortices, $\mathbf{u}^{vi}$, and the velocity induced by vortices with a Gaussian distribution of vorticity, $\mathbf{u}^{v}$:
\begin{equation}
  \mathbf{u}^{m} = \mathbf{u}^{vi} - \mathbf{u}^{v} .
\end{equation}

The missing velocity is obtained by creating a vortex-based method that runs in conjunction with the CFD simulation. Previous works differed slightly in this vortex-based method, which are briefly compared in~\citep{kleine2022non}. In the present work, we use the non-iterative method with the free-vortex wake method described in~\citep{kleine2022non} (unless otherwise stated). In this method, the vortex sheet is formed by following passive particles that are advected with $\mathbf{u}^{s}$, the circulation is found by solving a small linear system inspired by the lifting line method and the missing velocity is calculated from the multiplication of an influence matrix and the vector of circulation. The details are provided in~\citep{kleine2022non}.

A topic that is not addressed in previous works is the magnitude of the circulation generated in the CFD simulation, which is treated in section \ref{sec:magnitudecorrection}

\subsection{Vorticity magnitude correction} \label{sec:magnitudecorrection}
One of the principal aspects of the smearing correction is that it acts on the velocity that is used to calculate the forces, but does not modify the velocity in the CFD simulation directly. Also, it does not correct the vorticity created by the body forces. In other words, the vorticity created by the actuator line is smeared and the velocity induced by the vorticity in the simulation is still affected by this smearing. Hence, when evaluating the velocity at the control point, two velocities are defined: $\mathbf{u}^s$ which is the velocity sampled from the simulation and $\mathbf{u}^{c}$ which is the corrected velocity used to calculate the forces. In most cases, the magnitude of these velocities is approximately the same. Because of this, previous works did not discuss in depth the ambiguity in the calculation of forces and application of these forces to the CFD simulation.

However, for non-planar wings, especially in the case of wings with winglets or near the intersection of a horizontal and a vertical tail, the bound vortex of one surface induces a velocity on the other surface that may be relevant compared to the freestream velocity. This induced velocity is not well represented in the calculated velocity field, $\mathbf{u}^{s}$, being considered almost only in the missing velocity, $\mathbf{u}^{m}$. For this case, the difference between $\mathbf{u}^s$ and $\mathbf{u}^{c}$ may not be negligible. Therefore, it is important to resolve this ambiguity. 

The corrected lift force is calculated from the corrected velocity $F_l^c=\sfrac{1}{2} \rho u_{r}^{c \, 2} c C_l$. The corrected circulation, which is used by the smearing correction, is calculated using the Kutta-Joukowski theorem, $F_l^c=\rho u_{r}^c \Gamma^c$. The numerical solver, however, does not know the corrected velocity, the only velocity known by the numerical solver is $\mathbf{u}^s$. The lift force imposed as body force, $F_l^s$, is usually the corrected force. Imposing a lift force $F_l^s=F_l^c=\rho u_{r}^c \Gamma^c$ would result in the creation of circulation equal to 
\begin{equation}
  \label{eq:vortx_corr}
  \Gamma^s = \frac{F_l^s}{\rho u_{r}^s} = \frac{u_{r}^c}{u_{r}^s} \Gamma^c .
\end{equation}
Thus, in the standard vortex-based smearing correction, the circulation created in the CFD simulation is different from the circulation calculated from the corrected velocity and the corrected lift force.

The reason for this difference is simple, the Kutta-Joukowski theorem binds the velocity, the lift force and the circulation. Since the velocity in the numerical simulation is different from the corrected velocity, either the circulation or the lift coefficient must be different. By imposing the same lift coefficient, the result is a difference in the circulation created in the numerical simulation. This difference is of first order with respect to the ratio $u^m/u^s$, but causes second-order errors in the corrected circulation and corrected velocity, when compared to the lifting line method. The estimate of the order of these errors are in appendix \ref{app:errorvelocity}.

Hence, the smearing correction corrects for the errors of the induced velocity caused by the distribution of the vorticity but does not correct for the errors related to the magnitude of the vorticity generated. To have the corrected circulation in the numerical simulation, a second-order correction is proposed by calculating two different lift forces. Instead of imposing the corrected lift force, $F_l^c$, as body force in the numerical simulation, a different lift force is calculated as
\begin{equation}
  \label{eq:lift_corr}
  F_l^s = \rho u_{r}^s \Gamma^c = \frac{u_{r}^s}{u_{r}^c} F_l^c ,
\end{equation}
so that $\Gamma^s = \Gamma^c$. Hence, the body force $\mathbf{f}$ is obtained from equation \eqref{eq:bodyforce}, using the following definition of $\mathbf{F}_{2D}$:
\begin{equation}
  \mathbf{F}_{2D}^s = (F_l^s,F_d) = \left( \frac{1}{2}\rho \, u_{r}^s u_{r}^c \, c \, C_l,\frac{1}{2}\rho \, u_{r}^{c \, 2} \, c \, C_d \right)
\end{equation}
with $C_l$ and $C_d$ being obtained with the angle of attack defined by $u_{r}^{c}$, which is also the velocity used to define the direction of the lift and drag forces. It is worth noting that, at this moment, the correction is not applied to the drag force. More studies are needed to understand the impact of the smearing of the drag force and the effect of the ambiguity of velocities in the definition of the drag force. Since drag is usually less relevant than the lift, eventual errors related to the drag force are usually neglected. Similarly to previous studies~\citep{dag2020new,meyer2019vortex,martinez2019filtered,kleine2022non}, in this work we focus only in the vorticity created by the lift forces.

By imposing $\mathbf{F}_{2D}^s$ as body force, the corrected total vorticity due to lift would be generated inside the CFD simulation, even though there are two distinct values of velocity ($\mathbf{u}^s$ and $\mathbf{u}^c$) and lift forces ($F_l^s$ and $F_l^c$) for each control point. This vorticity magnitude correction is then used as an additional step at the end of the standard vortex-based smearing correction. If an iterative smearing correction is used, this step is performed outside the iteration.

The force used to calculate the properties of the configuration is the corrected force. So, to calculate the lift coefficient of a wing, for example, $F_l^c$ should be used. In the case of a rotor, $F_l^c$ is used to calculate the power and thrust coefficients.

The cases analyzed by \cite{dag2020new,martinez2019filtered,meyer2019vortex,kleine2022non} would not be greatly affected by the differences discussed here, since $\mathbf{u}^s \approx \mathbf{u}^c$. However, there are configurations in which the difference in these velocities affects the results, such as a wing with winglets and a T-tail configuration. As shown in the appendix and in section \ref{sec:results}, the induced velocities, the circulation and the forces are all affected if the vorticity magnitude correction is not used, since these are directly connected.

The cost of this correction is negligible. It just requires the computation and storage of one extra small vector of size $N$, where $N$ is the number of control points. Therefore, we recommend the application of this correction for all cases, also for the cases where the difference is minimal, such as simulations of conventional wind turbines.

\subsection{Numerical method and simulation parameters} \label{sec:numericalmethods}

The numerical method, domain size and discretization employed in this work are the same as the ones used for the straight wing in uniform flow of~\citep{kleine2022non}. The actuator line method is implemented in the spectral-element code \texttt{Nek5000}~\citep{fischer2008nek5000}. In each spectral element, seventh-order Lagrange polynomials on Gauss-Lobatto-Legendre quadrature points are used for spatial discretization and a third-order implicit/explicit scheme is applied for temporal discretization. The $\mathbb{P}_N-\mathbb{P}_{N-2}$ formulation~\citep{maday1989spectral} is employed. Filtering of the higher modes is applied to stabilize the simulation~\citep{fischer2001filter}.

The choice of a spectral-element code is due to the intended use of this solver for vortex stability studies. Due to its low dissipation and dispersion, this code was previously used in several vortex stability studies~\citep{kleusberg2019wind,kleusberg2019tip,kleine2019tip,kleine2022stability}, with a different option of tip correction.

To reduce the computational cost, an adaptive mesh refinement (AMR) strategy with a spectral error indicator~\citep{offermans2019aspects,offermans2020adaptive,tanarro2020enabling} is employed. Maximum discretization is enforced around the actuator lines, in order to guarantee adequate and constant discretization for the chosen smearing parameter. Dirichlet boundary conditions are used for the inflow, upper, lower and lateral boundary conditions. The natural outflow boundary condition is imposed at the outlet.

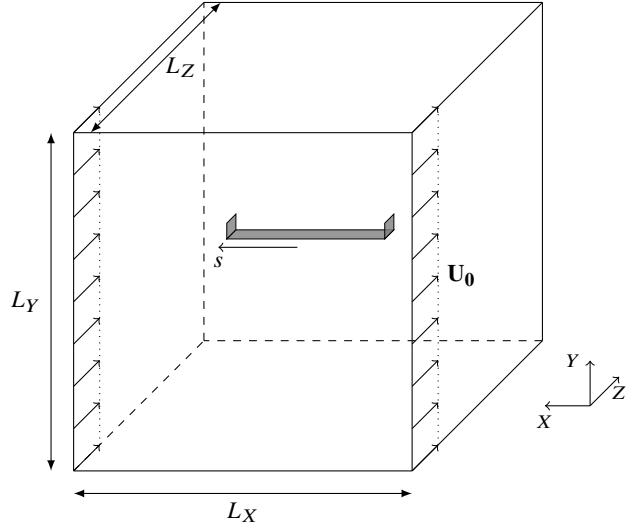
\begin{figure}
  \centering
  \begin{tikzpicture}
  \def\L{3.0}
  \def\lx{-1.5}
  \def\ly{1.5}
  \def\lz{-1.5}
  \def\lzm{-0.75}
  \def\rn{0.35}
  \def\Lx{\lx*\L}
  \def\Ly{\ly*\L}
  \def\Lz{\lz*\L}
  \def\Lzm{\lzm*\L}
  \def\R{\rn*\L}
  \def\Vd{0.4*\Lzm}
  \def\Vu{0.4*\Lzm}
  \def\xV{0.0*\Lx}
  \def\xVl{1.0*\Lx}
  \def\chord{0.3*\R}
  \def\cm{0.25*\chord}
  \def\cp{-0.75*\chord}
  
  \draw[-] (0,0,0)--(\Lx,0,0);
  \draw[-] (\Lx,0,0)--(\Lx,\Ly,0);
  \draw[-] (\Lx,\Ly,0)--(0,\Ly,0);
  \draw[-] (0,\Ly,0)--(0,0,0);
  \draw[dashed] (0,0,\Lz)--(\Lx,0,\Lz);
  \draw[dashed] (\Lx,0,\Lz)--(\Lx,\Ly,\Lz);
  \draw[-] (\Lx,\Ly,\Lz)--(0,\Ly,\Lz);
  \draw[-] (0,\Ly,\Lz)--(0,0,\Lz);
  \draw[-] (0,0,0)--(0,0,\Lz);
  \draw[dashed] (\Lx,0,0)--(\Lx,0,\Lz);
  \draw[-] (\Lx,\Ly,0)--(\Lx,\Ly,\Lz);
  \draw[-] (0,\Ly,0)--(0,\Ly,\Lz);
  
  \draw[very thin,fill=gray,fill opacity=0.8] (0.5*\Lx-\R,0.5*\Ly,\Lzm+\cm) -- (0.5*\Lx+\R,0.5*\Ly,\Lzm+\cm) -- (0.5*\Lx+\R,0.5*\Ly,\Lzm+\cp) -- (0.5*\Lx-\R,0.5*\Ly,\Lzm+\cp) -- cycle;
  \draw[very thin,fill=gray,fill opacity=0.8] (0.5*\Lx-\R,0.5*\Ly,\Lzm+\cm) -- (0.5*\Lx-\R,0.5*\Ly+0.2*\R,\Lzm+\cm) -- (0.5*\Lx-\R,0.5*\Ly+0.2*\R,\Lzm+\cp) -- (0.5*\Lx-\R,0.5*\Ly,\Lzm+\cp) -- cycle;
  \draw[very thin,fill=gray,fill opacity=0.8] (0.5*\Lx+\R,0.5*\Ly,\Lzm+\cm) -- (0.5*\Lx+\R,0.5*\Ly+0.2*\R,\Lzm+\cm) -- (0.5*\Lx+\R,0.5*\Ly+0.2*\R,\Lzm+\cp) -- (0.5*\Lx+\R,0.5*\Ly,\Lzm+\cp) -- cycle;
  \draw[->] (0.5*\Lx,0.5*\Ly,\Lzm+0.35*\R) -- (0.5*\Lx-\R,0.5*\Ly,\Lzm+0.35*\R) node[below]{\small{$s$}};
  
  
  \draw[->] (0.5*\L,0,\Lzm)--++(-\L/5,0,0) node[below]{\scriptsize{$X$}};
  \draw[->] (0.5*\L,0,\Lzm)--++(0,\L/5,0) node[left]{\scriptsize{$Y$}};
  \draw[->] (0.5*\L,0,\Lzm)--++(0,0,-\L/3) node[below]{\scriptsize{$Z$}};
  
  \draw[latex-latex] (0,-\L/10,0)--(\Lx,-\L/10,0) node[midway,below]{\small{$L_X$}};
  \draw[latex-latex] (\Lx-\L/10,0,0)--(\Lx-\L/10,\Ly,0) node[midway,left]{\small{$L_Y$}};
  \draw[latex-latex] (0.95*\Lx,\Ly,0)--(0.95*\Lx,\Ly,\Lz) node[midway,right]{\small{$L_Z$}};
  
  \draw[->] (\xV,0,0)--(\xV,0,\Vd);
  \draw[->] (\xV,0.125*\Ly,0)--(\xV,0.125*\Ly,0.125*\Vu+0.875*\Vd);
  \draw[->] (\xV,0.25*\Ly,0)--(\xV,0.25*\Ly,0.25*\Vu+0.75*\Vd);
  \draw[->] (\xV,0.375*\Ly,0)--(\xV,0.375*\Ly,0.375*\Vu+0.625*\Vd);
  \draw[->] (\xV,0.5*\Ly,0)--(\xV,0.5*\Ly,0.5*\Vu+0.5*\Vd);
  \draw[->] (\xV,0.625*\Ly,0)--(\xV,0.625*\Ly,0.625*\Vu+0.375*\Vd);
  \draw[->] (\xV,0.75*\Ly,0)--(\xV,0.75*\Ly,0.75*\Vu+0.25*\Vd);
  \draw[->] (\xV,0.875*\Ly,0)--(\xV,0.875*\Ly,0.875*\Vu+0.125*\Vd);
  \draw[->] (\xV,\Ly,0)--(\xV,\Ly,\Vu);
  \draw[dotted] (\xV,0,\Vd)--(\xV,\Ly,\Vu) node[midway,right]{\small{$\mathbf{U_{0}}$}};
  
  \draw[->] (\xVl,0,0)--(\xVl,0,\Vd);
  \draw[->] (\xVl,0.125*\Ly,0)--(\xVl,0.125*\Ly,0.125*\Vu+0.875*\Vd);
  \draw[->] (\xVl,0.25*\Ly,0)--(\xVl,0.25*\Ly,0.25*\Vu+0.75*\Vd);
  \draw[->] (\xVl,0.375*\Ly,0)--(\xVl,0.375*\Ly,0.375*\Vu+0.625*\Vd);
  \draw[->] (\xVl,0.5*\Ly,0)--(\xVl,0.5*\Ly,0.5*\Vu+0.5*\Vd);
  \draw[->] (\xVl,0.625*\Ly,0)--(\xVl,0.625*\Ly,0.625*\Vu+0.375*\Vd);
  \draw[->] (\xVl,0.75*\Ly,0)--(\xVl,0.75*\Ly,0.75*\Vu+0.25*\Vd);
  \draw[->] (\xVl,0.875*\Ly,0)--(\xVl,0.875*\Ly,0.875*\Vu+0.125*\Vd);
  \draw[->] (\xVl,\Ly,0)--(\xVl,\Ly,\Vu);
  \draw[dotted] (\xVl,0,\Vd)--(\xVl,\Ly,\Vu);
  
\end{tikzpicture}
  \caption{Reference system and schematic view of the computational domain. The center of the wing is located at the center of the domain, which coincides with the origin of the coordinate system. The dimensions of the domain are $L_X=L_Y=L_Z=12$.}
  \label{fig:domain}
\end{figure}

The global coordinate system is denoted as $(X,Y,Z)$, see figure \ref{fig:domain}, while the local coordinate system of each cross-section is denoted as $(x,y,z)$, in order to differentiate them. For the cases studied, the $Z$-direction coincides with the local $z$-direction for every cross-section, because the sweep angle is null. The inflow velocity is $U_0=1$, in the direction of the global $Z$-direction.

The span of the surfaces is denoted $R$, in order to be consistent with the notation of~\citep{kleine2022non}. The parameters of the simulated geometries are shown in tables \ref{tab:param_winglet} and \ref{tab:param_ttail}. An ideal symmetrical airfoil with $d C_l / d \alpha = 2\pi$ and without drag is used for both configurations, however, it is worth noting that the method allows the use of experimental airfoil data. The angle of attack is chosen so that a unity lift coefficient, $C_{l 0}=1$, would be expected for a two-dimensional simulation (or an infinite wing). The Reynolds number based on the chord is $Re_c=c U_0/\nu = 10^4$ for the wing with winglets and $Re_c=c U_0/\nu = 2 \cdot 10^4$ for the T-tail (Reynolds number based on the span is $Re=R U_0/\nu = 10^5$).

\begin{table}
    \centering
    \begin{tabular}{c c c c c}
        \hline
        Span of main wing, $R$ & Span of winglets, $R_{wl}$  & Chord, $c$ & $\alpha_g$ (rad) & $\alpha_g$ (deg) \\
        $1$ & $0.2$ & $0.1$ & $1/(2\pi)$ & $9.189$ \\
        \hline
    \end{tabular}
    \caption{Parameters of the simulation of the wing with winglets.}
    \label{tab:param_winglet}
\end{table}

\begin{table}
    \centering
    \begin{tabular}{c c c c c}
        \hline
        Span of horizontal tail, $R$ & Span of vertical tail, $R_{v}$  & Chord, $c$ & $\alpha_g$ (rad) & $\alpha_g$ (deg) \\
        $1$ & $0.5$ & $0.2$ & $-1/(2\pi)$ & $-9.189$ \\
        \hline
    \end{tabular}
    \caption{Parameters of the simulation of the T-tail.}
    \label{tab:param_ttail}
\end{table}

Compared to real aircraft wings and tail geometries, these cases are highly idealized. However, compared to cases that usually apply the ALM, the configurations have a high degree of complexity due to the non-planar geometries.

\section{Results and discussion} \label{sec:results}

\begin{figure}
  \centering
  \input{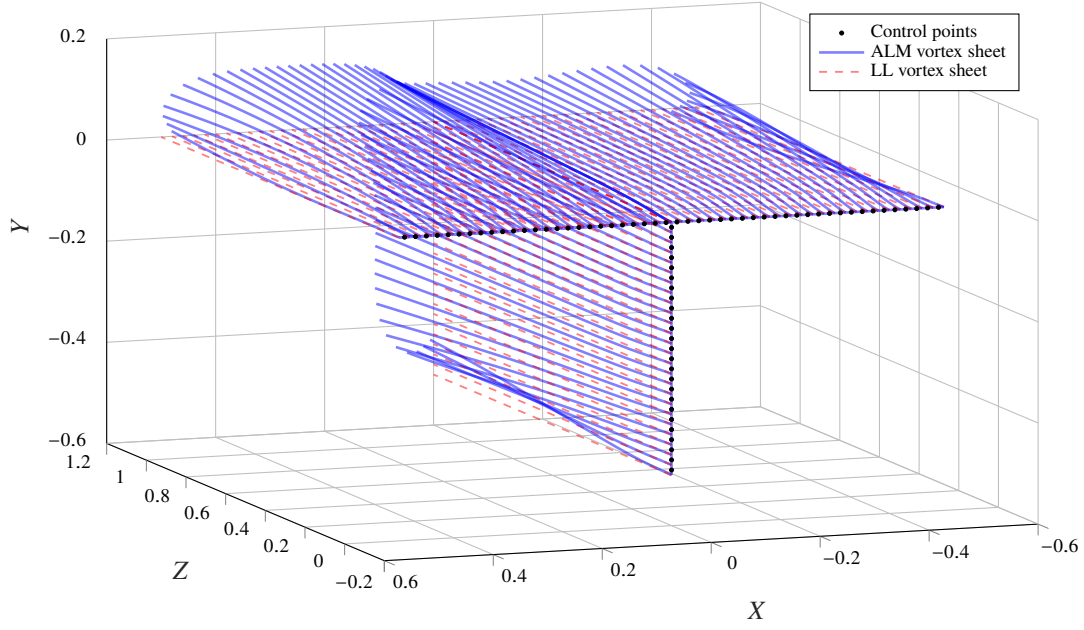}
  \caption{Comparison of the vortex sheet created by the ALM and LL for the T-tail case (ALM$^{VMC}$, $\varepsilon = 3.5 \Delta x$).}
  \label{fig:tail_VSheet}
\end{figure}

The results of the actuator line method with the vorticity magnitude correction (ALM$^{VMC}$) were compared to the results of the ALM without vorticity magnitude correction (ALM$^1$) and the non-linear iterative lifting line method (LL), described in~\citep{kleine2022non}. The same set of control points was used for all methods, with a uniform distance of $0.02$ between control points in the spanwise direction. The control points and the vortex sheet for the T-tail configuration can be seen in figure \ref{fig:tail_VSheet}. The vortex sheet formed by the LL is, by definition, aligned to the undisturbed velocity, while the vortex sheet created by the ALM is formed by passive particles that are advected by the CFD velocity. Hence, as discussed at~\citep{kleine2022non}, some differences are expected due to the different directions of the vortex sheet. Since the cases studied here are more three-dimensional than the planar straight wing studied in \citep{kleine2022non}, the effects of these different directions on the induced velocity are expected to be higher.

Two values of smearing parameter were used for the comparison, $\varepsilon = 3.5 \Delta x = R/16$ and $\varepsilon = 7 \Delta x = R/8$, where $\Delta x = R/56 = 0.01786$ is the average grid spacing in the region of the actuator lines. A value of $\varepsilon = 3.5 \Delta x$ is within the recommended values of the smearing parameter, based on parametric studies of the current implementation of the actuator line~\citep{kleusberg2019wind}, and provides reasonable resolution for the vortex core. For discussion on lower smearing parameters, the reader is referred to~\citep{kleine2022non}.

The velocity differences are normalized with the velocity at infinity (thus the normalized velocity at infinity is $U_0=1$, by construction), while the circulation and lift coefficient differences are normalized with the absolute value of the corresponding quantities for a two-dimensional airfoil in this angle of attack ($C_{lref} = \|C_{l} (\alpha_g)\| = 1$ and $\Gamma_{ref}=0.05$, for the wing with winglets; $\Gamma_{ref}=0.1$, for the T-tail). The lift coefficient $C_{l U_0}^c$ is defined using the velocity at infinity $C_{l U_0}^c =2 F_l^c / (\rho U_{0}^2 c)$, so it is directly proportional to the lift coefficient $F_l^c$, as opposed to the local lift coefficient $C_{l}$ which is defined based on the local velocity. For the configuration of the wing with winglets, the lift coefficient of the winglets is considered positive when it induces a positive lift in the main wing, in other words, when the lift of the winglets is in the direction of $X=0$. The angle of attack and the circulation are considered positive in the direction that makes $C_{l}(\alpha_g)$ positive. The results are shown as a function of the spanwise distance $s$ from the center of the wing. The distances $-0.5 \leq s \leq 0.5$ correspond to the main wing and the distances $0.5 \leq s \leq 0.7$ correspond to the winglet located at $X=0.5$ ($-0.7 \leq s \leq -0.5$ correspond to the winglet located at $X=-0.5$).

\begin{figure}
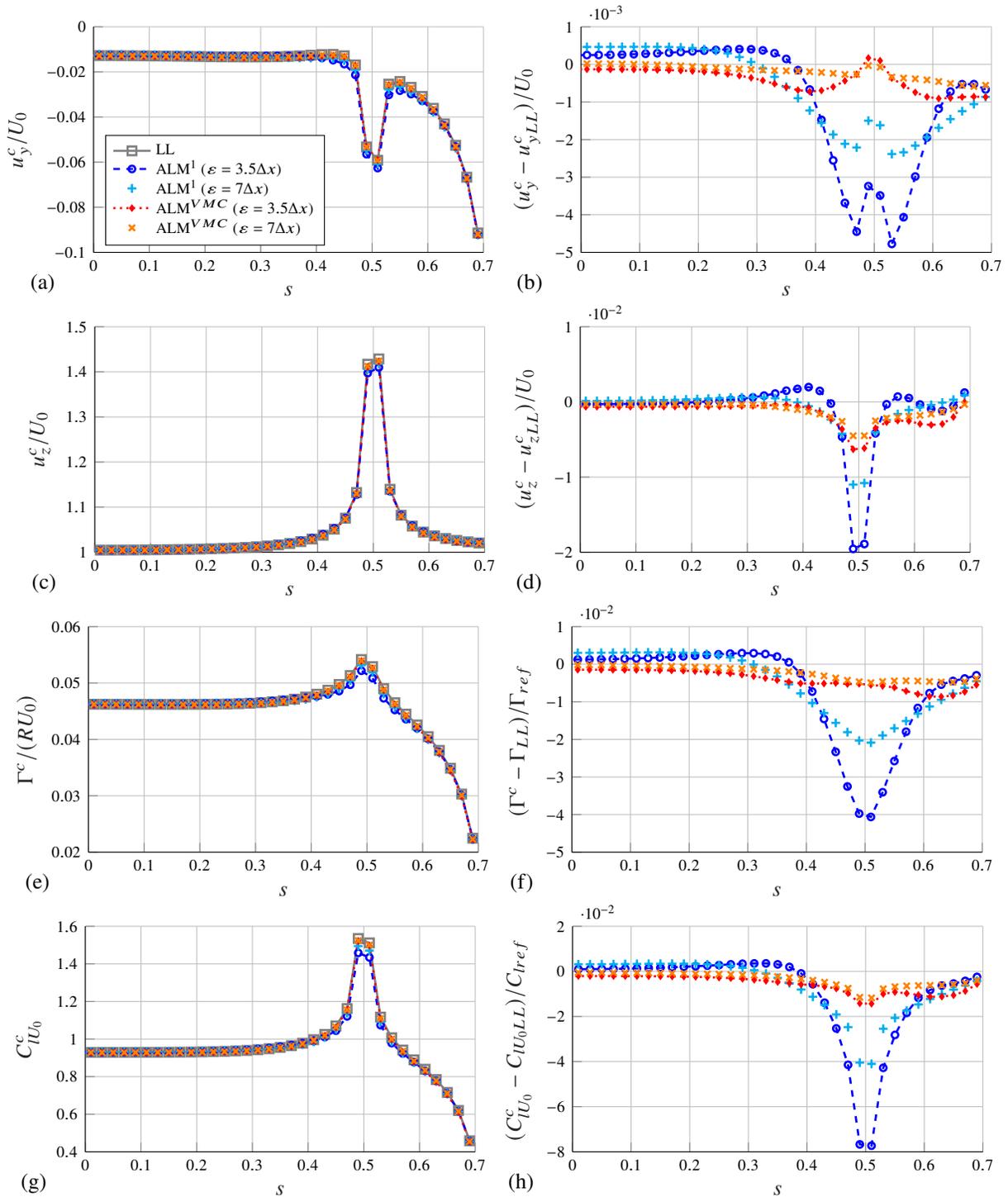

  \centering
  \input{LLACL4uyWL}
%
%
\definecolor{mycolor1}{rgb}{0.00000,1.00000,1.00000}%
\begin{tikzpicture}

\begin{axis}[%
width=0.40\textwidth,
height=0.22\textwidth,
scale only axis,
xmin=0,
xmax=0.7,
xlabel style={font=\color{white!15!black}},
xlabel={$s$},
ymin=-0.005,
ymax=0.001,
ytick={-0.005,-0.004,...,0.001},
ylabel style={font=\color{white!15!black}},
ylabel={$(u^c_y - u^c_{yLL})/U_0$},
axis background/.style={fill=white},
axis x line*=bottom,
axis y line*=left,
xmajorgrids,
ymajorgrids,
tick label style={font=\scriptsize},
legend style={legend cell align=left, align=left, draw=white!15!black}
]
\addplot [color=blue, dashed, line width=1.0pt, mark size=1.5pt, mark=o, mark options={solid, blue}]
  table[row sep=crcr]{%
-0.69	-0.000666112683373368\\
-0.67	-0.000525161905462757\\
-0.65	-0.000527593788176579\\
-0.63	-0.000722229601784953\\
-0.61	-0.00117881312220901\\
-0.59	-0.00194507183376689\\
-0.57	-0.00298268726382958\\
-0.55	-0.00406581695564983\\
-0.53	-0.00477647082787806\\
-0.51	-0.00348721272208741\\
-0.49	-0.00323838700792028\\
-0.47	-0.00444985099893266\\
-0.45	-0.00368511121753577\\
-0.43	-0.00255318300449342\\
-0.41	-0.00147453421765369\\
-0.39	-0.000669631366739464\\
-0.37	-0.000152224019963069\\
-0.35	0.000145741700089946\\
-0.33	0.00029892898698031\\
-0.31	0.000373269273028502\\
-0.29	0.000402462763772891\\
-0.27	0.000405286911674184\\
-0.25	0.000398416986388428\\
-0.23	0.000383877024028074\\
-0.21	0.000366242780413676\\
-0.19	0.000347780978756511\\
-0.17	0.000329343682978978\\
-0.15	0.000312063941184415\\
-0.13	0.000296386741138001\\
-0.11	0.000282912997979792\\
-0.09	0.000271441620911425\\
-0.07	0.00026202321740869\\
-0.05	0.000254952831552142\\
-0.03	0.000250532810357963\\
-0.01	0.000248601968327438\\
0.01	0.000248601399898095\\
0.03	0.000250531315401634\\
0.05	0.000254949339133123\\
0.07	0.000262017171834742\\
0.09	0.000271434437111027\\
0.11	0.000282904228296067\\
0.13	0.000296376459001331\\
0.15	0.000312057580491788\\
0.17	0.000329339718023163\\
0.19	0.000347779819193256\\
0.21	0.00036624361790286\\
0.23	0.000383877471075782\\
0.25	0.000398419004439236\\
0.27	0.000405289923292401\\
0.29	0.000402466175246478\\
0.31	0.000373273632280917\\
0.33	0.000298934109706791\\
0.35	0.000145747226654658\\
0.37	-0.00015221708741031\\
0.39	-0.000669625659301502\\
0.41	-0.0014745304665575\\
0.43	-0.00255318125662033\\
0.45	-0.00368511120035725\\
0.47	-0.00444985188877621\\
0.49	-0.00323838821957952\\
0.51	-0.0034872133593189\\
0.53	-0.00477647133896364\\
0.55	-0.00406581662077212\\
0.57	-0.00298268669613335\\
0.59	-0.00194507090570699\\
0.61	-0.00117881345140314\\
0.63	-0.00072223195494945\\
0.65	-0.000527594627933274\\
0.67	-0.000525162640527072\\
0.69	-0.00066611366174045\\
};
\addlegendentry{ALM$^1$ ($\varepsilon = 3.5 \Delta x$)}

\addplot [color=cyan, line width=1.0pt, only marks, mark size=2.0pt, mark=+, mark options={solid, cyan}]
  table[row sep=crcr]{%
-0.69	-0.000863648539090767\\
-0.67	-0.00105757713121606\\
-0.65	-0.00124382522291118\\
-0.63	-0.00145402456927676\\
-0.61	-0.00168813446281711\\
-0.59	-0.00193181616060219\\
-0.57	-0.00216017531536788\\
-0.55	-0.00233829914619664\\
-0.53	-0.00238431210865926\\
-0.51	-0.00161810379777551\\
-0.49	-0.00149650718307649\\
-0.47	-0.00221101888320756\\
-0.45	-0.00211245002252807\\
-0.43	-0.00186529993821792\\
-0.41	-0.00155707197572119\\
-0.39	-0.00122598695036366\\
-0.37	-0.00089920256440147\\
-0.35	-0.000596600768333054\\
-0.33	-0.00033130575381859\\
-0.31	-0.000110613769761897\\
-0.29	6.35447026818906e-05\\
-0.27	0.000195527084612982\\
-0.25	0.000294388842810727\\
-0.23	0.000361265030952374\\
-0.21	0.000406513618576475\\
-0.19	0.00043565231568931\\
-0.17	0.000453160634715479\\
-0.15	0.000462853528644514\\
-0.13	0.000467467066685601\\
-0.11	0.000468877149776193\\
-0.09	0.000468621175031525\\
-0.07	0.00046736159882059\\
-0.05	0.000465940047370042\\
-0.03	0.000464975442503563\\
-0.01	0.000464573430056339\\
0.01	0.000464572731514396\\
0.03	0.000464985488067234\\
0.05	0.000465956558190723\\
0.07	0.000467380053866043\\
0.09	0.000468642171391127\\
0.11	0.000468902155358766\\
0.13	0.000467493438966331\\
0.15	0.000462871898407387\\
0.17	0.000453173172754664\\
0.19	0.000435662650189357\\
0.21	0.00040652193327716\\
0.23	0.000361266472760682\\
0.25	0.000294379503814536\\
0.27	0.000195518178687801\\
0.29	6.35304429088789e-05\\
0.31	-0.000110629619644583\\
0.33	-0.000331324809074709\\
0.35	-0.000596625973390342\\
0.37	-0.00089923197261391\\
0.39	-0.0012260121526135\\
0.41	-0.0015570942732189\\
0.43	-0.00186531671764543\\
0.45	-0.00211246178633005\\
0.47	-0.00221102579019451\\
0.49	-0.00149650988516811\\
0.51	-0.00161809845347421\\
0.53	-0.00238430836737964\\
0.55	-0.00233829987501512\\
0.57	-0.00216018278467685\\
0.59	-0.00193183186549089\\
0.61	-0.00168815708167814\\
0.63	-0.00145405054063424\\
0.65	-0.00124385126688198\\
0.67	-0.00105760089947207\\
0.69	-0.000863665587617052\\
};
\addlegendentry{ALM$^1$ ($\varepsilon = 7 \Delta x$)}

\addplot [color=red, dotted, line width=1.0pt, mark size=1.0pt, mark=diamond, mark options={solid, red}]
  table[row sep=crcr]{%
-0.69	-0.000866422092229771\\
-0.67	-0.000857731048556562\\
-0.65	-0.000868233151596683\\
-0.63	-0.000894649269572556\\
-0.61	-0.00090804504848091\\
-0.59	-0.000863865683313796\\
-0.57	-0.000744285722701181\\
-0.55	-0.000557591454684335\\
-0.53	-0.000364894220541163\\
-0.51	9.95384680580816e-05\\
-0.49	0.00017084714095341\\
-0.47	-0.000262234801349064\\
-0.45	-0.000437239293957869\\
-0.43	-0.00059814938784712\\
-0.41	-0.00069486692843039\\
-0.39	-0.000723150846671064\\
-0.37	-0.000683764540196669\\
-0.35	-0.000603792799486955\\
-0.33	-0.000514549369774891\\
-0.31	-0.000428338709273998\\
-0.29	-0.000355753106560209\\
-0.27	-0.000299977804117417\\
-0.25	-0.000254392621409572\\
-0.23	-0.000220665665060526\\
-0.21	-0.000195352192405325\\
-0.19	-0.00017644417453049\\
-0.17	-0.000162803360822621\\
-0.15	-0.000152879730587585\\
-0.13	-0.000145771369456899\\
-0.11	-0.000140392123802907\\
-0.09	-0.000136628028385174\\
-0.07	-0.00013416182513671\\
-0.05	-0.000132527194190058\\
-0.03	-0.000131292943805737\\
-0.01	-0.000130471031306562\\
0.01	-0.000130473594751405\\
0.03	-0.000131297722936467\\
0.05	-0.000132534224377976\\
0.07	-0.000134173151908256\\
0.09	-0.000136644741125173\\
0.11	-0.000140411383428534\\
0.13	-0.000145788678379768\\
0.15	-0.000152898438362912\\
0.17	-0.000162818103921537\\
0.19	-0.000176453971301444\\
0.21	-0.00019535677677514\\
0.23	-0.000220665133714817\\
0.25	-0.000254391232684763\\
0.27	-0.000299972256090499\\
0.29	-0.000355745266248521\\
0.31	-0.000428327781908284\\
0.33	-0.000514534059413209\\
0.35	-0.000603774678216942\\
0.37	-0.000683749710561109\\
0.39	-0.000723136945600503\\
0.41	-0.000694856393572399\\
0.43	-0.000598142528808136\\
0.45	-0.000437234007276154\\
0.47	-0.000262230547523713\\
0.49	0.000170849752934486\\
0.51	9.95408982877935e-05\\
0.53	-0.00036489545466694\\
0.55	-0.000557595044546624\\
0.57	-0.000744289738414956\\
0.59	-0.000863870379411786\\
0.61	-0.000908050038018741\\
0.63	-0.000894652864895346\\
0.65	-0.000868237900947179\\
0.67	-0.000857736096412764\\
0.69	-0.000866426779194157\\
};
\addlegendentry{ALM$^2$ ($\varepsilon = 3.5 \Delta x$)}

\addplot [color=orange, line width=1.0pt, only marks, mark size=2.0pt, mark=x, mark options={solid, orange}]
  table[row sep=crcr]{%
-0.69	-0.000555178467894671\\
-0.67	-0.000586840461026555\\
-0.65	-0.000556360163349284\\
-0.63	-0.000506560212129754\\
-0.61	-0.000455389422236309\\
-0.59	-0.000412843800521295\\
-0.57	-0.000382726420343579\\
-0.55	-0.000364159413468134\\
-0.53	-0.000336193498262964\\
-0.51	-7.3649001286917e-05\\
-0.49	-3.15778076675904e-05\\
-0.47	-0.000266355265934964\\
-0.45	-0.000266079370724969\\
-0.43	-0.00023915901602232\\
-0.41	-0.000213576893635391\\
-0.39	-0.000194457454793264\\
-0.37	-0.000181116670551269\\
-0.35	-0.000170673383620553\\
-0.33	-0.00016037245148539\\
-0.31	-0.000149073290737498\\
-0.29	-0.000136857386317009\\
-0.27	-0.000122876975304017\\
-0.25	-0.000104770076849472\\
-0.23	-8.84556868133261e-05\\
-0.21	-7.14763838911246e-05\\
-0.19	-5.49447144190893e-05\\
-0.17	-3.96643041918211e-05\\
-0.15	-2.59396179736852e-05\\
-0.13	-1.39881196998988e-05\\
-0.11	-3.968629279507e-06\\
-0.09	4.27789247192494e-06\\
-0.07	1.05749055180904e-05\\
-0.05	1.51759445587429e-05\\
-0.03	1.83468570278626e-05\\
-0.01	2.00310382403399e-05\\
0.01	2.00307393911948e-05\\
0.03	1.83501596217328e-05\\
0.05	1.5185684660024e-05\\
0.07	1.05876458154439e-05\\
0.09	4.28962015582707e-06\\
0.11	-3.95348683263422e-06\\
0.13	-1.39652585557683e-05\\
0.15	-2.5921009822712e-05\\
0.17	-3.96445285617362e-05\\
0.19	-5.49264457667432e-05\\
0.21	-7.14651059346393e-05\\
0.23	-8.84515286192177e-05\\
0.25	-0.000104771471958663\\
0.27	-0.0001228758017475\\
0.29	-0.000136857130253522\\
0.31	-0.000149077478651383\\
0.33	-0.000160378520201409\\
0.35	-0.000170676085914942\\
0.37	-0.000181125565266609\\
0.39	-0.000194468359095203\\
0.41	-0.000213582682478399\\
0.43	-0.000239163486498136\\
0.45	-0.000266081677472654\\
0.47	-0.000266351207721012\\
0.49	-3.15720769285127e-05\\
0.51	-7.36359523626079e-05\\
0.53	-0.00033617366868384\\
0.55	-0.000364134981026425\\
0.57	-0.000382699218171755\\
0.59	-0.000412813950975383\\
0.61	-0.000455355671022742\\
0.63	-0.000506522767584144\\
0.65	-0.000556327490818978\\
0.67	-0.000586812164146563\\
0.69	-0.000555157998347752\\
};
\addlegendentry{ALM$^2$ ($\varepsilon = 7 \Delta x$)}

\legend{}
\end{axis}
\node [left] at (-0.5,-0.5) {(b)};
\end{tikzpicture}%
  \input{LLACL4uzWL}
%
%
\definecolor{mycolor1}{rgb}{0.00000,1.00000,1.00000}%
\begin{tikzpicture}

\begin{axis}[%
width=0.38\textwidth,
height=0.22\textwidth,
scale only axis,
xmin=0,
xmax=0.7,
xlabel style={font=\color{white!15!black}},
xlabel={$s$},
ymin=-0.02,
ymax=0.01,
ytick={-0.02,-0.01,...,0.01},
ylabel style={font=\color{white!15!black}},
ylabel={$(u^c_z - u^c_{zLL})/U_0$},
axis background/.style={fill=white},
axis x line*=bottom,
axis y line*=left,
xmajorgrids,
ymajorgrids,
tick label style={font=\scriptsize},
legend style={legend cell align=left, align=left, draw=white!15!black}
]
\addplot [color=blue, dashed, line width=1.0pt, mark size=1.5pt, mark=o, mark options={solid, blue}]
  table[row sep=crcr]{%
-0.69	0.00124017458127041\\
-0.67	-0.000538809918849172\\
-0.65	-0.0011909266052336\\
-0.63	-0.000940849222559052\\
-0.61	-0.000346064720968903\\
-0.59	0.00051943904686997\\
-0.57	0.000703934127922301\\
-0.55	-0.000314333106730444\\
-0.53	-0.0041713812592245\\
-0.51	-0.0189059015646746\\
-0.49	-0.0195494757160802\\
-0.47	-0.00464429607757783\\
-0.45	-0.000215967639997344\\
-0.43	0.00146518049226058\\
-0.41	0.00196787625964812\\
-0.39	0.00172730394511195\\
-0.37	0.00149065786010578\\
-0.35	0.00111128657691424\\
-0.33	0.000838970120167554\\
-0.31	0.000590994250773564\\
-0.29	0.000419685355166849\\
-0.27	0.000277405920080675\\
-0.25	0.000139067150903272\\
-0.23	5.99867233102482e-05\\
-0.21	-2.00337138267608e-05\\
-0.19	-8.46975817967721e-05\\
-0.17	-0.000137059764178068\\
-0.15	-0.000179205056186981\\
-0.13	-0.000213454377980927\\
-0.11	-0.000241029286568679\\
-0.09	-0.000262397593832259\\
-0.07	-0.000277630834203678\\
-0.05	-0.000288456934578548\\
-0.03	-0.00029668710630666\\
-0.01	-0.00030154789286583\\
0.01	-0.000301544879814465\\
0.03	-0.000296688875098884\\
0.05	-0.000288459666322047\\
0.07	-0.000277633716156799\\
0.09	-0.000262408324040761\\
0.11	-0.000241040378183861\\
0.13	-0.000213453446083259\\
0.15	-0.000179213631083335\\
0.17	-0.000137062995145125\\
0.19	-8.47013294247665e-05\\
0.21	-2.00386367432831e-05\\
0.23	5.99927812885807e-05\\
0.25	0.00013907247296539\\
0.27	0.00027740838768945\\
0.29	0.00041969312581585\\
0.31	0.00059100415546706\\
0.33	0.000838982687256896\\
0.35	0.00111130297448086\\
0.37	0.00149066826786774\\
0.39	0.00172731505279668\\
0.41	0.0019678801398309\\
0.43	0.0014651795510813\\
0.45	-0.000215970472295934\\
0.47	-0.00464430113949721\\
0.49	-0.0195494841763451\\
0.51	-0.0189059066511973\\
0.53	-0.00417137906198833\\
0.55	-0.000314325687875108\\
0.57	0.000703944757121905\\
0.59	0.000519446929219319\\
0.61	-0.000346060317093687\\
0.63	-0.000940842438559741\\
0.65	-0.00119092939373285\\
0.67	-0.00053881416024797\\
0.69	0.00124017165769211\\
};
\addlegendentry{ALM$^1$ ($\varepsilon = 3.5 \Delta x$)}

\addplot [color=cyan, line width=1.0pt, only marks, mark size=2.0pt, mark=+, mark options={solid, cyan}]
  table[row sep=crcr]{%
-0.69	0.000880458771732769\\
-0.67	0.000299214895960707\\
-0.65	-8.45584466879039e-05\\
-0.63	-0.000395619752225188\\
-0.61	-0.000706982112349902\\
-0.59	-0.00107897316098254\\
-0.57	-0.00159813016016902\\
-0.55	-0.00241181350294704\\
-0.53	-0.00401992156694647\\
-0.51	-0.0107959296149622\\
-0.49	-0.0109971140315325\\
-0.47	-0.00421342002722674\\
-0.45	-0.00234705486217182\\
-0.43	-0.00128541845505165\\
-0.41	-0.000562622385090188\\
-0.39	-5.90650132122339e-05\\
-0.37	0.000281066179836545\\
-0.35	0.000488732850270027\\
-0.33	0.000593958961037439\\
-0.31	0.00062439622565001\\
-0.29	0.000609906624677475\\
-0.27	0.00056862353636704\\
-0.25	0.000499104289929487\\
-0.23	0.000442296309435311\\
-0.21	0.000382904759115277\\
-0.19	0.000327151989673452\\
-0.17	0.000280195485045801\\
-0.15	0.000240730023805514\\
-0.13	0.000208134309909291\\
-0.11	0.00018167602692996\\
-0.09	0.00016045349362151\\
-0.07	0.00014595080306873\\
-0.05	0.000135501246855045\\
-0.03	0.000127274600405224\\
-0.01	0.000122698343893355\\
0.01	0.000122715904302483\\
0.03	0.000127256060486957\\
0.05	0.000135464892903639\\
0.07	0.000145922410878797\\
0.09	0.000160421851079587\\
0.11	0.000181627543133204\\
0.13	0.000208077106384482\\
0.15	0.000240702024696699\\
0.17	0.000280180287014071\\
0.19	0.000327127029109605\\
0.21	0.000382871613584509\\
0.23	0.000442279095343372\\
0.25	0.00049911946700552\\
0.27	0.000568627508728582\\
0.29	0.000609933877818318\\
0.31	0.000624414061739713\\
0.33	0.000593966582568595\\
0.35	0.000488753836282715\\
0.37	0.000281110632273138\\
0.39	-5.90458741135183e-05\\
0.41	-0.000562608040018897\\
0.43	-0.0012854015294522\\
0.45	-0.00234703770854106\\
0.47	-0.00421341616780224\\
0.49	-0.0109971183700759\\
0.51	-0.0107959589930932\\
0.53	-0.00401994211544426\\
0.55	-0.00241182657701351\\
0.57	-0.0015981344035628\\
0.59	-0.00107897184452257\\
0.61	-0.000706982364592844\\
0.63	-0.000395621659810885\\
0.65	-8.45553394614659e-05\\
0.67	0.000299225807933134\\
0.69	0.000880470060147388\\
};
\addlegendentry{ALM$^1$ ($\varepsilon = 7 \Delta x$)}

\addplot [color=red, dotted, line width=1.0pt, mark size=1.0pt, mark=diamond, mark options={solid, red}]
  table[row sep=crcr]{%
-0.69	-3.78084145463073e-05\\
-0.67	-0.00200831368484181\\
-0.65	-0.00293399894901773\\
-0.63	-0.00306650472076314\\
-0.61	-0.00293028124462538\\
-0.59	-0.00254164041841281\\
-0.57	-0.00249138774323793\\
-0.55	-0.00268829217367915\\
-0.53	-0.00350242253757388\\
-0.51	-0.00615988828812308\\
-0.49	-0.00631096797273101\\
-0.47	-0.00362760524582412\\
-0.45	-0.00224870992779549\\
-0.43	-0.0013865092201934\\
-0.41	-0.000767401093889723\\
-0.39	-0.000570985450568495\\
-0.37	-0.00039694145277884\\
-0.35	-0.000438781303570487\\
-0.33	-0.000465246818378433\\
-0.31	-0.000521432332065712\\
-0.29	-0.000543691912329553\\
-0.27	-0.000563235525383577\\
-0.25	-0.000600293000760376\\
-0.23	-0.000597755378207452\\
-0.21	-0.000609918375729334\\
-0.19	-0.000619105108285362\\
-0.17	-0.00062602354532367\\
-0.15	-0.000631200392821185\\
-0.13	-0.000635485877865106\\
-0.11	-0.000639182467729222\\
-0.09	-0.000641762574549815\\
-0.07	-0.000642634018841291\\
-0.05	-0.000643012791698623\\
-0.03	-0.000644351435611435\\
-0.01	-0.000645801047653307\\
0.01	-0.000645792385445244\\
0.03	-0.000644341441389763\\
0.05	-0.000643005926959181\\
0.07	-0.000642622254115876\\
0.09	-0.000641740901306617\\
0.11	-0.000639153800874333\\
0.13	-0.0006354640132239\\
0.15	-0.000631177189246573\\
0.17	-0.000626008028749208\\
0.19	-0.000619095181697596\\
0.21	-0.000609917090721238\\
0.23	-0.000597763795212553\\
0.25	-0.000600282128438451\\
0.27	-0.000563230779551414\\
0.29	-0.000543682148860199\\
0.31	-0.000521420495908561\\
0.33	-0.000465240982682979\\
0.35	-0.000438779946085936\\
0.37	-0.000396928570647284\\
0.39	-0.000570970439239905\\
0.41	-0.000767378389963491\\
0.43	-0.00138648535126581\\
0.45	-0.0022486947406295\\
0.47	-0.0036275944547568\\
0.49	-0.00631095255854258\\
0.51	-0.00615987908639959\\
0.53	-0.0035023992219016\\
0.55	-0.00268826325920186\\
0.57	-0.00249136458163698\\
0.59	-0.00254162012619832\\
0.61	-0.00293026150515959\\
0.63	-0.00306649232110168\\
0.65	-0.00293399230744536\\
0.67	-0.00200831154297003\\
0.69	-3.78045356016141e-05\\
};
\addlegendentry{ALM$^2$ ($\varepsilon = 3.5 \Delta x$)}

\addplot [color=orange, line width=1.0pt, only marks, mark size=2.0pt, mark=x, mark options={solid, orange}]
  table[row sep=crcr]{%
-0.69	-0.000328140151495453\\
-0.67	-0.00094790255516012\\
-0.65	-0.00134058035480715\\
-0.63	-0.00160496449704545\\
-0.61	-0.00178589989099023\\
-0.59	-0.00191060923737945\\
-0.57	-0.00201934593875047\\
-0.55	-0.00216628086123626\\
-0.53	-0.00252246678647933\\
-0.51	-0.00449553662415991\\
-0.49	-0.00450272296099474\\
-0.47	-0.00259852893212081\\
-0.45	-0.00199869981293024\\
-0.43	-0.00160602968969753\\
-0.41	-0.00128898885388706\\
-0.39	-0.00102363186816806\\
-0.37	-0.000802301646985068\\
-0.35	-0.000627880108644382\\
-0.33	-0.000498514180594749\\
-0.31	-0.000409359000375769\\
-0.29	-0.000348685823547907\\
-0.27	-0.000310306397276095\\
-0.25	-0.000301983452074894\\
-0.23	-0.000289502620069946\\
-0.21	-0.00028788608871878\\
-0.19	-0.000291246679937657\\
-0.17	-0.000294254563945459\\
-0.15	-0.000297092805891763\\
-0.13	-0.000299469739728165\\
-0.11	-0.000301324891717431\\
-0.09	-0.000302783242884769\\
-0.07	-0.000302162976923146\\
-0.05	-0.000301519384979883\\
-0.03	-0.00030237092357195\\
-0.01	-0.00030322078694909\\
0.01	-0.000303212128827536\\
0.03	-0.000302351815337621\\
0.05	-0.000301500447239175\\
0.07	-0.000302137599239637\\
0.09	-0.000302734750396524\\
0.11	-0.000301281443068531\\
0.13	-0.000299462964615919\\
0.15	-0.00029705424730278\\
0.17	-0.000294233355886602\\
0.19	-0.000291237553983451\\
0.21	-0.000287873109252171\\
0.23	-0.000289493783037071\\
0.25	-0.000301967931964593\\
0.27	-0.000310322540317674\\
0.29	-0.000348715896491388\\
0.31	-0.000409389386087951\\
0.33	-0.000498558710593577\\
0.35	-0.000627946370717562\\
0.37	-0.00080231929177961\\
0.39	-0.0010236537446596\\
0.41	-0.00128901518186136\\
0.43	-0.00160603602999973\\
0.45	-0.0019986867507209\\
0.47	-0.00259850826709762\\
0.49	-0.00450267570587998\\
0.51	-0.0044955154444063\\
0.53	-0.00252247709952508\\
0.55	-0.0021662909845382\\
0.57	-0.0020193453537066\\
0.59	-0.0019105978499219\\
0.61	-0.00178588625529526\\
0.63	-0.00160495207350442\\
0.65	-0.0013405583277026\\
0.67	-0.000947886652898178\\
0.69	-0.000328133475160884\\
};
\addlegendentry{ALM$^2$ ($\varepsilon = 7 \Delta x$)}

\legend{}
\end{axis}
\node [left] at (-0.5,-0.5) {(d)};
\end{tikzpicture}%
  \input{LLACL4GammaWL}
%
%
\definecolor{mycolor1}{rgb}{0.00000,1.00000,1.00000}%
\begin{tikzpicture}

\begin{axis}[%
width=0.40\textwidth,
height=0.22\textwidth,
scale only axis,
xmin=0,
xmax=0.7,
xlabel style={font=\color{white!15!black}},
xlabel={$s$},
ymin=-0.05,
ymax=0.01,
ytick={-0.05,-0.04,...,0.01},
ylabel style={font=\color{white!15!black}},
ylabel={$(\Gamma^c-\Gamma_{LL})/\Gamma_{ref}$},
axis background/.style={fill=white},
axis x line*=bottom,
axis y line*=left,
xmajorgrids,
ymajorgrids,
tick label style={font=\scriptsize},
legend style={legend cell align=left, align=left, draw=white!15!black}
]
\addplot [color=blue, dashed, line width=1.0pt, mark size=1.5pt, mark=o, mark options={solid, blue}]
  table[row sep=crcr]{%
-0.69	-0.00295597555701858\\
-0.67	-0.00380390234474237\\
-0.65	-0.00445307011114995\\
-0.63	-0.00542514175446867\\
-0.61	-0.00769501628868384\\
-0.59	-0.0116333241588989\\
-0.57	-0.0179526004360264\\
-0.55	-0.0257601378367192\\
-0.53	-0.0340636672706921\\
-0.51	-0.0406346779277475\\
-0.49	-0.0397397381781791\\
-0.47	-0.0325292838880833\\
-0.45	-0.0233224451197973\\
-0.43	-0.0145470256497753\\
-0.41	-0.00728115915629395\\
-0.39	-0.00247438218684356\\
-0.37	0.000533603963395892\\
-0.35	0.00202278253480284\\
-0.33	0.00271111068342996\\
-0.31	0.00292933528040627\\
-0.29	0.00294111267487454\\
-0.27	0.00281656047441204\\
-0.25	0.00263515451183517\\
-0.23	0.00246490116355572\\
-0.21	0.0022742999990985\\
-0.19	0.00209386043492532\\
-0.17	0.00192587194838362\\
-0.15	0.00177535589388841\\
-0.13	0.00164278318609079\\
-0.11	0.00153070166284324\\
-0.09	0.00143738382902485\\
-0.07	0.00136307648921167\\
-0.05	0.00130790274836616\\
-0.03	0.0012719480467751\\
-0.01	0.00125497548606413\\
0.01	0.00125497493451823\\
0.03	0.00127193690400756\\
0.05	0.00130787811768381\\
0.07	0.00136303569893468\\
0.09	0.00143732805415014\\
0.11	0.00153063558269315\\
0.13	0.00164271964508306\\
0.15	0.00177530743607829\\
0.17	0.00192584385625847\\
0.19	0.0020938494168701\\
0.21	0.00227430032786357\\
0.23	0.00246491002406666\\
0.25	0.00263517248696405\\
0.27	0.00281658182518402\\
0.29	0.00294114183543656\\
0.31	0.00292937251773282\\
0.33	0.00271115537030041\\
0.35	0.00202283358417826\\
0.37	0.000533657839284636\\
0.39	-0.00247433529381355\\
0.41	-0.00728113175802169\\
0.43	-0.0145470156333137\\
0.45	-0.0233224478440765\\
0.47	-0.032529294523705\\
0.49	-0.0397397542019033\\
0.51	-0.0406346869893372\\
0.53	-0.0340636682724199\\
0.55	-0.0257601283245823\\
0.57	-0.0179525862590614\\
0.59	-0.0116333104762698\\
0.61	-0.00769501394483816\\
0.63	-0.00542514967260732\\
0.65	-0.00445307813710207\\
0.67	-0.00380391116245178\\
0.69	-0.00295598455811636\\
};
\addlegendentry{ALM$^1$ ($\varepsilon = 3.5 \Delta x$)}

\addplot [color=cyan, line width=1.0pt, only marks, mark size=2.0pt, mark=+, mark options={solid, cyan}]
  table[row sep=crcr]{%
-0.69	-0.00454332763866075\\
-0.67	-0.00628590829642265\\
-0.65	-0.00781713120644792\\
-0.63	-0.00945070435689635\\
-0.61	-0.0112389744301326\\
-0.59	-0.0131481182579955\\
-0.57	-0.0151076342981211\\
-0.55	-0.0170448339263544\\
-0.53	-0.0189405992441876\\
-0.51	-0.0208585841572634\\
-0.49	-0.020310604779598\\
-0.47	-0.018070912508928\\
-0.45	-0.0155950526972577\\
-0.43	-0.0129847654020042\\
-0.41	-0.0103289027114193\\
-0.39	-0.00774885926775151\\
-0.37	-0.00535938470347297\\
-0.35	-0.00325417728655873\\
-0.33	-0.00148544133007543\\
-0.31	-7.12176951308097e-05\\
-0.29	0.00100628552525157\\
-0.27	0.00179253981712114\\
-0.25	0.00234288871906178\\
-0.23	0.0027053950771872\\
-0.21	0.00292971060832289\\
-0.19	0.00305665324743949\\
-0.17	0.00311946674370239\\
-0.15	0.00314076633151295\\
-0.13	0.00313708668560402\\
-0.11	0.00311945825192136\\
-0.09	0.00309661983406454\\
-0.07	0.00307420958121709\\
-0.05	0.00305483876227134\\
-0.03	0.00304055857890584\\
-0.01	0.0030334590276677\\
0.01	0.00303347220626578\\
0.03	0.00304060303328033\\
0.05	0.00305490594552307\\
0.07	0.00307429691672021\\
0.09	0.00309671985450188\\
0.11	0.00311956657182315\\
0.13	0.00313719485522824\\
0.15	0.00314085352197473\\
0.17	0.00311953016566118\\
0.19	0.0030566930899041\\
0.21	0.00292972960109205\\
0.23	0.00270538690503411\\
0.25	0.00234284533842062\\
0.27	0.00179248794905117\\
0.29	0.00100622336862548\\
0.31	-7.12992361960889e-05\\
0.33	-0.00148555318003305\\
0.35	-0.00325431432898529\\
0.37	-0.0053595246329901\\
0.39	-0.00774899813876549\\
0.41	-0.0103290281653371\\
0.43	-0.0129848536789607\\
0.45	-0.0155951092945669\\
0.47	-0.0180709519324017\\
0.49	-0.0203106259981882\\
0.51	-0.0208585801357231\\
0.53	-0.0189405963679753\\
0.55	-0.0170448515586316\\
0.57	-0.015107685275366\\
0.59	-0.0131482151557466\\
0.61	-0.0112391160455069\\
0.63	-0.00945086845974098\\
0.65	-0.00781729061332809\\
0.67	-0.00628604555879813\\
0.69	-0.00454342240799853\\
};
\addlegendentry{ALM$^1$ ($\varepsilon = 7 \Delta x$)}

\addplot [color=red, dotted, line width=1.0pt, mark size=1.0pt, mark=diamond, mark options={solid, red}]
  table[row sep=crcr]{%
-0.69	-0.00547652755006834\\
-0.67	-0.00734446920648143\\
-0.65	-0.00832000521778639\\
-0.63	-0.00862606219526184\\
-0.61	-0.00858540172352251\\
-0.59	-0.00793108325316691\\
-0.57	-0.00714058715403573\\
-0.55	-0.00617366109451073\\
-0.53	-0.00578092520111667\\
-0.51	-0.00549780663981892\\
-0.49	-0.00520889631224888\\
-0.47	-0.00527168570506836\\
-0.45	-0.00499335349112599\\
-0.43	-0.00514057204488025\\
-0.41	-0.00512760718684075\\
-0.39	-0.005108034677644\\
-0.37	-0.00468659779402445\\
-0.35	-0.00422669281369572\\
-0.33	-0.00369341064967013\\
-0.31	-0.00320895699288949\\
-0.29	-0.00277607139987093\\
-0.27	-0.00244590793388741\\
-0.25	-0.00219717334698107\\
-0.23	-0.00198321184559607\\
-0.21	-0.00183670965553107\\
-0.19	-0.00172739706913258\\
-0.17	-0.0016488420640855\\
-0.15	-0.00159185052597915\\
-0.13	-0.00155161587528528\\
-0.11	-0.00152162901655757\\
-0.09	-0.00150064676108752\\
-0.07	-0.00148608696531729\\
-0.05	-0.00147624104513652\\
-0.03	-0.00146985734673044\\
-0.01	-0.00146616173358995\\
0.01	-0.00146616914536184\\
0.03	-0.00146987731926601\\
0.05	-0.00147627826119869\\
0.07	-0.00148614622165212\\
0.09	-0.00150072987974418\\
0.11	-0.00152172111050169\\
0.13	-0.00155170253897094\\
0.15	-0.00159194462051728\\
0.17	-0.00164891898583252\\
0.19	-0.00172744856700785\\
0.21	-0.00183673711328394\\
0.23	-0.00198321693044306\\
0.25	-0.00219715376900613\\
0.27	-0.00244586840501127\\
0.29	-0.00277601248254491\\
0.31	-0.00320887664887486\\
0.33	-0.00369330882587991\\
0.35	-0.00422657784299222\\
0.37	-0.00468649193454384\\
0.39	-0.00510793250726602\\
0.41	-0.00512751842694646\\
0.43	-0.00514050516663273\\
0.45	-0.00499330515460927\\
0.47	-0.00527164825224613\\
0.49	-0.0052088645852491\\
0.51	-0.00549778226475861\\
0.53	-0.00578090961835632\\
0.55	-0.00617365466331779\\
0.57	-0.0071405891307727\\
0.59	-0.0079310923413059\\
0.61	-0.00858541318130681\\
0.63	-0.00862607225951862\\
0.65	-0.00832002821882053\\
0.67	-0.00734449853345887\\
0.69	-0.00547655283016011\\
};
\addlegendentry{ALM$^2$ ($\varepsilon = 3.5 \Delta x$)}

\addplot [color=orange, line width=1.0pt, only marks, mark size=2.0pt, mark=x, mark options={solid, orange}]
  table[row sep=crcr]{%
-0.69	-0.00383041162318711\\
-0.67	-0.00459691348292049\\
-0.65	-0.00478208143669707\\
-0.63	-0.00474175201448743\\
-0.61	-0.00461234946578661\\
-0.59	-0.00447932423468327\\
-0.57	-0.00440595880380528\\
-0.55	-0.00444069909708979\\
-0.53	-0.00462151234347587\\
-0.51	-0.00491643265113366\\
-0.49	-0.004666709282625\\
-0.47	-0.00426855990313913\\
-0.45	-0.00367007712639463\\
-0.43	-0.0031088813410767\\
-0.41	-0.00263116103915616\\
-0.39	-0.00224558305095663\\
-0.37	-0.00194031639481738\\
-0.35	-0.00170019870291866\\
-0.33	-0.00150609796015096\\
-0.31	-0.00134600699399598\\
-0.29	-0.00120869612164446\\
-0.27	-0.00108265030757461\\
-0.25	-0.00096081204904325\\
-0.23	-0.000846071823944562\\
-0.21	-0.000738036182439083\\
-0.19	-0.000637788670001419\\
-0.17	-0.000545034792674052\\
-0.15	-0.000461863948532817\\
-0.13	-0.000389346226842703\\
-0.11	-0.000328415818448696\\
-0.09	-0.000278199757798138\\
-0.07	-0.000238122853614961\\
-0.05	-0.000208650215262494\\
-0.03	-0.000189633051297977\\
-0.01	-0.000179929511434167\\
0.01	-0.000179922727948034\\
0.03	-0.000189593236535523\\
0.05	-0.000208570206152749\\
0.07	-0.000238017593513921\\
0.09	-0.000278077734780767\\
0.11	-0.000328277428370842\\
0.13	-0.000389196110972569\\
0.15	-0.000461708719937025\\
0.17	-0.00054488959425925\\
0.19	-0.000637665002524758\\
0.21	-0.000737952493437947\\
0.23	-0.000846036916867614\\
0.25	-0.00096080527735851\\
0.27	-0.00108265909122856\\
0.29	-0.0012087245862604\\
0.31	-0.00134606363396327\\
0.33	-0.00150618053585264\\
0.35	-0.00170028190228521\\
0.37	-0.00194038980993067\\
0.39	-0.00224567330120861\\
0.41	-0.0026312236658819\\
0.43	-0.00310891571549723\\
0.45	-0.00367007853028774\\
0.47	-0.00426851380359482\\
0.49	-0.00466662624561926\\
0.51	-0.00491632997971342\\
0.53	-0.00462139849642146\\
0.55	-0.00444055624359091\\
0.57	-0.00440578797605224\\
0.59	-0.00447912614735426\\
0.61	-0.00461212487063291\\
0.63	-0.00474150572979426\\
0.65	-0.00478185554906113\\
0.67	-0.00459672120917796\\
0.69	-0.0038302776557309\\
};
\addlegendentry{ALM$^2$ ($\varepsilon = 7 \Delta x$)}

\legend{}
\end{axis}
\node [left] at (-0.5,-0.5) {(f)};
\end{tikzpicture}%
  \input{LLACL4ClWL}
%
%
\definecolor{mycolor1}{rgb}{0.00000,1.00000,1.00000}%
\begin{tikzpicture}

\begin{axis}[%
width=0.40\textwidth,
height=0.22\textwidth,
scale only axis,
xmin=0,
xmax=0.7,
xlabel style={font=\color{white!15!black}},
xlabel={$s$},
ymin=-0.08,
ymax=0.02,
ylabel style={font=\color{white!15!black}},
ylabel={$(C_{l U_0}^c-C_{l U_0 LL})/C_{l ref}$},
axis background/.style={fill=white},
axis x line*=bottom,
axis y line*=left,
xmajorgrids,
ymajorgrids,
tick label style={font=\scriptsize},
legend style={legend cell align=left, align=left, draw=white!15!black}
]
\addplot [color=blue, dashed, line width=1.0pt, mark size=1.5pt, mark=o, mark options={solid, blue}]
  table[row sep=crcr]{%
-0.69	-0.00246310334742644\\
-0.67	-0.00421633860633452\\
-0.65	-0.00539569147054864\\
-0.63	-0.00630160354399556\\
-0.61	-0.00825190587484492\\
-0.59	-0.0117170596975485\\
-0.57	-0.0183800920553344\\
-0.55	-0.0281626397680819\\
-0.53	-0.0427630244965311\\
-0.51	-0.077290543229585\\
-0.49	-0.0767060362664502\\
-0.47	-0.0414528795738915\\
-0.45	-0.0252930899514205\\
-0.43	-0.0138855754255065\\
-0.41	-0.00568086985765215\\
-0.39	-0.000910004149112909\\
-0.37	0.00195229396363961\\
-0.35	0.0031058981066312\\
-0.33	0.00354062237160235\\
-0.31	0.00352176660502201\\
-0.29	0.00336750521102613\\
-0.27	0.00310437109168538\\
-0.25	0.00278916208730684\\
-0.23	0.00254147231702395\\
-0.21	0.00227326546214124\\
-0.19	0.00203018932666066\\
-0.17	0.00181152274245366\\
-0.15	0.00162024861266252\\
-0.13	0.00145458943971266\\
-0.11	0.00131590313194163\\
-0.09	0.00120196717850463\\
-0.07	0.00111292671120244\\
-0.05	0.00104727978430363\\
-0.03	0.00100341775674773\\
-0.01	0.000981808071728096\\
0.01	0.000981810307594255\\
0.03	0.00100340492535633\\
0.05	0.00104725250962945\\
0.07	0.00111288305709145\\
0.09	0.00120190119083008\\
0.11	0.00131582643657546\\
0.13	0.00145452640952104\\
0.15	0.0016201919269121\\
0.17	0.0018114914779146\\
0.19	0.00203017475957712\\
0.21	0.00227326123042637\\
0.23	0.00254148687210309\\
0.25	0.00278918517553783\\
0.27	0.00310439496916348\\
0.29	0.00336754197249001\\
0.31	0.00352181362764847\\
0.33	0.00354067958794035\\
0.35	0.00310596560649645\\
0.37	0.00195235900758228\\
0.39	-0.000909945271634904\\
0.41	-0.00568083767132188\\
0.43	-0.0138855657848407\\
0.45	-0.0252930956325981\\
0.47	-0.0414528965996144\\
0.49	-0.0767060674850961\\
0.51	-0.0772905611798649\\
0.53	-0.0427630235551573\\
0.55	-0.0281626227667654\\
0.57	-0.0183800677798553\\
0.59	-0.0117170387785616\\
0.61	-0.0082518999140373\\
0.63	-0.00630160657081202\\
0.65	-0.00539570163085668\\
0.67	-0.00421635017748589\\
0.69	-0.00246311385054016\\
};
\addlegendentry{LL}

\addplot [color=cyan, line width=1.0pt, only marks, mark size=2.0pt, mark=+, mark options={solid, cyan}]
  table[row sep=crcr]{%
-0.69	-0.00424517187399243\\
-0.67	-0.00625178005329452\\
-0.65	-0.00808121209997659\\
-0.63	-0.0100365183537995\\
-0.61	-0.0122115041443529\\
-0.59	-0.0146398974178485\\
-0.57	-0.0173815735592024\\
-0.55	-0.0206487864447638\\
-0.53	-0.025447288431891\\
-0.51	-0.0409984983997911\\
-0.49	-0.0404667592466501\\
-0.47	-0.0247125813147895\\
-0.45	-0.0190709485663965\\
-0.43	-0.0148858448571283\\
-0.41	-0.0112539603342422\\
-0.39	-0.00803197385810894\\
-0.37	-0.00522228222173438\\
-0.35	-0.00286090843182274\\
-0.33	-0.000956267517331688\\
-0.31	0.000508633457607988\\
-0.29	0.00158520532125017\\
-0.27	0.00233958087026742\\
-0.25	0.00282822752920275\\
-0.23	0.00313870785951198\\
-0.21	0.00330746037885932\\
-0.19	0.00338167069962658\\
-0.17	0.0033997216585917\\
-0.15	0.00338320220982657\\
-0.13	0.00334816993909881\\
-0.11	0.00330502117683762\\
-0.09	0.00326168318614661\\
-0.07	0.00322517758471241\\
-0.05	0.00319563609628171\\
-0.03	0.00317340772277785\\
-0.01	0.00316190724354204\\
0.01	0.00316193677838816\\
0.03	0.0031734351993784\\
0.05	0.00319566989644948\\
0.07	0.00322523903868355\\
0.09	0.00326175439686416\\
0.11	0.00330508514047745\\
0.13	0.00334822570094517\\
0.15	0.00338326398737621\\
0.17	0.00339977141852266\\
0.19	0.00338168768733516\\
0.21	0.00330744878129829\\
0.23	0.00313868364410119\\
0.25	0.00282819780207177\\
0.27	0.00233953210963156\\
0.29	0.00158516769864803\\
0.31	0.000508567319584485\\
0.33	-0.000956374158635698\\
0.35	-0.00286102860213555\\
0.37	-0.0052223838207357\\
0.39	-0.00803209875950706\\
0.41	-0.0112540768323679\\
0.43	-0.0148859212673869\\
0.45	-0.0190709925052099\\
0.47	-0.0247126219009124\\
0.49	-0.0404667936885483\\
0.51	-0.0409985231704029\\
0.53	-0.0254473049165873\\
0.55	-0.0206488174232015\\
0.57	-0.0173816311228053\\
0.59	-0.0146399974377275\\
0.61	-0.0122116510008453\\
0.63	-0.0100366888348381\\
0.65	-0.00808137354611271\\
0.67	-0.00625191398567393\\
0.69	-0.00424526365636618\\
};
\addlegendentry{ALM$^1$ ($\varepsilon = 3.5 \Delta x$)}

\addplot [color=red, dotted, line width=1.0pt, mark size=1.0pt, mark=diamond, mark options={solid, red}]
  table[row sep=crcr]{%
-0.69	-0.00560612725465037\\
-0.67	-0.0087166418368706\\
-0.65	-0.0105611564312906\\
-0.63	-0.0111964405854975\\
-0.61	-0.0112450085075769\\
-0.59	-0.0104287995069305\\
-0.57	-0.00975358971547646\\
-0.55	-0.0091653961404019\\
-0.53	-0.0100000846386132\\
-0.51	-0.014338645458525\\
-0.49	-0.0141839885346562\\
-0.47	-0.00967417869478959\\
-0.45	-0.00759636475706249\\
-0.43	-0.00674794834255632\\
-0.41	-0.00605454682827622\\
-0.39	-0.00579760139067498\\
-0.37	-0.00516945505524236\\
-0.35	-0.00471820040071269\\
-0.33	-0.00418619234890771\\
-0.31	-0.00373717156612796\\
-0.29	-0.00331282770837982\\
-0.27	-0.00299250604335255\\
-0.25	-0.00277231826659119\\
-0.23	-0.00255181475317989\\
-0.21	-0.00241369117673862\\
-0.19	-0.00231082626827939\\
-0.17	-0.00223721322652637\\
-0.15	-0.00218396195500747\\
-0.13	-0.00214693840711522\\
-0.11	-0.00211982057030646\\
-0.09	-0.00210083233699332\\
-0.07	-0.00208680426070551\\
-0.05	-0.00207712276867633\\
-0.03	-0.00207185874658411\\
-0.01	-0.00206944321842983\\
0.01	-0.00206944266401776\\
0.03	-0.00207186957675765\\
0.05	-0.00207715380767848\\
0.07	-0.00208685291985444\\
0.09	-0.0021008958346258\\
0.11	-0.00211988665381257\\
0.13	-0.00214700533883294\\
0.15	-0.00218403515562182\\
0.17	-0.0022372762871854\\
0.19	-0.00231086893944277\\
0.21	-0.00241371764366172\\
0.23	-0.00255182764751882\\
0.25	-0.00277228847458821\\
0.27	-0.00299246172992451\\
0.29	-0.00331275907444006\\
0.31	-0.00373707916917942\\
0.33	-0.00418608348522775\\
0.35	-0.00471808198048351\\
0.37	-0.00516933465757174\\
0.39	-0.00579748209260889\\
0.41	-0.00605443309240783\\
0.43	-0.00674785499310482\\
0.45	-0.00759629784597604\\
0.47	-0.00967412539874646\\
0.49	-0.0141839271647963\\
0.51	-0.0143386010984949\\
0.53	-0.0100000442188755\\
0.55	-0.00916536247956135\\
0.57	-0.00975357135966137\\
0.59	-0.0104287918660996\\
0.61	-0.0112450045305773\\
0.63	-0.0111964415973682\\
0.65	-0.0105611753927927\\
0.67	-0.00871667050064584\\
0.69	-0.00560615133228815\\
};
\addlegendentry{ALM$^1$ ($\varepsilon = 7 \Delta x$)}

\addplot [color=orange, line width=1.0pt, only marks, mark size=2.0pt, mark=x, mark options={solid, orange}]
  table[row sep=crcr]{%
-0.69	-0.00405554986855239\\
-0.67	-0.00527330312565455\\
-0.65	-0.00583664410751061\\
-0.63	-0.00610021473712152\\
-0.61	-0.00621787120125494\\
-0.59	-0.00629725617198851\\
-0.57	-0.00644926689093039\\
-0.55	-0.00681133137661782\\
-0.53	-0.00772701591445313\\
-0.51	-0.011758730975967\\
-0.49	-0.0114687827611939\\
-0.47	-0.00748996007950753\\
-0.45	-0.00592834014392851\\
-0.43	-0.00482825836628842\\
-0.41	-0.00396517841546218\\
-0.39	-0.00328185137773696\\
-0.37	-0.00274087652578836\\
-0.35	-0.00232065085562871\\
-0.33	-0.00199517774522573\\
-0.31	-0.00174507810504199\\
-0.29	-0.00154656090619187\\
-0.27	-0.0013813970716785\\
-0.25	-0.00124915220340116\\
-0.23	-0.00112081283110188\\
-0.21	-0.00100969828204878\\
-0.19	-0.000911333911867374\\
-0.17	-0.00082041194419824\\
-0.15	-0.000739138874659484\\
-0.13	-0.000668270177743291\\
-0.11	-0.00060864930238036\\
-0.09	-0.000559483848057263\\
-0.07	-0.000518619132301557\\
-0.05	-0.000488405847452245\\
-0.03	-0.00047008724634412\\
-0.01	-0.000461125843060017\\
0.01	-0.00046111102446178\\
0.03	-0.000470029585149634\\
0.05	-0.000488307960998524\\
0.07	-0.000518489913874376\\
0.09	-0.000559316406573918\\
0.11	-0.00060847004870046\\
0.13	-0.000668112983416957\\
0.15	-0.000738947123715805\\
0.17	-0.000820246248283363\\
0.19	-0.000911200976292825\\
0.21	-0.00100960198044975\\
0.23	-0.00112076947057482\\
0.25	-0.00124913101752622\\
0.27	-0.00138142088880444\\
0.29	-0.00154661758330399\\
0.31	-0.00174516373995348\\
0.33	-0.00199530311862772\\
0.35	-0.0023207976116435\\
0.37	-0.0027409682133458\\
0.39	-0.00328196492256294\\
0.41	-0.00396526854507784\\
0.43	-0.0048283006053067\\
0.45	-0.00592832870331805\\
0.47	-0.00748988687192242\\
0.49	-0.0114686145137184\\
0.51	-0.0117585624459589\\
0.53	-0.00772689652941128\\
0.55	-0.00681118645790735\\
0.57	-0.00644908598296134\\
0.59	-0.0062970399693395\\
0.61	-0.00621762787541325\\
0.63	-0.00609995193443813\\
0.65	-0.00583639731763075\\
0.67	-0.00527309702454792\\
0.69	-0.00405541021013217\\
};
\addlegendentry{ALM$^2$ ($\varepsilon = 3.5 \Delta x$)}

\legend{}
\end{axis}
\node [left] at (-0.5,-0.5) {(h)};
\end{tikzpicture}%
  \caption{Comparison of the results for the wing with winglet. LL - Non-linear iterative lifting line method; ALM$^1$ - ALM with smearing correction without vorticity magnitude correction; ALM$^{VMC}$ - ALM with smearing correction with vorticity magnitude correction. (a) Corrected velocity in local $y$-direction. (c) Corrected velocity in local $z$-direction. (e) Circulation. (f) Lift coefficient, non-dimensionalized with $U_0$. (b,d,f,h) Normalized differences between quantities obtained by the ALM and the LL.}
  \label{fig:res_resultswinglet}
\end{figure}

The results for the wing with winglets, presented in figure \ref{fig:res_resultswinglet}, show that both formulations of the ALM method are, overall, in good agreement with the LL. Such agreement is the result of the smearing correction, as discussed in previous works~\citep{dag2020new,meyer2019vortex,kleine2022non}. The ALM in both formulations captures the continuity of circulation (``continuity'' in the context of a discretized method) and the increase in $u_z$ and forces near the intersection due to velocity induced by the bound vortices. At the point of intersection of the lines, the ideal case is singular, because the induced velocities by ideal vortices tend to infinity in the region near the point of intersection. If a higher spanwise resolution is sought, desingularization of the vortices and a smooth transition between the surfaces are recommended (instead of the abrupt change of dihedral angle used here), avoiding the causes of the singularity.

Some ALM use control points at the boundaries of each actuator line segment, which is not recommended for this case, because the velocity is not well defined at the point of intersection. If the implementation of the ALM requires a point at the intersection (which is not the case for this work), the local velocity should not be used to calculate the local circulation and forces, because the local velocity is ill defined according to the LL. Hence, such method would not be able to emulate the LL, even with the smearing and vorticity magnitude corrections. How to accurately simulate non-planar wings for such implementations of the ALM could be a topic of further research.

Despite the overall good agreement, there are noticeable differences between the results of the ALM without the vorticity magnitude correction and the LL. These differences near the intersection of the surfaces were the main motivation for the development of the vorticity magnitude correction. The vorticity magnitude correction clearly improves the agreement of all quantities. For the circulation, the maximum difference reduces from about 4\% to less than 1\% for the lower value of the smearing parameter $\varepsilon$. By observing the lift coefficient results, it is possible to note that the maximum differences in the forces are reduced from around 8\% to around 1\%.

Another important advantage of the vorticity magnitude correction is that it makes the ALM much less dependent on the smearing parameter. The smearing correction itself had the goal of making the ALM insensitive to changes of $\varepsilon$. However, we note that the maximum differences in circulation and force for $\varepsilon=3.5 \Delta x$ are approximately double the maximum differences for $\varepsilon=7 \Delta x$, for ALM$^1$. The use of the vorticity magnitude correction brings back this independence from the smearing parameter for non-planar wings.

As explained in~\citep{kleine2022non}, it is not reasonable to expect absolutely no effect of the smearing parameter. The vorticity generated in the CFD simulation is more spread if the smearing parameter is increased, which indirectly affects the results. The fact that the differences to the LL, in general, are lower for the larger smearing parameter was also discussed in~\citep{kleine2022non}: for a larger smearing parameter, less information about the induced velocity is comes from the CFD solution ($\mathbf{u}^s$ is closer to the undisturbed velocity) and more information comes from the missing velocity. Because the missing velocity is designed to emulate a LL, the agreement between ALM with smearing correction and LL is better for larger values of the smearing parameter.

These results show that the theoretical derivation of section \ref{sec:magnitudecorrection} translates to a simple method that actually improves the result of the ALM. In order to explore another case relevant to airplane aerodynamics, a T-tail configuration is investigated. For this configuration, we use the following convention: the lift is positive if it is in the positive $Y$ or $X$ global directions shown in figure \ref{fig:tail_VSheet} (analogous to the previous case, the angle of attack and the circulation are considered positive in the direction that makes $C_{l}(\alpha_g)$ positive).

\begin{figure}
  \centering
%
%
\definecolor{mycolor1}{rgb}{0.00000,1.00000,1.00000}%
\begin{tikzpicture}

\begin{axis}[%
width=0.48\textwidth,
height=0.22\textwidth,
scale only axis,
xmin=-0.5,
xmax=0.5,
xtick={-0.5,-0.4,...,0.5},
xlabel style={font=\color{white!15!black}},
xlabel={$X$},
ymin=-0.05,
ymax=0.2,
ytick={-0.05,0.0,...,0.2},
ylabel style={font=\color{white!15!black}},
ylabel={$u^c_y/U_0$},
yticklabel style={
        /pgf/number format/fixed,
        /pgf/number format/precision=3
},
axis background/.style={fill=white},
axis x line*=bottom,
axis y line*=left,
xmajorgrids,
ymajorgrids,
tick label style={font=\scriptsize},
legend style={legend cell align=left, align=left, draw=white!15!black}
]
\addplot [color=gray, line width=1.0pt, mark size=2.0pt, mark=square, mark options={solid, gray}]
  table[row sep=crcr]{%
-0.49	0.109108310104959\\
-0.47	0.0881118626939952\\
-0.45	0.0748316667016047\\
-0.43	0.0652178715658019\\
-0.41	0.0577583328287034\\
-0.39	0.0517062367859812\\
-0.37	0.0466330115962875\\
-0.35	0.0422678982972051\\
-0.33	0.0384276639424957\\
-0.31	0.0349814921129503\\
-0.29	0.0318317210110706\\
-0.27	0.0289025035992132\\
-0.25	0.0261327673681646\\
-0.23	0.0234716961304345\\
-0.21	0.0208758547078038\\
-0.19	0.0183076155007745\\
-0.17	0.0157350344436749\\
-0.15	0.0131340890585086\\
-0.13	0.0104959287841114\\
-0.11	0.00784657566717808\\
-0.09	0.00530180332872212\\
-0.07	0.00323797092403204\\
-0.05	0.0029417485560179\\
-0.03	0.010170172638527\\
-0.01	0.0843970814983183\\
0.01	-0.0190351621568056\\
0.03	0.0551631830089291\\
0.05	0.0627501389893969\\
0.07	0.0629400847121241\\
0.09	0.061531092971162\\
0.11	0.0598261496797643\\
0.13	0.058214963988823\\
0.15	0.0568286140904224\\
0.17	0.0557120822861434\\
0.19	0.0548802268220036\\
0.21	0.0543386308438622\\
0.23	0.0540923192323329\\
0.25	0.0541499840785625\\
0.27	0.0545266228939303\\
0.29	0.0552458845894951\\
0.31	0.0563428873953831\\
0.33	0.0578682086341033\\
0.35	0.0598940081582606\\
0.37	0.0625239475428266\\
0.39	0.0659101585513143\\
0.41	0.0702843065960714\\
0.43	0.076019880031406\\
0.45	0.0837743393769817\\
0.47	0.0948847701569018\\
0.49	0.112933492683784\\
};
\addlegendentry{LL}

\addplot [color=blue, dashed, line width=1.0pt, mark size=1.5pt, mark=o, mark options={solid, blue}]
  table[row sep=crcr]{%
-0.49	0.11080031516959\\
-0.47	0.089458545744942\\
-0.45	0.0759312133250008\\
-0.43	0.0661951088551849\\
-0.41	0.0587219313676405\\
-0.39	0.0527311594116796\\
-0.37	0.0477608489476649\\
-0.35	0.0435206049795425\\
-0.33	0.0398204361212401\\
-0.31	0.0365300439948346\\
-0.29	0.0335569569511535\\
-0.27	0.0308311461195749\\
-0.25	0.0282992024282965\\
-0.23	0.0259250492051109\\
-0.21	0.0237030610576614\\
-0.19	0.0216242192651024\\
-0.17	0.019723064159237\\
-0.15	0.018061779817508\\
-0.13	0.0167838433709487\\
-0.11	0.0160438910991078\\
-0.09	0.0160668263046868\\
-0.07	0.0169519348098708\\
-0.05	0.0192408956918632\\
-0.03	0.0270922165261881\\
-0.01	0.0922473588076267\\
0.01	-0.0017041385002622\\
0.03	0.0616859782993222\\
0.05	0.0658081249101606\\
0.07	0.0642805095626537\\
0.09	0.0621715778207642\\
0.11	0.060298107325822\\
0.13	0.058635553412089\\
0.15	0.0572420298022886\\
0.17	0.0560689387181997\\
0.19	0.0551666396835451\\
0.21	0.0545305023790283\\
0.23	0.0541697629885049\\
0.25	0.054114718532719\\
0.27	0.0543638918978153\\
0.29	0.0549448200188278\\
0.31	0.0558965117700266\\
0.33	0.0572698928197789\\
0.35	0.0591384851896071\\
0.37	0.0616086639029688\\
0.39	0.0648424774391237\\
0.41	0.0690916656401664\\
0.43	0.0747679068705344\\
0.45	0.0825762607660673\\
0.47	0.0939074880617226\\
0.49	0.112421527992977\\
};
\addlegendentry{ALM$^1$ ($\varepsilon = 3.5 \Delta x$)}

\addplot [color=cyan, line width=1.0pt, only marks, mark size=2.0pt, mark=+, mark options={solid, cyan}]
  table[row sep=crcr]{%
-0.49	0.110223643780424\\
-0.47	0.0892518673532003\\
-0.45	0.0759620351790793\\
-0.43	0.0663405095008679\\
-0.41	0.0588892508504825\\
-0.39	0.0528699984782557\\
-0.37	0.0478590798788989\\
-0.35	0.0435886906901729\\
-0.33	0.0398799263354788\\
-0.31	0.0366086492364042\\
-0.29	0.0336852764733449\\
-0.27	0.0310432945993643\\
-0.25	0.0286282200064234\\
-0.23	0.0264240623852558\\
-0.21	0.0243764777180057\\
-0.19	0.0224640785529786\\
-0.17	0.0206566274285561\\
-0.15	0.0189129545540803\\
-0.13	0.0171931768354043\\
-0.11	0.0154815940592946\\
-0.09	0.013833424088474\\
-0.07	0.01253640713173\\
-0.05	0.0127236507020455\\
-0.03	0.0196537584444078\\
-0.01	0.0883038070609455\\
0.01	-0.0082099095968586\\
0.03	0.0603027901957013\\
0.05	0.0665300451440085\\
0.07	0.065814543671485\\
0.09	0.0636416658126222\\
0.11	0.0612836299979459\\
0.13	0.0591314193632746\\
0.15	0.0573185668315237\\
0.17	0.055875061151681\\
0.19	0.0547949780251638\\
0.21	0.0540629398128717\\
0.23	0.0536669905501026\\
0.25	0.0536056343995783\\
0.27	0.0538722258005485\\
0.29	0.0544924106648562\\
0.31	0.0554962186057763\\
0.33	0.0569335852610043\\
0.35	0.0588777139391996\\
0.37	0.0614347570534321\\
0.39	0.0647596573387839\\
0.41	0.0690890925543906\\
0.43	0.0748019529368349\\
0.45	0.0825655761166302\\
0.47	0.0937396115936327\\
0.49	0.111976141140248\\
};
\addlegendentry{ALM$^1$ ($\varepsilon = 7 \Delta x$)}

\addplot [color=red, dotted, line width=1.0pt, mark size=1.0pt, mark=diamond, mark options={solid, red}]
  table[row sep=crcr]{%
-0.49	0.110727691735257\\
-0.47	0.0894320744733334\\
-0.45	0.0759569465818229\\
-0.43	0.0662814581303969\\
-0.41	0.058872487289066\\
-0.39	0.0529431387012597\\
-0.37	0.0480275571506805\\
-0.35	0.0438340072647996\\
-0.33	0.0401727820281763\\
-0.31	0.0369143798506275\\
-0.29	0.0339668121425923\\
-0.27	0.0312586802432597\\
-0.25	0.0287330931989051\\
-0.23	0.026345529930327\\
-0.21	0.0240735778841797\\
-0.19	0.0218794131161072\\
-0.17	0.0197306883835399\\
-0.15	0.0175709195066176\\
-0.13	0.0153443411390913\\
-0.11	0.0129775401150688\\
-0.09	0.0105130794749845\\
-0.07	0.00824776687052\\
-0.05	0.0075079542499715\\
-0.03	0.0141023910532995\\
-0.01	0.0868353688485021\\
0.01	-0.0155751615622299\\
0.03	0.0577996223894468\\
0.05	0.0651290769789523\\
0.07	0.0650574140842905\\
0.09	0.0633329251535625\\
0.11	0.0613136952400765\\
0.13	0.0594042306412886\\
0.15	0.05779403938389\\
0.17	0.0564860588203135\\
0.19	0.0555013267245102\\
0.21	0.0548208077667197\\
0.23	0.0544402891741008\\
0.25	0.0543650565126139\\
0.27	0.0546068142657865\\
0.29	0.0551870419678773\\
0.31	0.0561369666352059\\
0.33	0.0575073974330652\\
0.35	0.0593712641570984\\
0.37	0.0618335076307202\\
0.39	0.0650545678335026\\
0.41	0.0692873374806728\\
0.43	0.0749467622264743\\
0.45	0.0827426511993188\\
0.47	0.0940687275091042\\
0.49	0.11258220218072\\
};
\addlegendentry{ALM$^2$ ($\varepsilon = 3.5 \Delta x$)}

\addplot [color=orange, line width=1.0pt, only marks, mark size=2.0pt, mark=x, mark options={solid, orange}]
  table[row sep=crcr]{%
-0.49	0.110162757421716\\
-0.47	0.0892560437390273\\
-0.45	0.0760148280012944\\
-0.43	0.066433055013818\\
-0.41	0.059013689101182\\
-0.39	0.0530169060871595\\
-0.37	0.0480179102608387\\
-0.35	0.0437430871948971\\
-0.33	0.0400080781677084\\
-0.31	0.0366791427735499\\
-0.29	0.0336549120331073\\
-0.27	0.030855376831997\\
-0.25	0.0282100919720929\\
-0.23	0.0256824852653373\\
-0.21	0.0232056369324707\\
-0.19	0.0207479818064608\\
-0.17	0.0182761478409073\\
-0.15	0.0157641865637074\\
-0.13	0.01320575385526\\
-0.11	0.0106331433015509\\
-0.09	0.00816734785597\\
-0.07	0.0061833119537834\\
-0.05	0.0059432027132762\\
-0.03	0.0130784387102079\\
-0.01	0.0860079183007621\\
0.01	-0.0159574988359357\\
0.03	0.0570648447672704\\
0.05	0.0643656411678567\\
0.07	0.0643727682331348\\
0.09	0.0628048424681621\\
0.11	0.0609452285730319\\
0.13	0.0591757033078123\\
0.15	0.0576264901470807\\
0.17	0.0563447773548585\\
0.19	0.055347325104156\\
0.21	0.0546408955052342\\
0.23	0.0542339600192069\\
0.25	0.0541434590737881\\
0.27	0.0543730440082574\\
0.29	0.054956157978167\\
0.31	0.0559271118947092\\
0.33	0.0573354874533655\\
0.35	0.0592548556271449\\
0.37	0.0617912676422799\\
0.39	0.0650998631888555\\
0.41	0.0694160627332831\\
0.43	0.0751176579660217\\
0.45	0.0828699288747681\\
0.47	0.0940300485323304\\
0.49	0.112241411392435\\
};
\addlegendentry{ALM$^2$ ($\varepsilon = 7 \Delta x$)}
\legend{}
\end{axis}
\end{tikzpicture}%
%
%
\definecolor{mycolor1}{rgb}{0.00000,1.00000,1.00000}%
\begin{tikzpicture}

\begin{axis}[%
width=0.24\textwidth,
height=0.22\textwidth,
scale only axis,
xmin=-0.5,
xmax=0,
xtick={-0.5,-0.4,...,0.0},
xlabel style={font=\color{white!15!black}},
xlabel={$Y$},
ymin=-0.05,
ymax=0.2,
ytick={-0.05,0.0,...,0.2},
ylabel style={font=\color{white!15!black}},
ylabel={$u^c_y/U_0$},
yticklabel style={
        /pgf/number format/fixed,
        /pgf/number format/precision=3
},
axis background/.style={fill=white},
axis x line*=bottom,
axis y line*=left,
xmajorgrids,
ymajorgrids,
tick label style={font=\scriptsize},
legend style={legend cell align=left, align=left, draw=white!15!black}
]
\addplot [color=gray, line width=1.0pt, mark size=2.0pt, mark=square, mark options={solid, gray}]
  table[row sep=crcr]{%
-0.01	0.193075855492976\\
-0.03	0.0726362166992187\\
-0.05	0.0536495824252593\\
-0.07	0.0471586004673107\\
-0.09	0.0442984106654714\\
-0.11	0.042917086805576\\
-0.13	0.0422831407237593\\
-0.15	0.0420983735776715\\
-0.17	0.0422282667328561\\
-0.19	0.0426102439966215\\
-0.21	0.0432177137393622\\
-0.23	0.0440446967917522\\
-0.25	0.0450991113188161\\
-0.27	0.0464002145568087\\
-0.29	0.0479784377947623\\
-0.31	0.0498770327695711\\
-0.33	0.0521556186525242\\
-0.35	0.0548963296019732\\
-0.37	0.058214163952009\\
-0.39	0.0622749045394465\\
-0.41	0.067328050682292\\
-0.43	0.0737729513271775\\
-0.45	0.0823098747412744\\
-0.47	0.0943609189385187\\
-0.49	0.113737487976043\\
};
\addlegendentry{LL}

\addplot [color=blue, dashed, line width=1.0pt, mark size=1.5pt, mark=o, mark options={solid, blue}]
  table[row sep=crcr]{%
-0.01	0.184587885794883\\
-0.03	0.0796844473561589\\
-0.05	0.0608961841754528\\
-0.07	0.0521539533864692\\
-0.09	0.0468398320076046\\
-0.11	0.0435580833248337\\
-0.13	0.0417764274925749\\
-0.15	0.0409597671430133\\
-0.17	0.040815467903201\\
-0.19	0.041107608605528\\
-0.21	0.0417366688460421\\
-0.23	0.0426369735236483\\
-0.25	0.0437890438688154\\
-0.27	0.045205814185078\\
-0.29	0.0468923656651425\\
-0.31	0.0488965051797367\\
-0.33	0.051277194157309\\
-0.35	0.0541154426397505\\
-0.37	0.057528366276362\\
-0.39	0.0616916082132203\\
-0.41	0.0668759248396336\\
-0.43	0.0735153326517229\\
-0.45	0.08234413900905\\
-0.47	0.0948010402313096\\
-0.49	0.11466599816041\\
};
\addlegendentry{ALM$^1$ ($\varepsilon = 3.5 \Delta x$)}

\addplot [color=cyan, line width=1.0pt, only marks, mark size=2.0pt, mark=+, mark options={solid, cyan}]
  table[row sep=crcr]{%
-0.01	0.188304954277021\\
-0.03	0.0770088125829945\\
-0.05	0.0590042368043078\\
-0.07	0.0522786846527246\\
-0.09	0.0487610336999447\\
-0.11	0.0465542530252002\\
-0.13	0.0450649682963295\\
-0.15	0.0440769583414622\\
-0.17	0.0435044327367585\\
-0.19	0.0433105662465738\\
-0.21	0.0434759273187583\\
-0.23	0.0439845923351772\\
-0.25	0.0448158168938999\\
-0.27	0.0459970624348926\\
-0.29	0.0475162241867505\\
-0.31	0.0494067128067685\\
-0.33	0.0517181257377784\\
-0.35	0.0545244006916403\\
-0.37	0.0579341721719723\\
-0.39	0.0621093468579673\\
-0.41	0.0672977143216632\\
-0.43	0.0738954079391812\\
-0.45	0.082594960163371\\
-0.47	0.0948039162031942\\
-0.49	0.11428916782029\\
};
\addlegendentry{ALM$^1$ ($\varepsilon = 7 \Delta x$)}

\addplot [color=red, dotted, line width=1.0pt, mark size=1.0pt, mark=diamond, mark options={solid, red}]
  table[row sep=crcr]{%
-0.01	0.184363562837141\\
-0.03	0.0694832858447686\\
-0.05	0.0520296894692711\\
-0.07	0.0465803765257796\\
-0.09	0.0444984655909833\\
-0.11	0.0435720756750375\\
-0.13	0.0431688450000212\\
-0.15	0.0429983055680835\\
-0.17	0.04305716325479\\
-0.19	0.0433278273097547\\
-0.21	0.0438403926857941\\
-0.23	0.0445927977443358\\
-0.25	0.0455885061406686\\
-0.27	0.0468549502467328\\
-0.29	0.0483992887973839\\
-0.31	0.050269549977139\\
-0.33	0.0525230479199895\\
-0.35	0.0552381685595796\\
-0.37	0.0585292407426873\\
-0.39	0.0625696519310254\\
-0.41	0.0676290478335915\\
-0.43	0.0741419532276823\\
-0.45	0.082843982768252\\
-0.47	0.0951705147780734\\
-0.49	0.114880483579367\\
};
\addlegendentry{ALM$^2$ ($\varepsilon = 3.5 \Delta x$)}

\addplot [color=orange, line width=1.0pt, only marks, mark size=2.0pt, mark=x, mark options={solid, orange}]
  table[row sep=crcr]{%
-0.01	0.18818370905625\\
-0.03	0.0713960146936365\\
-0.05	0.0529470127068181\\
-0.07	0.046639766149846\\
-0.09	0.0439042623756973\\
-0.11	0.042650138884839\\
-0.13	0.0421566774818661\\
-0.15	0.0421176325652147\\
-0.17	0.0423872037789413\\
-0.19	0.0428934193286191\\
-0.21	0.043606797163367\\
-0.23	0.0445194618498039\\
-0.25	0.0456287410370119\\
-0.27	0.0469799447498944\\
-0.29	0.048589252057951\\
-0.31	0.0505089568236898\\
-0.33	0.0528058357384857\\
-0.35	0.0555672777231528\\
-0.37	0.0589119157778681\\
-0.39	0.0630072078763981\\
-0.41	0.0681054596273891\\
-0.43	0.0746025776582658\\
-0.45	0.0831893004107327\\
-0.47	0.0952648980385299\\
-0.49	0.114568294486268\\
};
\addlegendentry{ALM$^2$ ($\varepsilon = 7 \Delta x$)}
\legend{}
\end{axis}
\end{tikzpicture}%
%
%
\definecolor{mycolor1}{rgb}{0.00000,1.00000,1.00000}%
\begin{tikzpicture}

\begin{axis}[%
width=0.48\textwidth,
height=0.22\textwidth,
scale only axis,
xmin=-0.5,
xmax=0.5,
xtick={-0.5,-0.4,...,0.5},
xlabel style={font=\color{white!15!black}},
xlabel={$X$},
ymin=0,
ymax=2.5,
ytick={0,0.5,...,2.5},
ylabel style={font=\color{white!15!black}},
ylabel={$u^c_z/U_0$},
yticklabel style={
        /pgf/number format/fixed,
        /pgf/number format/precision=3
},
axis background/.style={fill=white},
axis x line*=bottom,
axis y line*=left,
xmajorgrids,
ymajorgrids,
tick label style={font=\scriptsize},
legend style={legend cell align=left, align=left, draw=white!15!black}
]
\addplot [color=gray, line width=1.0pt, mark size=2.0pt, mark=square, mark options={solid, gray}]
  table[row sep=crcr]{%
-0.49	1.00906123876027\\
-0.47	1.00966995992994\\
-0.45	1.01034028929593\\
-0.43	1.01108089497857\\
-0.41	1.01190209655226\\
-0.39	1.01281628343082\\
-0.37	1.0138384700592\\
-0.35	1.01498704362627\\
-0.33	1.01628478749587\\
-0.31	1.01776030720278\\
-0.29	1.01945005693273\\
-0.27	1.02140128341233\\
-0.25	1.02367640977896\\
-0.23	1.02635975031911\\
-0.21	1.02956813429394\\
-0.19	1.03346836221761\\
-0.17	1.0383072015491\\
-0.15	1.04446578477028\\
-0.13	1.05256503064677\\
-0.11	1.06368782307694\\
-0.09	1.07990162543629\\
-0.07	1.10568838243155\\
-0.05	1.152841540336\\
-0.03	1.26504842225429\\
-0.01	1.8336927418451\\
0.01	0.166307258154898\\
0.03	0.734951577745706\\
0.05	0.847158459663997\\
0.07	0.89431161756845\\
0.09	0.920098374563707\\
0.11	0.936312176923056\\
0.13	0.947434969353229\\
0.15	0.95553421522972\\
0.17	0.961692798450904\\
0.19	0.966531637782388\\
0.21	0.970431865706063\\
0.23	0.973640249680892\\
0.25	0.976323590221043\\
0.27	0.978598716587669\\
0.29	0.980549943067269\\
0.31	0.982239692797224\\
0.33	0.983715212504131\\
0.35	0.985012956373734\\
0.37	0.986161529940795\\
0.39	0.987183716569185\\
0.41	0.988097903447743\\
0.43	0.988919105021434\\
0.45	0.989659710704073\\
0.47	0.990330040070062\\
0.49	0.99093876123973\\
};
\addlegendentry{LL}

\addplot [color=blue, dashed, line width=1.0pt, mark size=1.5pt, mark=o, mark options={solid, blue}]
  table[row sep=crcr]{%
-0.49	1.01322547589829\\
-0.47	1.00929699437015\\
-0.45	1.00709548418765\\
-0.43	1.00636832495287\\
-0.41	1.00679464605449\\
-0.39	1.00797281018233\\
-0.37	1.00958841036297\\
-0.35	1.01146063914342\\
-0.33	1.01352594648698\\
-0.31	1.0157914726874\\
-0.29	1.01829305393221\\
-0.27	1.02108922922393\\
-0.25	1.02423064960913\\
-0.23	1.02787604728552\\
-0.21	1.0321900085552\\
-0.19	1.03733355543926\\
-0.17	1.04372289361809\\
-0.15	1.05172902737758\\
-0.13	1.0621809269519\\
-0.11	1.07563944077414\\
-0.09	1.09402250118569\\
-0.07	1.11909371431757\\
-0.05	1.15931596555066\\
-0.03	1.25009419774295\\
-0.01	1.73700168453959\\
0.01	0.232890465537977\\
0.03	0.72427766349478\\
0.05	0.818323703984649\\
0.07	0.861897294896165\\
0.09	0.889831339887535\\
0.11	0.910409889474263\\
0.13	0.925125619390769\\
0.15	0.936388517270579\\
0.17	0.944764561952359\\
0.19	0.951351265087628\\
0.21	0.956528695783387\\
0.23	0.960687031624055\\
0.25	0.964146454146508\\
0.27	0.966956787898795\\
0.29	0.969290902932512\\
0.31	0.97122439098004\\
0.33	0.972822185306621\\
0.35	0.974142476609606\\
0.37	0.975243653960159\\
0.39	0.976231939447482\\
0.41	0.977294000796237\\
0.43	0.978734471063921\\
0.45	0.980938386987414\\
0.47	0.984226477521334\\
0.49	0.988825694809427\\
};
\addlegendentry{ALM$^1$ ($\varepsilon = 3.5 \Delta x$)}

\addplot [color=cyan, line width=1.0pt, only marks, mark size=2.0pt, mark=+, mark options={solid, cyan}]
  table[row sep=crcr]{%
-0.49	1.01110236006013\\
-0.47	1.01041407496132\\
-0.45	1.01024148620339\\
-0.43	1.01043924388333\\
-0.41	1.01097130695083\\
-0.39	1.01182348093711\\
-0.37	1.01297913030208\\
-0.35	1.01442269290012\\
-0.33	1.0161478305082\\
-0.31	1.01816486730979\\
-0.29	1.02048892108007\\
-0.27	1.02313815301728\\
-0.25	1.02613100746911\\
-0.23	1.02962113021018\\
-0.21	1.03354448614722\\
-0.19	1.03806262363034\\
-0.17	1.04332412713329\\
-0.15	1.0495598416537\\
-0.13	1.05722080749323\\
-0.11	1.06719846735609\\
-0.09	1.08130897128992\\
-0.07	1.10360669791589\\
-0.05	1.14495917354169\\
-0.03	1.24573821232954\\
-0.01	1.76861144790786\\
0.01	0.213462751259539\\
0.03	0.738690244897096\\
0.05	0.841000662927588\\
0.07	0.8835731829\\
0.09	0.906989352799739\\
0.11	0.922137688273924\\
0.13	0.933037586848222\\
0.15	0.94146582330039\\
0.17	0.948266676576778\\
0.19	0.953885835888381\\
0.21	0.958576931108419\\
0.23	0.962515986421961\\
0.25	0.965872038137202\\
0.27	0.96867428302623\\
0.29	0.971075823820784\\
0.31	0.973151302086761\\
0.33	0.974966396581522\\
0.35	0.97659014963528\\
0.37	0.97809016741477\\
0.39	0.979530706649099\\
0.41	0.980963369396393\\
0.43	0.982433140261142\\
0.45	0.983982902796335\\
0.47	0.985666864389717\\
0.49	0.987601774905074\\
};
\addlegendentry{ALM$^1$ ($\varepsilon = 7 \Delta x$)}

\addplot [color=red, dotted, line width=1.0pt, mark size=1.0pt, mark=diamond, mark options={solid, red}]
  table[row sep=crcr]{%
-0.49	1.01238374365385\\
-0.47	1.0084140525104\\
-0.45	1.00614334379041\\
-0.43	1.00532806212881\\
-0.41	1.00564920673703\\
-0.39	1.00670357958442\\
-0.37	1.00817577545696\\
-0.35	1.00988265568399\\
-0.33	1.01175726993169\\
-0.31	1.01380156930099\\
-0.29	1.01604785295552\\
-0.27	1.01854647971362\\
-0.25	1.02134762894137\\
-0.23	1.02459257523759\\
-0.21	1.02843149452884\\
-0.19	1.03304087832728\\
-0.17	1.03875084398868\\
-0.15	1.04583993605526\\
-0.13	1.05492930531748\\
-0.11	1.06662438472141\\
-0.09	1.0829933697852\\
-0.07	1.10748084433896\\
-0.05	1.15185632750238\\
-0.03	1.25898640817962\\
-0.01	1.81663012973391\\
0.01	0.162802793373615\\
0.03	0.724661024895598\\
0.05	0.835428503315169\\
0.07	0.882518791908517\\
0.09	0.90843961196503\\
0.11	0.925276295453153\\
0.13	0.936708659843269\\
0.15	0.945366572085305\\
0.17	0.951907982958485\\
0.19	0.957142957108074\\
0.21	0.961324239945943\\
0.23	0.964737891566462\\
0.25	0.967588464350145\\
0.27	0.969919299437029\\
0.29	0.971877283338217\\
0.31	0.973497917644935\\
0.33	0.97483513085774\\
0.35	0.975938779439263\\
0.37	0.97685940780849\\
0.39	0.977693774358243\\
0.41	0.978625601596225\\
0.43	0.979955754221122\\
0.45	0.982065672528963\\
0.47	0.985268730468091\\
0.49	0.98978067658173\\
};
\addlegendentry{ALM$^2$ ($\varepsilon = 3.5 \Delta x$)}

\addplot [color=orange, line width=1.0pt, only marks, mark size=2.0pt, mark=x, mark options={solid, orange}]
  table[row sep=crcr]{%
-0.49	1.00993200533781\\
-0.47	1.00917408746932\\
-0.45	1.00891464260068\\
-0.43	1.00900735190842\\
-0.41	1.00941663012645\\
-0.39	1.01012695368581\\
-0.37	1.01112364623739\\
-0.35	1.0123816736939\\
-0.33	1.01389502956046\\
-0.31	1.01566558910059\\
-0.29	1.01770278317028\\
-0.27	1.0200235897552\\
-0.25	1.02265147732172\\
-0.23	1.02573273617309\\
-0.21	1.02924666023188\\
-0.19	1.03339146213665\\
-0.17	1.03837992250514\\
-0.15	1.04454853637107\\
-0.13	1.05248592082882\\
-0.11	1.06324571184173\\
-0.09	1.07885660739482\\
-0.07	1.10371628060909\\
-0.05	1.14941433749546\\
-0.03	1.25884559017133\\
-0.01	1.8170111715354\\
0.01	0.171078550651556\\
0.03	0.730760464330477\\
0.05	0.841182070649315\\
0.07	0.887635419692955\\
0.09	0.913141181975577\\
0.11	0.929298208706409\\
0.13	0.940482382104629\\
0.15	0.94870724649602\\
0.17	0.95501654038944\\
0.19	0.960000955517497\\
0.21	0.964014897689853\\
0.23	0.967300240404219\\
0.25	0.97005911537564\\
0.27	0.97234288432074\\
0.29	0.974298237561902\\
0.31	0.975995890133993\\
0.33	0.977491058654342\\
0.35	0.978845410893149\\
0.37	0.980117888662888\\
0.39	0.981363130791196\\
0.41	0.982628325831901\\
0.43	0.983952122010848\\
0.45	0.985374276243587\\
0.47	0.986942489459043\\
0.49	0.988764312173011\\
};
\addlegendentry{ALM$^2$ ($\varepsilon = 7 \Delta x$)}
\legend{}
\end{axis}
\end{tikzpicture}%
%
%
\definecolor{mycolor1}{rgb}{0.00000,1.00000,1.00000}%
\begin{tikzpicture}

\begin{axis}[%
width=0.24\textwidth,
height=0.22\textwidth,
scale only axis,
xmin=-0.5,
xmax=0,
xtick={-0.5,-0.4,...,0.0},
xlabel style={font=\color{white!15!black}},
xlabel={$Y$},
ymin=0,
ymax=2.5,
ytick={0,0.5,...,2.5},
ylabel style={font=\color{white!15!black}},
ylabel={$u^c_z/U_0$},
yticklabel style={
        /pgf/number format/fixed,
        /pgf/number format/precision=3
},
axis background/.style={fill=white},
axis x line*=bottom,
axis y line*=left,
xmajorgrids,
ymajorgrids,
tick label style={font=\scriptsize},
legend style={legend cell align=left, align=left, draw=white!15!black}
]
\addplot [color=gray, line width=1.0pt, mark size=2.0pt, mark=square, mark options={solid, gray}]
  table[row sep=crcr]{%
-0.01	2.26656932051349\\
-0.03	1.42071583429682\\
-0.05	1.25088910662217\\
-0.07	1.17771442253696\\
-0.09	1.13680022689333\\
-0.11	1.11058023537504\\
-0.13	1.09229569018037\\
-0.15	1.07879087887049\\
-0.17	1.06839386977009\\
-0.19	1.06013562269697\\
-0.21	1.05341528689192\\
-0.23	1.04784012029457\\
-0.25	1.04314225338877\\
-0.27	1.03913245639199\\
-0.29	1.03567304679604\\
-0.31	1.03266129231838\\
-0.33	1.03001885510672\\
-0.35	1.02768485899863\\
-0.37	1.0256112073773\\
-0.39	1.02375934238706\\
-0.41	1.02209795228638\\
-0.43	1.02060131746232\\
-0.45	1.0192480958695\\
-0.47	1.01802041665554\\
-0.49	1.01690319372822\\
};
\addlegendentry{LL}

\addplot [color=blue, dashed, line width=1.0pt, mark size=1.5pt, mark=o, mark options={solid, blue}]
  table[row sep=crcr]{%
-0.01	2.10048396297957\\
-0.03	1.37392850542885\\
-0.05	1.23093120084917\\
-0.07	1.16764094321067\\
-0.09	1.13064051509352\\
-0.11	1.10550527143438\\
-0.13	1.08785928175205\\
-0.15	1.07433782378444\\
-0.17	1.06392162126047\\
-0.19	1.05549929047072\\
-0.21	1.04863252684628\\
-0.23	1.04283317962964\\
-0.25	1.03785910225895\\
-0.27	1.03359856555384\\
-0.29	1.02979868966469\\
-0.31	1.02641090698287\\
-0.33	1.02335729132988\\
-0.35	1.02058575881674\\
-0.37	1.01807152563127\\
-0.39	1.0158638545185\\
-0.41	1.01411761080531\\
-0.43	1.01311320490945\\
-0.45	1.01320948553366\\
-0.47	1.01468586077796\\
-0.49	1.01773269516189\\
};
\addlegendentry{ALM$^1$ ($\varepsilon = 3.5 \Delta x$)}

\addplot [color=cyan, line width=1.0pt, only marks, mark size=2.0pt, mark=+, mark options={solid, cyan}]
  table[row sep=crcr]{%
-0.01	2.16544954836603\\
-0.03	1.38558815620665\\
-0.05	1.23037943273798\\
-0.07	1.1640378811158\\
-0.09	1.12712476661086\\
-0.11	1.10343705784248\\
-0.13	1.08679642565333\\
-0.15	1.07433742201362\\
-0.17	1.06455552238857\\
-0.19	1.05661267337263\\
-0.21	1.05001752892899\\
-0.23	1.04444037850601\\
-0.25	1.03957550582391\\
-0.27	1.03544154845488\\
-0.29	1.03180417112648\\
-0.31	1.02860590211642\\
-0.33	1.02580911907106\\
-0.35	1.02338182708002\\
-0.37	1.02129715576607\\
-0.39	1.01954938031922\\
-0.41	1.01814371058549\\
-0.43	1.01708701679933\\
-0.45	1.01637842354513\\
-0.47	1.01603116318286\\
-0.49	1.01614864224135\\
};
\addlegendentry{ALM$^1$ ($\varepsilon = 7 \Delta x$)}

\addplot [color=red, dotted, line width=1.0pt, mark size=1.0pt, mark=diamond, mark options={solid, red}]
  table[row sep=crcr]{%
-0.01	2.20514230018613\\
-0.03	1.39571006597088\\
-0.05	1.23537197742814\\
-0.07	1.16671379816055\\
-0.09	1.12840930065861\\
-0.11	1.10325071580152\\
-0.13	1.08573558598684\\
-0.15	1.07234230949053\\
-0.17	1.06199534020713\\
-0.19	1.05364813012524\\
-0.21	1.04688926925394\\
-0.23	1.04124634166346\\
-0.25	1.03642245877816\\
-0.27	1.03231052282693\\
-0.29	1.02865016735017\\
-0.31	1.02538772443443\\
-0.33	1.02244550035614\\
-0.35	1.01977144012257\\
-0.37	1.01734263744016\\
-0.39	1.01520928680245\\
-0.41	1.01352556234493\\
-0.43	1.01256992866195\\
-0.45	1.01269906498697\\
-0.47	1.01418748100523\\
-0.49	1.01721313862638\\
};
\addlegendentry{ALM$^2$ ($\varepsilon = 3.5 \Delta x$)}

\addplot [color=orange, line width=1.0pt, only marks, mark size=2.0pt, mark=x, mark options={solid, orange}]
  table[row sep=crcr]{%
-0.01	2.22415392254697\\
-0.03	1.4035038836385\\
-0.05	1.23907188479795\\
-0.07	1.16835933514471\\
-0.09	1.12890943385535\\
-0.11	1.10366837132228\\
-0.13	1.0860854673301\\
-0.15	1.07308590139005\\
-0.17	1.06303889398554\\
-0.19	1.05500327180878\\
-0.21	1.04841767527016\\
-0.23	1.04290874907189\\
-0.25	1.03814766613655\\
-0.27	1.03410836803075\\
-0.29	1.03057070522887\\
-0.31	1.02746671837762\\
-0.33	1.02475527854112\\
-0.35	1.02240380589022\\
-0.37	1.02039002192372\\
-0.39	1.01870410732435\\
-0.41	1.01735583666655\\
-0.43	1.01634745692469\\
-0.45	1.01567682542299\\
-0.47	1.01535563598671\\
-0.49	1.01548232825247\\
};
\addlegendentry{ALM$^2$ ($\varepsilon = 7 \Delta x$)}
\legend{}
\end{axis}
\end{tikzpicture}%
  \caption{Comparison of the velocity results for the T-tail (legend in figure \ref{fig:res_resultswinglet}).}
  \label{fig:res_resultsutail}
\end{figure}
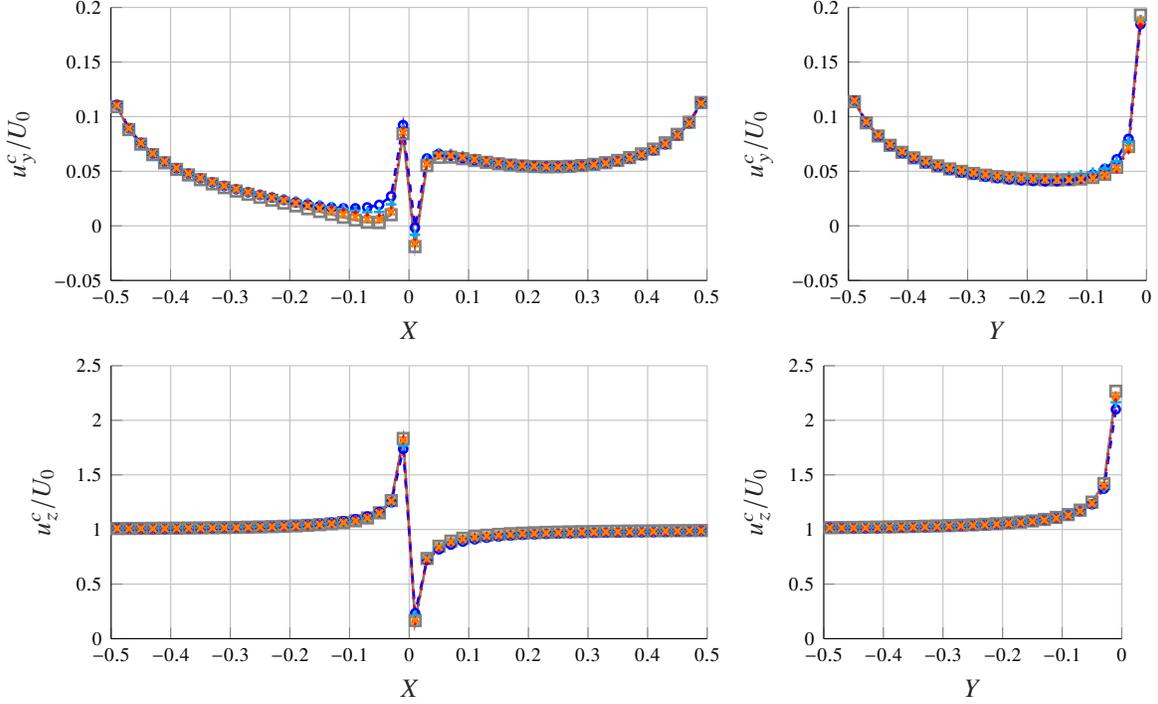

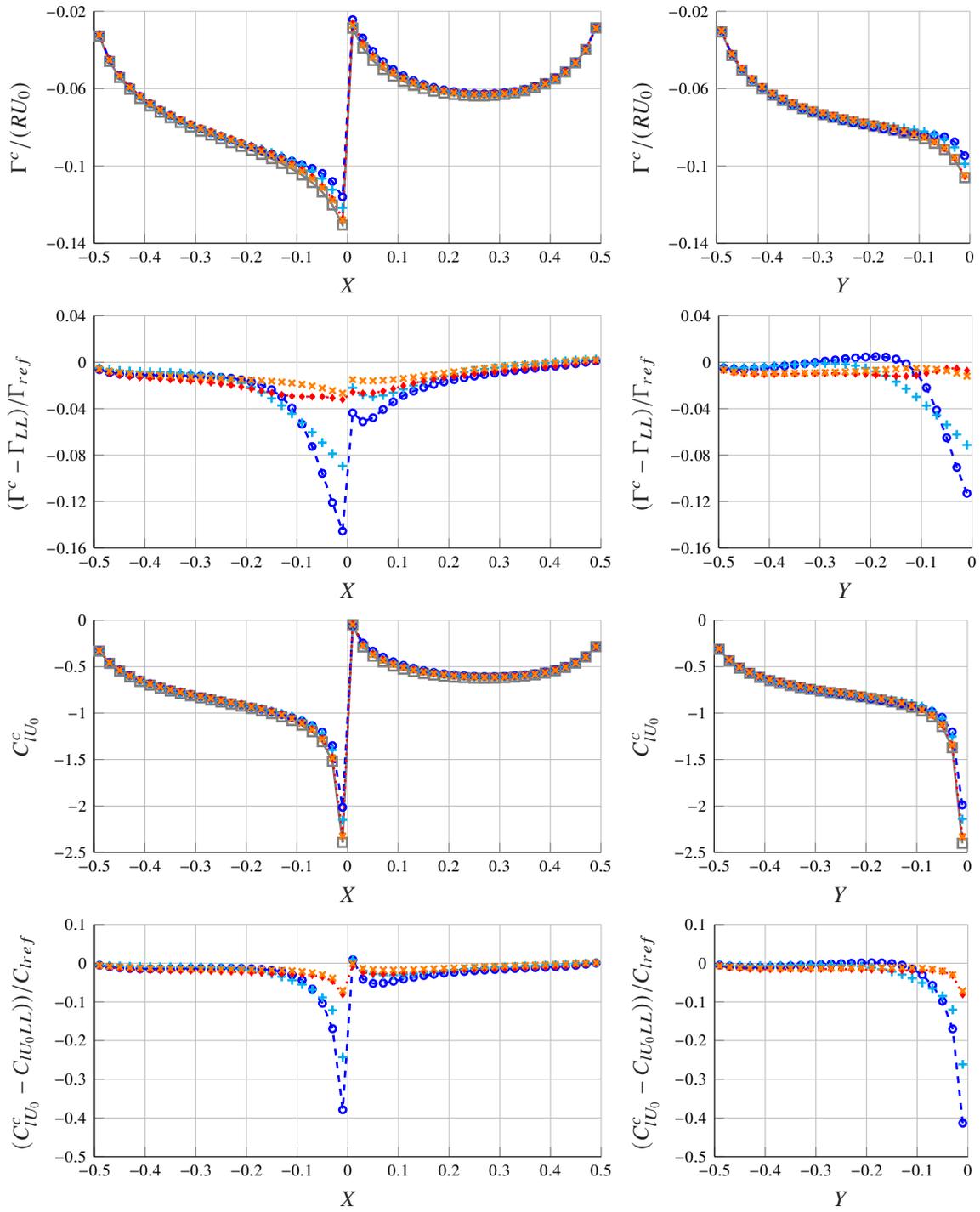
\begin{figure}
  \centering
%
%
\definecolor{mycolor1}{rgb}{0.00000,1.00000,1.00000}%
\begin{tikzpicture}

\begin{axis}[%
width=0.48\textwidth,
height=0.22\textwidth,
scale only axis,
xmin=-0.5,
xmax=0.5,
xtick={-0.5,-0.4,...,0.5},
xlabel style={font=\color{white!15!black}},
xlabel={$X$},
ymin=-0.14,
ymax=-0.02,
ytick={-0.14,-0.10,...,-0.02},
ylabel style={font=\color{white!15!black}},
ylabel={$\Gamma^c/(RU_0)$},
yticklabel style={
        /pgf/number format/fixed,
        /pgf/number format/precision=3
},
axis background/.style={fill=white},
axis x line*=bottom,
axis y line*=left,
xmajorgrids,
ymajorgrids,
tick label style={font=\scriptsize},
legend style={legend cell align=left, align=left, draw=white!15!black}
]
\addplot [color=gray, line width=1.0pt, mark size=2.0pt, mark=square, mark options={solid, gray}]
  table[row sep=crcr]{%
-0.49	-0.0328067827830563\\
-0.47	-0.0459184335387533\\
-0.45	-0.0542497827945412\\
-0.43	-0.0603122506881733\\
-0.41	-0.0650446016285001\\
-0.39	-0.0689114355608355\\
-0.37	-0.0721803183585615\\
-0.35	-0.0750212874002682\\
-0.33	-0.0775505242597766\\
-0.31	-0.0798522672198918\\
-0.29	-0.0819909615059229\\
-0.27	-0.0840185903997347\\
-0.25	-0.0859794844481914\\
-0.23	-0.0879137994440231\\
-0.21	-0.0898603653663066\\
-0.19	-0.0918594085045743\\
-0.17	-0.0939556222704886\\
-0.15	-0.0962021978206325\\
-0.13	-0.0986668094586104\\
-0.11	-0.101441450484771\\
-0.09	-0.104660195447924\\
-0.07	-0.108534796694118\\
-0.05	-0.113436134686281\\
-0.03	-0.120118713700095\\
-0.01	-0.130516400032784\\
0.01	-0.0287254147700067\\
0.03	-0.0390094250907907\\
0.05	-0.0454849372831686\\
0.07	-0.0500734177571155\\
0.09	-0.0535255054134021\\
0.11	-0.0562067802951508\\
0.13	-0.0583216932531573\\
0.15	-0.0599948037139895\\
0.17	-0.0613060620955611\\
0.19	-0.0623081054896691\\
0.21	-0.0630354396783287\\
0.23	-0.0635095257979071\\
0.25	-0.0637415741706683\\
0.27	-0.0637338933779707\\
0.29	-0.0634801720068965\\
0.31	-0.0629647871342371\\
0.33	-0.0621609964768559\\
0.35	-0.0610275749352918\\
0.37	-0.0595029537871117\\
0.39	-0.0574948977413971\\
0.41	-0.0548613842330836\\
0.43	-0.051372086883589\\
0.45	-0.0466203321333587\\
0.47	-0.0397779068568055\\
0.49	-0.0286246103375325\\
};
\addlegendentry{LL}

\addplot [color=blue, dashed, line width=1.0pt, mark size=1.5pt, mark=o, mark options={solid, blue}]
  table[row sep=crcr]{%
-0.49	-0.0321666821093764\\
-0.47	-0.0450387451690718\\
-0.45	-0.0532392251323422\\
-0.43	-0.0592333672671439\\
-0.41	-0.0639358213124792\\
-0.39	-0.067790846164531\\
-0.37	-0.0710543558461963\\
-0.35	-0.0738890244574598\\
-0.33	-0.0764067388884283\\
-0.31	-0.0786893187815081\\
-0.29	-0.0807979213846852\\
-0.27	-0.0827820195079605\\
-0.25	-0.0846799602070804\\
-0.23	-0.0865300921117974\\
-0.21	-0.0883523150500595\\
-0.19	-0.0901683034655641\\
-0.17	-0.0919979631897019\\
-0.15	-0.093839297068366\\
-0.13	-0.09568517103789\\
-0.11	-0.0974946197651031\\
-0.09	-0.0993183283011149\\
-0.07	-0.101270363999607\\
-0.05	-0.103857529285778\\
-0.03	-0.108015338985565\\
-0.01	-0.115961755141279\\
0.01	-0.0243598478374724\\
0.03	-0.0338816855740531\\
0.05	-0.0407008851269684\\
0.07	-0.0460015241486536\\
0.09	-0.0501043437765783\\
0.11	-0.053326836594337\\
0.13	-0.0558328797445099\\
0.15	-0.0578266473268386\\
0.17	-0.0593948070871183\\
0.19	-0.0606154675130418\\
0.21	-0.0615294089273831\\
0.23	-0.0621697980293326\\
0.25	-0.0625496905209404\\
0.27	-0.0626749732364857\\
0.29	-0.0625457405145487\\
0.31	-0.0621452322580619\\
0.33	-0.0614482165426837\\
0.35	-0.0604147341355057\\
0.37	-0.0589844235266533\\
0.39	-0.0570671043827505\\
0.41	-0.0545249435556773\\
0.43	-0.0511326835199581\\
0.45	-0.0464914222407014\\
0.47	-0.0397717159740011\\
0.49	-0.0287284042684757\\
};
\addlegendentry{ALM$^1$ ($\varepsilon = 3.5 \Delta x$)}

\addplot [color=cyan, line width=1.0pt, only marks, mark size=2.0pt, mark=+, mark options={solid, cyan}]
  table[row sep=crcr]{%
-0.49	-0.0323132343266042\\
-0.47	-0.0452787469420423\\
-0.45	-0.0535340788152141\\
-0.43	-0.0595492570198915\\
-0.41	-0.0642486389770209\\
-0.39	-0.0680888867845826\\
-0.37	-0.0713318069038161\\
-0.35	-0.0741424826426509\\
-0.33	-0.0766315862285427\\
-0.31	-0.0788773933031448\\
-0.29	-0.0809371689069997\\
-0.27	-0.082854134261035\\
-0.25	-0.0846640703171155\\
-0.23	-0.0863922281986369\\
-0.21	-0.0880661242645054\\
-0.19	-0.0897152297551676\\
-0.17	-0.0913732810543406\\
-0.15	-0.0930890923240421\\
-0.13	-0.0949326980258619\\
-0.11	-0.0970030768585842\\
-0.09	-0.0994472184470712\\
-0.07	-0.102490075923653\\
-0.05	-0.106507569745264\\
-0.03	-0.112239820858093\\
-0.01	-0.1215804898373\\
0.01	-0.0265217425812692\\
0.03	-0.0361807361497388\\
0.05	-0.0425146976572084\\
0.07	-0.0472098445680033\\
0.09	-0.0509014259336063\\
0.11	-0.05388365267205\\
0.13	-0.0563138741659356\\
0.15	-0.0582860193666376\\
0.17	-0.0598655195317766\\
0.19	-0.0611004539329109\\
0.21	-0.0620256610538403\\
0.23	-0.062666126612933\\
0.25	-0.0630396374543769\\
0.27	-0.0631532437789646\\
0.29	-0.0630062456859689\\
0.31	-0.0625873909949372\\
0.33	-0.0618721416153277\\
0.35	-0.0608218232906645\\
0.37	-0.0593771367752628\\
0.39	-0.0574481439915824\\
0.41	-0.0548929134911548\\
0.43	-0.0514806855750925\\
0.45	-0.0468018642590958\\
0.47	-0.0400197165839948\\
0.49	-0.0288829952306614\\
};
\addlegendentry{ALM$^1$ ($\varepsilon = 7 \Delta x$)}

\addplot [color=red, dotted, line width=1.0pt, mark size=1.0pt, mark=diamond, mark options={solid, red}]
  table[row sep=crcr]{%
-0.49	-0.0321278968584084\\
-0.47	-0.0449671310879927\\
-0.45	-0.0531281497492769\\
-0.43	-0.0590756583101948\\
-0.41	-0.0637275079887231\\
-0.39	-0.0675317813113258\\
-0.37	-0.0707467225887556\\
-0.35	-0.0735356212768014\\
-0.33	-0.0760098544541667\\
-0.31	-0.0782502314632288\\
-0.29	-0.0803172588797738\\
-0.27	-0.0822604524338607\\
-0.25	-0.0841202947081678\\
-0.23	-0.0859386803918889\\
-0.21	-0.0877445904358492\\
-0.19	-0.0895793067997848\\
-0.17	-0.091496070484325\\
-0.15	-0.0935580588081571\\
-0.13	-0.0958623381012527\\
-0.11	-0.0985155394564479\\
-0.09	-0.101697882658304\\
-0.07	-0.105567664265313\\
-0.05	-0.110469260269268\\
-0.03	-0.117044751792656\\
-0.01	-0.127294491238416\\
0.01	-0.0261554620920273\\
0.03	-0.0363389397830904\\
0.05	-0.0428308800617644\\
0.07	-0.0475762221210254\\
0.09	-0.0512386232465428\\
0.11	-0.0541783314382108\\
0.13	-0.0565104064109075\\
0.15	-0.0583791067423178\\
0.17	-0.0598481484970024\\
0.19	-0.0609852092303038\\
0.21	-0.0618273331613553\\
0.23	-0.0624056751880726\\
0.25	-0.0627373438226829\\
0.27	-0.0628193564590716\\
0.29	-0.0626529915535258\\
0.31	-0.0622223397100376\\
0.33	-0.0615011430767096\\
0.35	-0.0604489798164669\\
0.37	-0.0590055994434671\\
0.39	-0.0570808787439695\\
0.41	-0.0545359701649552\\
0.43	-0.0511432022756479\\
0.45	-0.0465003569975121\\
0.47	-0.0397754175944492\\
0.49	-0.028723799310649\\
};
\addlegendentry{ALM$^2$ ($\varepsilon = 3.5 \Delta x$)}

\addplot [color=orange, line width=1.0pt, only marks, mark size=2.0pt, mark=x, mark options={solid, orange}]
  table[row sep=crcr]{%
-0.49	-0.0322344080799251\\
-0.47	-0.045152448916951\\
-0.45	-0.0533687430347008\\
-0.43	-0.0593485760310271\\
-0.41	-0.0640157390723182\\
-0.39	-0.0678277385861064\\
-0.37	-0.0710472700500544\\
-0.35	-0.0738421116278686\\
-0.33	-0.0763263885366618\\
-0.31	-0.0785835486580263\\
-0.29	-0.0806776774858224\\
-0.27	-0.0826603449848827\\
-0.25	-0.0845778546458779\\
-0.23	-0.0864676463509281\\
-0.21	-0.0883694978107225\\
-0.19	-0.0903229856760644\\
-0.17	-0.0923702736397719\\
-0.15	-0.0945612269510275\\
-0.13	-0.09695872922888\\
-0.11	-0.0996479393068741\\
-0.09	-0.102755836387528\\
-0.07	-0.106486786217644\\
-0.05	-0.111207144548673\\
-0.03	-0.1176728485269\\
-0.01	-0.127849099918117\\
0.01	-0.0272230059652962\\
0.03	-0.0374052297112132\\
0.05	-0.0438806134847053\\
0.07	-0.0485135619334823\\
0.09	-0.0520371692872896\\
0.11	-0.0548096513279305\\
0.13	-0.0570297722599736\\
0.15	-0.0588172438496003\\
0.17	-0.0602467483501143\\
0.19	-0.0613667354426723\\
0.21	-0.0622083201055121\\
0.23	-0.0627903331866375\\
0.25	-0.0631223705390622\\
0.27	-0.0632072737061104\\
0.29	-0.0630388379794519\\
0.31	-0.0626027479448563\\
0.33	-0.0618736413387064\\
0.35	-0.0608118784700118\\
0.37	-0.059357358955941\\
0.39	-0.0574190635578364\\
0.41	-0.0548554076029371\\
0.43	-0.0514356890601761\\
0.45	-0.0467512906085035\\
0.47	-0.0399663732355581\\
0.49	-0.028834125868028\\
};
\addlegendentry{ALM$^2$ ($\varepsilon = 7 \Delta x$)}
\legend{}
\end{axis}
\end{tikzpicture}%
%
%
\definecolor{mycolor1}{rgb}{0.00000,1.00000,1.00000}%
\begin{tikzpicture}

\begin{axis}[%
width=0.24\textwidth,
height=0.22\textwidth,
scale only axis,
xmin=-0.5,
xmax=0,
xtick={-0.5,-0.4,...,0.0},
xlabel style={font=\color{white!15!black}},
xlabel={$Y$},
ymin=-0.14,
ymax=-0.02,
ytick={-0.14,-0.10,...,-0.02},
ylabel style={font=\color{white!15!black}},
ylabel={$\Gamma^c/(RU_0)$},
yticklabel style={
        /pgf/number format/fixed,
        /pgf/number format/precision=3
},
axis background/.style={fill=white},
axis x line*=bottom,
axis y line*=left,
xmajorgrids,
ymajorgrids,
tick label style={font=\scriptsize},
legend style={legend cell align=left, align=left, draw=white!15!black}
]
\addplot [color=gray, line width=1.0pt, mark size=2.0pt, mark=square, mark options={solid, gray}]
  table[row sep=crcr]{%
-0.01	-0.106018456837891\\
-0.03	-0.0965985453480206\\
-0.05	-0.0914844966924059\\
-0.07	-0.0882272318018229\\
-0.09	-0.0859256965021193\\
-0.11	-0.0841685593144709\\
-0.13	-0.0827374196886608\\
-0.15	-0.0815032566165577\\
-0.17	-0.0803830598092924\\
-0.19	-0.0793191106945825\\
-0.21	-0.078268004686519\\
-0.23	-0.0771942620539505\\
-0.25	-0.0760662095854821\\
-0.27	-0.0748529640204278\\
-0.29	-0.0735218350611387\\
-0.31	-0.0720356490782731\\
-0.33	-0.0703494951543498\\
-0.35	-0.0684062188866103\\
-0.37	-0.0661295348374056\\
-0.39	-0.0634125885169794\\
-0.41	-0.0600973072443751\\
-0.43	-0.0559332359512392\\
-0.45	-0.0504837746489463\\
-0.47	-0.042865202443718\\
-0.49	-0.0307130424884618\\
};
\addlegendentry{LL}

\addplot [color=blue, dashed, line width=1.0pt, mark size=1.5pt, mark=o, mark options={solid, blue}]
  table[row sep=crcr]{%
-0.01	-0.094719249687069\\
-0.03	-0.0875315852867613\\
-0.05	-0.0849693380179494\\
-0.07	-0.084103458923286\\
-0.09	-0.0837250659930462\\
-0.11	-0.0832633982121866\\
-0.13	-0.0826131375095383\\
-0.15	-0.0817721381083149\\
-0.17	-0.0808213405509687\\
-0.19	-0.0797971104927355\\
-0.21	-0.0787178661514818\\
-0.23	-0.0775759011631345\\
-0.25	-0.076359234912641\\
-0.27	-0.0750486809704016\\
-0.29	-0.0736158664696526\\
-0.31	-0.0720261733687039\\
-0.33	-0.0702351199728762\\
-0.35	-0.0681870896412556\\
-0.37	-0.0658067112357126\\
-0.39	-0.0629895724414202\\
-0.41	-0.0595828229217791\\
-0.43	-0.0553449193171122\\
-0.45	-0.0498562488181822\\
-0.47	-0.0422538175509208\\
-0.49	-0.0302152675572737\\
};
\addlegendentry{ALM$^1$ ($\varepsilon = 3.5 \Delta x$)}

\addplot [color=cyan, line width=1.0pt, only marks, mark size=2.0pt, mark=+, mark options={solid, cyan}]
  table[row sep=crcr]{%
-0.01	-0.0988880801207299\\
-0.03	-0.0903652501809973\\
-0.05	-0.086095140634226\\
-0.07	-0.0836655600613052\\
-0.09	-0.0821738153036522\\
-0.11	-0.081185076255014\\
-0.13	-0.0804524473127224\\
-0.15	-0.0798245935982321\\
-0.17	-0.0792047312796872\\
-0.19	-0.0785321078602576\\
-0.21	-0.0777697692700541\\
-0.23	-0.0768947398535615\\
-0.25	-0.0758894689705845\\
-0.27	-0.074738726953787\\
-0.29	-0.0734267350651143\\
-0.31	-0.0719269857688059\\
-0.33	-0.0702048812278178\\
-0.35	-0.0682112369059191\\
-0.37	-0.0658757724371181\\
-0.39	-0.0630971652867102\\
-0.41	-0.0597219463016558\\
-0.43	-0.055504860029799\\
-0.45	-0.0500163638646443\\
-0.47	-0.042386199440853\\
-0.49	-0.0302914095301771\\
};
\addlegendentry{ALM$^1$ ($\varepsilon = 7 \Delta x$)}

\addplot [color=red, dotted, line width=1.0pt, mark size=1.0pt, mark=diamond, mark options={solid, red}]
  table[row sep=crcr]{%
-0.01	-0.105301052179627\\
-0.03	-0.0960721395153656\\
-0.05	-0.0909489433833633\\
-0.07	-0.0874919603152838\\
-0.09	-0.0849647588127891\\
-0.11	-0.0830293501255855\\
-0.13	-0.0815309289147216\\
-0.15	-0.080299049370003\\
-0.17	-0.0792282941738664\\
-0.19	-0.0782250808637755\\
-0.21	-0.0772295163512892\\
-0.23	-0.0761957089452567\\
-0.25	-0.0750918658016357\\
-0.27	-0.0738902028561342\\
-0.29	-0.0725602971833465\\
-0.31	-0.0710668816387261\\
-0.33	-0.0693665243468871\\
-0.35	-0.0674052466863441\\
-0.37	-0.0651095854750614\\
-0.39	-0.0623766588235813\\
-0.41	-0.059054248778268\\
-0.43	-0.0549002987682971\\
-0.45	-0.0494941619793486\\
-0.47	-0.0419743960949455\\
-0.49	-0.030030310663026\\
};
\addlegendentry{ALM$^2$ ($\varepsilon = 3.5 \Delta x$)}

\addplot [color=orange, line width=1.0pt, only marks, mark size=2.0pt, mark=x, mark options={solid, orange}]
  table[row sep=crcr]{%
-0.01	-0.104820697962927\\
-0.03	-0.0956570418391918\\
-0.05	-0.0907457251442942\\
-0.07	-0.087619299796528\\
-0.09	-0.0853859920568181\\
-0.11	-0.0836471120724954\\
-0.13	-0.0821982781948205\\
-0.15	-0.0809235640537411\\
-0.17	-0.0797511144112685\\
-0.19	-0.0786318625701592\\
-0.21	-0.0775281535343264\\
-0.23	-0.0764076261792178\\
-0.25	-0.0752391250690048\\
-0.27	-0.0739917942231268\\
-0.29	-0.0726335926237587\\
-0.31	-0.0711251581926219\\
-0.33	-0.0694208024986193\\
-0.35	-0.067462872703787\\
-0.37	-0.0651752801555559\\
-0.39	-0.062452856016921\\
-0.41	-0.0591397767826544\\
-0.43	-0.0549904672786147\\
-0.45	-0.0495763767580638\\
-0.47	-0.0420322151150573\\
-0.49	-0.0300516881568947\\
};
\addlegendentry{ALM$^2$ ($\varepsilon = 7 \Delta x$)}
\legend{}
\end{axis}
\end{tikzpicture}%
%
%
\definecolor{mycolor1}{rgb}{0.00000,1.00000,1.00000}%
\begin{tikzpicture}

\begin{axis}[%
width=0.48\textwidth,
height=0.22\textwidth,
scale only axis,
xmin=-0.5,
xmax=0.5,
xtick={-0.5,-0.4,...,0.5},
xlabel style={font=\color{white!15!black}},
xlabel={$X$},
ymin=-0.16,
ymax=0.04,
ytick={-0.16,-0.12,...,0.04},
ylabel style={font=\color{white!15!black}},
ylabel={$(\Gamma^c-\Gamma_{LL})/\Gamma_{ref}$},
yticklabel style={
        /pgf/number format/fixed,
        /pgf/number format/precision=3
},
axis background/.style={fill=white},
axis x line*=bottom,
axis y line*=left,
xmajorgrids,
ymajorgrids,
tick label style={font=\scriptsize},
legend style={legend cell align=left, align=left, draw=white!15!black}
]
\addplot [color=blue, dashed, line width=1.0pt, mark size=1.5pt, mark=o, mark options={solid, blue}]
  table[row sep=crcr]{%
-0.49	-0.00640100673604122\\
-0.47	-0.00879688369577412\\
-0.45	-0.0101055766207939\\
-0.43	-0.0107888342090175\\
-0.41	-0.0110878031588967\\
-0.39	-0.0112058939617195\\
-0.37	-0.0112596251223203\\
-0.35	-0.0113226294267445\\
-0.33	-0.0114378537121296\\
-0.31	-0.0116294843824611\\
-0.29	-0.0119304012109656\\
-0.27	-0.0123657089162793\\
-0.25	-0.0129952424095727\\
-0.23	-0.0138370733206197\\
-0.21	-0.0150805031606872\\
-0.19	-0.0169110503881013\\
-0.17	-0.0195765908055512\\
-0.15	-0.0236290075198696\\
-0.13	-0.0298163842036757\\
-0.11	-0.0394683071920089\\
-0.09	-0.0534186714617732\\
-0.07	-0.0726443269365153\\
-0.05	-0.0957860539936888\\
-0.03	-0.12103374713098\\
-0.01	-0.145546448897826\\
0.01	-0.0436556693201775\\
0.03	-0.0512773951613086\\
0.05	-0.0478405215563419\\
0.07	-0.0407189360798012\\
0.09	-0.0342116163641903\\
0.11	-0.0287994370047309\\
0.13	-0.0248881350835289\\
0.15	-0.021681563868944\\
0.17	-0.0191125500821669\\
0.19	-0.0169263797642701\\
0.21	-0.0150603075076736\\
0.23	-0.01339727768416\\
0.25	-0.0119188364958692\\
0.27	-0.0105892014135969\\
0.29	-0.00934431492237266\\
0.31	-0.00819554876078218\\
0.33	-0.00712779934087906\\
0.35	-0.00612840799713548\\
0.37	-0.0051853026039709\\
0.39	-0.0042779335859595\\
0.41	-0.00336440677366455\\
0.43	-0.00239403363602611\\
0.45	-0.00128909892642025\\
0.47	-6.1908828036419e-05\\
0.49	0.00103793930930904\\
};
\addlegendentry{ALM$^1$ ($\varepsilon = 3.5 \Delta x$)}

\addplot [color=cyan, line width=1.0pt, only marks, mark size=2.0pt, mark=+, mark options={solid, cyan}]
  table[row sep=crcr]{%
-0.49	-0.00493548456393663\\
-0.47	-0.00639686596635312\\
-0.45	-0.00715703979242379\\
-0.43	-0.00762993668191523\\
-0.41	-0.00795962651384981\\
-0.39	-0.00822548776155618\\
-0.37	-0.00848511454645049\\
-0.35	-0.00878804757513346\\
-0.33	-0.00918938031125169\\
-0.31	-0.00974873916631665\\
-0.29	-0.0105379259879853\\
-0.27	-0.0116445613856197\\
-0.25	-0.0131541413092029\\
-0.23	-0.0152157124520615\\
-0.21	-0.0179424110158895\\
-0.19	-0.0214417874915303\\
-0.17	-0.025823412158425\\
-0.15	-0.0311310549622209\\
-0.13	-0.0373411143230662\\
-0.11	-0.0443837362566163\\
-0.09	-0.0521297700023626\\
-0.07	-0.0604472076975014\\
-0.05	-0.0692856494019715\\
-0.03	-0.0787889284106987\\
-0.01	-0.0893591019442632\\
0.01	-0.0220367218847675\\
0.03	-0.0282868894071718\\
0.05	-0.029702396256088\\
0.07	-0.0286357318877339\\
0.09	-0.0262407947948535\\
0.11	-0.0232312762282597\\
0.13	-0.020078190869841\\
0.15	-0.0170878434714976\\
0.17	-0.0144054256361408\\
0.19	-0.0120765155661529\\
0.21	-0.0100977862436887\\
0.23	-0.00843399184874322\\
0.25	-0.00701936716208389\\
0.27	-0.0058064959893738\\
0.29	-0.00473926320871556\\
0.31	-0.00377396139255239\\
0.33	-0.00288854861494065\\
0.35	-0.00205751644602914\\
0.37	-0.00125817011834052\\
0.39	-0.000467537498091333\\
0.41	0.000315292580675028\\
0.43	0.00108598691490611\\
0.45	0.00181532125715641\\
0.47	0.00241809727160714\\
0.49	0.00258384893098315\\
};
\addlegendentry{ALM$^1$ ($\varepsilon = 7 \Delta x$)}

\addplot [color=red, dotted, line width=1.0pt, mark size=1.0pt, mark=diamond, mark options={solid, red}]
  table[row sep=crcr]{%
-0.49	-0.00678885924567529\\
-0.47	-0.00951302450648044\\
-0.45	-0.0112163304513156\\
-0.43	-0.0123659237783219\\
-0.41	-0.0131709363962112\\
-0.39	-0.013796542493465\\
-0.37	-0.0143359576963633\\
-0.35	-0.0148566612329105\\
-0.33	-0.015406698054276\\
-0.31	-0.0160203575647346\\
-0.29	-0.0167370262595108\\
-0.27	-0.0175813796566603\\
-0.25	-0.0185918973980364\\
-0.23	-0.0197511905190049\\
-0.21	-0.0211577493020711\\
-0.19	-0.0228010170451974\\
-0.17	-0.0245955178587264\\
-0.15	-0.0264413901216258\\
-0.13	-0.0280447135702582\\
-0.11	-0.0292591102797687\\
-0.09	-0.0296231278926955\\
-0.07	-0.0296713242845389\\
-0.05	-0.02966874416662\\
-0.03	-0.0307396190707509\\
-0.01	-0.0322190879398621\\
0.01	-0.0256995267767531\\
0.03	-0.026704853073843\\
0.05	-0.0265405722109021\\
0.07	-0.0249719563579464\\
0.09	-0.0228688216658874\\
0.11	-0.0202844885670004\\
0.13	-0.0181128684203545\\
0.15	-0.0161569697148057\\
0.17	-0.0145791359838623\\
0.19	-0.0132289625920876\\
0.21	-0.0120810651683041\\
0.23	-0.011038506097039\\
0.25	-0.0100423034786663\\
0.27	-0.00914536918790866\\
0.29	-0.00827180453272854\\
0.31	-0.00742447424111641\\
0.33	-0.00659853400068267\\
0.35	-0.00578595118756401\\
0.37	-0.00497354343585792\\
0.39	-0.00414018997378579\\
0.41	-0.00325414068089864\\
0.43	-0.00228884607914059\\
0.45	-0.00119975135832383\\
0.47	-2.48926235598241e-05\\
0.49	0.000991889731047506\\
};
\addlegendentry{ALM$^2$ ($\varepsilon = 3.5 \Delta x$)}

\addplot [color=orange, line width=1.0pt, only marks, mark size=2.0pt, mark=x, mark options={solid, orange}]
  table[row sep=crcr]{%
-0.49	-0.00572374703063434\\
-0.47	-0.00765984621711669\\
-0.45	-0.00881039759736119\\
-0.43	-0.0096367465703218\\
-0.41	-0.0102886255606013\\
-0.39	-0.0108369697460092\\
-0.37	-0.0113304830837308\\
-0.35	-0.011791757722601\\
-0.33	-0.0122413572296996\\
-0.31	-0.012687185617154\\
-0.29	-0.0131328401994513\\
-0.27	-0.0135824541469134\\
-0.25	-0.0140162980214768\\
-0.23	-0.0144615309292387\\
-0.21	-0.0149086755540775\\
-0.19	-0.0153642282832815\\
-0.17	-0.0158534863052916\\
-0.15	-0.0164097086941087\\
-0.13	-0.0170808022952825\\
-0.11	-0.0179351117768467\\
-0.09	-0.0190435906017054\\
-0.07	-0.0204801047623214\\
-0.05	-0.022289901373439\\
-0.03	-0.024458651729055\\
-0.01	-0.0266730011435083\\
0.01	-0.0150240880453272\\
0.03	-0.0160419537938767\\
0.05	-0.0160432379827352\\
0.07	-0.0155985582344865\\
0.09	-0.0148833612593642\\
0.11	-0.0139712896705503\\
0.13	-0.0129192099303081\\
0.15	-0.0117755986424991\\
0.17	-0.0105931374532149\\
0.19	-0.00941370046885402\\
0.21	-0.00827119572718694\\
0.23	-0.00719192611184516\\
0.25	-0.0061920363153289\\
0.27	-0.00526619671797972\\
0.29	-0.00441334027392412\\
0.31	-0.00362039189337947\\
0.33	-0.00287355138115536\\
0.35	-0.0021569646525444\\
0.37	-0.00145594831153515\\
0.39	-0.000758341835516907\\
0.41	-5.97663014576038e-05\\
0.43	0.000636021765795349\\
0.45	0.00130958475129325\\
0.47	0.00188466378730325\\
0.49	0.00209515530470695\\
};
\addlegendentry{ALM$^2$ ($\varepsilon = 7 \Delta x$)}
\legend{}
\end{axis}
\end{tikzpicture}%
%
%
\definecolor{mycolor1}{rgb}{0.00000,1.00000,1.00000}%
\begin{tikzpicture}

\begin{axis}[%
width=0.24\textwidth,
height=0.22\textwidth,
scale only axis,
xmin=-0.5,
xmax=0,
xtick={-0.5,-0.4,...,0.0},
xlabel style={font=\color{white!15!black}},
xlabel={$Y$},
ymin=-0.16,
ymax=0.04,
ytick={-0.16,-0.12,...,0.04},
ylabel style={font=\color{white!15!black}},
ylabel={$(\Gamma^c-\Gamma_{LL})/\Gamma_{ref}$},
yticklabel style={
        /pgf/number format/fixed,
        /pgf/number format/precision=3
},
axis background/.style={fill=white},
axis x line*=bottom,
axis y line*=left,
xmajorgrids,
ymajorgrids,
tick label style={font=\scriptsize},
legend style={legend cell align=left, align=left, draw=white!15!black}
]
\addplot [color=blue, dashed, line width=1.0pt, mark size=1.5pt, mark=o, mark options={solid, blue}]
  table[row sep=crcr]{%
-0.01	-0.112992071494848\\
-0.03	-0.0906696006018649\\
-0.05	-0.0651515867368562\\
-0.07	-0.0412377287804894\\
-0.09	-0.0220063050881269\\
-0.11	-0.0090516110217718\\
-0.13	-0.00124282179107779\\
-0.15	0.0026888149172543\\
-0.17	0.00438280741624403\\
-0.19	0.00477999798096465\\
-0.21	0.00449861464909531\\
-0.23	0.00381639109138856\\
-0.25	0.00293025327124201\\
-0.27	0.00195716949950607\\
-0.29	0.000940314085027579\\
-0.31	-9.47570956808267e-05\\
-0.33	-0.00114375181460095\\
-0.35	-0.00219129245328821\\
-0.37	-0.00322823601654846\\
-0.39	-0.00423016075509172\\
-0.41	-0.00514484322535078\\
-0.43	-0.00588316634057361\\
-0.45	-0.00627525830689878\\
-0.47	-0.00611384892724857\\
-0.49	-0.00497774931129159\\
};
\addlegendentry{ALM$^1$ ($\varepsilon = 3.5 \Delta x$)}

\addplot [color=cyan, line width=1.0pt, only marks, mark size=2.0pt, mark=+, mark options={solid, cyan}]
  table[row sep=crcr]{%
-0.01	-0.0713037671631721\\
-0.03	-0.0623329516628578\\
-0.05	-0.0538935605754222\\
-0.07	-0.0456167173997792\\
-0.09	-0.0375188119802315\\
-0.11	-0.0298348305910387\\
-0.13	-0.0228497237566803\\
-0.15	-0.0167866301812694\\
-0.17	-0.0117832852946582\\
-0.19	-0.00787002834231749\\
-0.21	-0.0049823541640598\\
-0.23	-0.0029952220035354\\
-0.25	-0.00176740614876715\\
-0.27	-0.00114237066627318\\
-0.29	-0.000950999960131647\\
-0.31	-0.0010866330945434\\
-0.33	-0.00144613926514909\\
-0.35	-0.00194981980668177\\
-0.37	-0.00253762400257518\\
-0.39	-0.00315423230231908\\
-0.41	-0.00375360942674835\\
-0.43	-0.00428375921389483\\
-0.45	-0.00467410784246723\\
-0.47	-0.00479003002808322\\
-0.49	-0.0042163295823477\\
};
\addlegendentry{ALM$^1$ ($\varepsilon = 7 \Delta x$)}

\addplot [color=red, dotted, line width=1.0pt, mark size=1.0pt, mark=diamond, mark options={solid, red}]
  table[row sep=crcr]{%
-0.01	-0.007174046581788\\
-0.03	-0.00526405832592723\\
-0.05	-0.00535553308979224\\
-0.07	-0.00735271486452057\\
-0.09	-0.00960937689216472\\
-0.11	-0.0113920918875059\\
-0.13	-0.0120649077379643\\
-0.15	-0.0120420724641218\\
-0.17	-0.0115476563528941\\
-0.19	-0.0109402983067753\\
-0.21	-0.0103848833510696\\
-0.23	-0.0099855310857564\\
-0.25	-0.0097434378373115\\
-0.27	-0.00962761164179726\\
-0.29	-0.00961537877678442\\
-0.31	-0.0096876743943238\\
-0.33	-0.00982970807346415\\
-0.35	-0.0100097220014781\\
-0.37	-0.0101994936222356\\
-0.39	-0.0103592969327555\\
-0.41	-0.0104305846598363\\
-0.43	-0.0103293718281985\\
-0.45	-0.0098961266948064\\
-0.47	-0.00890806348667097\\
-0.49	-0.00682731825354975\\
};
\addlegendentry{ALM$^2$ ($\varepsilon = 3.5 \Delta x$)}

\addplot [color=orange, line width=1.0pt, only marks, mark size=2.0pt, mark=x, mark options={solid, orange}]
  table[row sep=crcr]{%
-0.01	-0.0119775887482217\\
-0.03	-0.00941503508717409\\
-0.05	-0.00738771548024282\\
-0.07	-0.00607932005222936\\
-0.09	-0.00539704445237315\\
-0.11	-0.00521447241913791\\
-0.13	-0.00539141493776489\\
-0.15	-0.00579692562747974\\
-0.17	-0.0063194539794917\\
-0.19	-0.00687248124341963\\
-0.21	-0.00739851152105093\\
-0.23	-0.00786635874639607\\
-0.25	-0.00827084516379463\\
-0.27	-0.00861169797199145\\
-0.29	-0.00888242437274921\\
-0.31	-0.00910490885543475\\
-0.33	-0.00928692655620636\\
-0.35	-0.00943346182711734\\
-0.37	-0.00954254681736837\\
-0.39	-0.00959732499944867\\
-0.41	-0.00957530461607357\\
-0.43	-0.00942768672512917\\
-0.45	-0.00907397890775168\\
-0.47	-0.00832987328562136\\
-0.49	-0.00661354331488806\\
};
\addlegendentry{ALM$^2$ ($\varepsilon = 7 \Delta x$)}
\legend{}
\end{axis}
\end{tikzpicture}%
%
%
\definecolor{mycolor1}{rgb}{0.00000,1.00000,1.00000}%
\begin{tikzpicture}

\begin{axis}[%
width=0.48\textwidth,
height=0.22\textwidth,
scale only axis,
xmin=-0.5,
xmax=0.5,
xtick={-0.5,-0.4,...,0.5},
xlabel style={font=\color{white!15!black}},
xlabel={$X$},
ymin=-2.5,
ymax=0,
ytick={-2.5,-2.0,...,0},
ylabel style={font=\color{white!15!black}},
ylabel={$C_{l U_0}^c$},
axis background/.style={fill=white},
axis x line*=bottom,
axis y line*=left,
xmajorgrids,
ymajorgrids,
tick label style={font=\scriptsize},
legend style={legend cell align=left, align=left, draw=white!15!black}
]
\addplot [color=gray, line width=1.0pt, mark size=2.0pt, mark=square, mark options={solid, gray}]
  table[row sep=crcr]{%
-0.49	-0.331040528748098\\
-0.47	-0.463624629511186\\
-0.45	-0.548107412428779\\
-0.43	-0.609805644039699\\
-0.41	-0.658187687572856\\
-0.39	-0.697946240506076\\
-0.37	-0.731791835330303\\
-0.35	-0.761456347074347\\
-0.33	-0.788134180675403\\
-0.31	-0.812704680165553\\
-0.29	-0.835856903751825\\
-0.27	-0.85816696064784\\
-0.25	-0.880151699545702\\
-0.23	-0.902311852469717\\
-0.21	-0.925173687171598\\
-0.19	-0.94933792461501\\
-0.17	-0.97554799229475\\
-0.15	-1.00479904043353\\
-0.13	-1.03853233321621\\
-0.11	-1.07902035635914\\
-0.09	-1.13022715182694\\
-0.07	-1.20005663794257\\
-0.05	-1.30773888241494\\
-0.03	-1.5195598924952\\
-0.01	-2.39326975431867\\
0.01	-0.0477724496976201\\
0.03	-0.286700385174295\\
0.05	-0.385329494067227\\
0.07	-0.447812392315467\\
0.09	-0.492487305285722\\
0.11	-0.526270928159886\\
0.13	-0.552560116599335\\
0.15	-0.573270876847081\\
0.17	-0.589575984186851\\
0.19	-0.602227552460477\\
0.21	-0.611715993326425\\
0.23	-0.618354305549893\\
0.25	-0.622324025406478\\
0.27	-0.623699062628174\\
0.29	-0.622454790472628\\
0.31	-0.618465131717756\\
0.33	-0.611487178586989\\
0.35	-0.601129520073314\\
0.37	-0.586795239426946\\
0.39	-0.567580268361176\\
0.41	-0.542084187409509\\
0.43	-0.508028381840022\\
0.45	-0.461382644120275\\
0.47	-0.393932560914034\\
0.49	-0.283652359088444\\
};
\addlegendentry{LL}

\addplot [color=blue, dashed, line width=1.0pt, mark size=1.5pt, mark=o, mark options={solid, blue}]
  table[row sep=crcr]{%
-0.49	-0.32592101788342\\
-0.47	-0.454574701293472\\
-0.45	-0.536169832124316\\
-0.43	-0.596105845979541\\
-0.41	-0.643702425885007\\
-0.39	-0.683313297131\\
-0.37	-0.717356541681261\\
-0.35	-0.747358399034265\\
-0.33	-0.774402123498781\\
-0.31	-0.799319390098364\\
-0.29	-0.822759621181857\\
-0.27	-0.84527828492984\\
-0.25	-0.867318106517733\\
-0.23	-0.889422090511263\\
-0.21	-0.911963768273923\\
-0.19	-0.935346068218599\\
-0.17	-0.960203803473259\\
-0.15	-0.986935126355088\\
-0.13	-1.01634963668577\\
-0.11	-1.04869058282623\\
-0.09	-1.08656485941568\\
-0.07	-1.13331027798613\\
-0.05	-1.20403691843648\\
-0.03	-1.35029348533093\\
-0.01	-2.01425764022569\\
0.01	-0.056731763033032\\
0.03	-0.2453974806284\\
0.05	-0.333064990725545\\
0.07	-0.396485892248252\\
0.09	-0.445844153568984\\
0.11	-0.485492794098624\\
0.13	-0.516524274560101\\
0.15	-0.541482085491071\\
0.17	-0.561141088999062\\
0.19	-0.576666017024104\\
0.21	-0.588546452736324\\
0.23	-0.597257187254666\\
0.25	-0.603070623237261\\
0.27	-0.606039908023952\\
0.29	-0.606250172979294\\
0.31	-0.603569653521493\\
0.33	-0.59778188300248\\
0.35	-0.588525587344724\\
0.37	-0.575241847268669\\
0.39	-0.557107299902244\\
0.41	-0.532869002307169\\
0.43	-0.500453199589851\\
0.45	-0.456052207415444\\
0.47	-0.391443759180701\\
0.49	-0.284073843115416\\
};
\addlegendentry{ALM$^1$ ($\varepsilon = 3.5 \Delta x$)}

\addplot [color=cyan, line width=1.0pt, only marks, mark size=2.0pt, mark=+, mark options={solid, cyan}]
  table[row sep=crcr]{%
-0.49	-0.326719874888056\\
-0.47	-0.457502832068514\\
-0.45	-0.540823473448111\\
-0.43	-0.601709062369932\\
-0.41	-0.649535305164107\\
-0.39	-0.688939344395093\\
-0.37	-0.722576317203033\\
-0.35	-0.752118169006583\\
-0.33	-0.778690200945359\\
-0.31	-0.803101906862383\\
-0.29	-0.825954841731799\\
-0.27	-0.847712258976808\\
-0.25	-0.868764277709374\\
-0.23	-0.889512636392566\\
-0.21	-0.910202571499351\\
-0.19	-0.931300267792476\\
-0.17	-0.953319486993244\\
-0.15	-0.977025729993084\\
-0.13	-1.00364823664413\\
-0.11	-1.03521534952306\\
-0.09	-1.07533169476646\\
-0.07	-1.13108734259251\\
-0.05	-1.21946819031471\\
-0.03	-1.39821433787948\\
-0.01	-2.15028646168494\\
0.01	-0.05661404139595\\
0.03	-0.267263568470077\\
0.05	-0.357548889138782\\
0.07	-0.41713352629165\\
0.09	-0.461670513641054\\
0.11	-0.496881469107592\\
0.13	-0.52542961257859\\
0.15	-0.54874295209914\\
0.17	-0.5676847724794\\
0.19	-0.582828575729543\\
0.21	-0.594563678229612\\
0.23	-0.603171486720907\\
0.25	-0.608882231114893\\
0.27	-0.611749231383692\\
0.29	-0.611838419353569\\
0.31	-0.609070010409363\\
0.33	-0.603232589594777\\
0.35	-0.593979935085206\\
0.37	-0.580761936491265\\
0.39	-0.562722210797539\\
0.41	-0.538479373742679\\
0.43	-0.505763315923346\\
0.45	-0.460522342499451\\
0.47	-0.394461085591113\\
0.49	-0.28524897354376\\
};
\addlegendentry{ALM$^1$ ($\varepsilon = 7 \Delta x$)}

\addplot [color=red, dotted, line width=1.0pt, mark size=1.0pt, mark=diamond, mark options={solid, red}]
  table[row sep=crcr]{%
-0.49	-0.325257604972404\\
-0.47	-0.453454868902089\\
-0.45	-0.534545342381352\\
-0.43	-0.593904170879722\\
-0.41	-0.64087517856187\\
-0.39	-0.679844859818239\\
-0.37	-0.713251319069571\\
-0.35	-0.742623485023882\\
-0.33	-0.76903522830453\\
-0.31	-0.793302074555873\\
-0.29	-0.816061784400665\\
-0.27	-0.837860942461584\\
-0.25	-0.859160635460368\\
-0.23	-0.880521338552455\\
-0.21	-0.902393002787611\\
-0.19	-0.92539085776399\\
-0.17	-0.950416204372406\\
-0.15	-0.978467542413775\\
-0.13	-1.01127989739264\\
-0.11	-1.05079076658232\\
-0.09	-1.10138132640137\\
-0.07	-1.16914165955441\\
-0.05	-1.27244716435663\\
-0.03	-1.47357751655711\\
-0.01	-2.31247008132856\\
0.01	-0.042581822905598\\
0.03	-0.263334133468337\\
0.05	-0.357821380256714\\
0.07	-0.419869100698186\\
0.09	-0.465471950197117\\
0.11	-0.501299258069808\\
0.13	-0.529337870563596\\
0.15	-0.551896560223871\\
0.17	-0.569699303195814\\
0.19	-0.583715635025476\\
0.21	-0.594361140592245\\
0.23	-0.602051195027227\\
0.25	-0.607039301667968\\
0.27	-0.609297062078677\\
0.29	-0.608910192240529\\
0.31	-0.605733181387174\\
0.33	-0.599534748590848\\
0.35	-0.589945035804313\\
0.37	-0.576401749297302\\
0.39	-0.558076197828767\\
0.41	-0.53370296611313\\
0.43	-0.50118075359316\\
0.45	-0.456664043675986\\
0.47	-0.391894751971211\\
0.49	-0.28430261515692\\
};
\addlegendentry{ALM$^2$ ($\varepsilon = 3.5 \Delta x$)}

\addplot [color=orange, line width=1.0pt, only marks, mark size=2.0pt, mark=x, mark options={solid, orange}]
  table[row sep=crcr]{%
-0.49	-0.32554560393036\\
-0.47	-0.45566681432769\\
-0.45	-0.538445063049029\\
-0.43	-0.598831495406018\\
-0.41	-0.646185516094336\\
-0.39	-0.685146269533811\\
-0.37	-0.718375747482234\\
-0.35	-0.747564005589137\\
-0.33	-0.77386945961622\\
-0.31	-0.798146062413691\\
-0.29	-0.821058969170357\\
-0.27	-0.84315501821883\\
-0.25	-0.864936680023084\\
-0.23	-0.886926954819848\\
-0.21	-0.909540104880545\\
-0.19	-0.933390022323354\\
-0.17	-0.959154375838446\\
-0.15	-0.987737912091484\\
-0.13	-1.0204769741485\\
-0.11	-1.05950244161899\\
-0.09	-1.10858813035066\\
-0.07	-1.17531199618153\\
-0.05	-1.27823086376174\\
-0.03	-1.48131946450987\\
-0.01	-2.32303242821965\\
0.01	-0.046572724049215\\
0.03	-0.273342630321543\\
0.05	-0.369115853124267\\
0.07	-0.430623559076267\\
0.09	-0.475172822696588\\
0.11	-0.509345107988686\\
0.13	-0.536354960659445\\
0.15	-0.558003454590392\\
0.17	-0.575366411790393\\
0.19	-0.589121246619549\\
0.21	-0.599697473419729\\
0.23	-0.607371043864954\\
0.25	-0.61232430925536\\
0.27	-0.614591428254499\\
0.29	-0.614186287413303\\
0.31	-0.611000247052741\\
0.33	-0.604809311749711\\
0.35	-0.59525428168163\\
0.37	-0.58177209336502\\
0.39	-0.56348951980217\\
0.41	-0.539024773357006\\
0.43	-0.506102553978504\\
0.45	-0.460675191468078\\
0.47	-0.39444511895751\\
0.49	-0.285101546310107\\
};
\addlegendentry{ALM$^2$ ($\varepsilon = 7 \Delta x$)}
\legend{}
\end{axis}
\end{tikzpicture}%
%
%
\definecolor{mycolor1}{rgb}{0.00000,1.00000,1.00000}%
\begin{tikzpicture}

\begin{axis}[%
width=0.24\textwidth,
height=0.22\textwidth,
scale only axis,
xmin=-0.5,
xmax=0,
xtick={-0.5,-0.4,...,0.0},
xlabel style={font=\color{white!15!black}},
xlabel={$Y$},
ymin=-2.5,
ymax=0,
ytick={-2.5,-2.0,...,0},
ylabel style={font=\color{white!15!black}},
ylabel={$C_{l U_0}^c$},
axis background/.style={fill=white},
axis x line*=bottom,
axis y line*=left,
xmajorgrids,
ymajorgrids,
tick label style={font=\scriptsize},
legend style={legend cell align=left, align=left, draw=white!15!black}
]
\addplot [color=gray, line width=1.0pt, mark size=2.0pt, mark=square, mark options={solid, gray}]
  table[row sep=crcr]{%
-0.01	-2.40298181676946\\
-0.03	-1.37239082945973\\
-0.05	-1.14436960337342\\
-0.07	-1.03906483353518\\
-0.09	-0.976803512795766\\
-0.11	-0.934759384146432\\
-0.13	-0.903737269425685\\
-0.15	-0.87924969836183\\
-0.17	-0.858807683336106\\
-0.19	-0.840890148079712\\
-0.21	-0.824487126113075\\
-0.23	-0.808872448366618\\
-0.25	-0.793478772737425\\
-0.27	-0.777821443707685\\
-0.29	-0.761445829238054\\
-0.31	-0.743884264701629\\
-0.33	-0.724613064562194\\
-0.35	-0.703000354111155\\
-0.37	-0.678231920678908\\
-0.39	-0.649192299192039\\
-0.41	-0.61425334672401\\
-0.43	-0.570855343017655\\
-0.45	-0.514554911832436\\
-0.47	-0.436376512517777\\
-0.49	-0.312321909956272\\
};
\addlegendentry{LL}

\addplot [color=blue, dashed, line width=1.0pt, mark size=1.5pt, mark=o, mark options={solid, blue}]
  table[row sep=crcr]{%
-0.01	-1.98956264953146\\
-0.03	-1.20262140150858\\
-0.05	-1.04591409281793\\
-0.07	-0.982026421044654\\
-0.09	-0.946629517406164\\
-0.11	-0.920481256411123\\
-0.13	-0.898714684344099\\
-0.15	-0.878509009014879\\
-0.17	-0.859875716714316\\
-0.19	-0.842257935066963\\
-0.21	-0.825461148903754\\
-0.23	-0.808987236725863\\
-0.25	-0.792501269956137\\
-0.27	-0.775702089977145\\
-0.29	-0.758095228289788\\
-0.31	-0.739284499338771\\
-0.33	-0.718756221316717\\
-0.35	-0.69590772623026\\
-0.37	-0.669959389045183\\
-0.39	-0.639888298548135\\
-0.41	-0.604239900264706\\
-0.43	-0.560706685848146\\
-0.45	-0.505148242157088\\
-0.47	-0.428743512328111\\
-0.49	-0.307510656861017\\
};
\addlegendentry{ALM$^1$ ($\varepsilon = 3.5 \Delta x$)}

\addplot [color=cyan, line width=1.0pt, only marks, mark size=2.0pt, mark=+, mark options={solid, cyan}]
  table[row sep=crcr]{%
-0.01	-2.14137148436218\\
-0.03	-1.25209020383441\\
-0.05	-1.05929690295036\\
-0.07	-0.973898812561284\\
-0.09	-0.926201423956527\\
-0.11	-0.895826216835502\\
-0.13	-0.874354321745296\\
-0.15	-0.857585480996098\\
-0.17	-0.843178340830936\\
-0.19	-0.829780204318149\\
-0.21	-0.816596209543199\\
-0.23	-0.803119711977749\\
-0.25	-0.788928330918031\\
-0.27	-0.773875831665755\\
-0.29	-0.757620115123839\\
-0.31	-0.739845220832372\\
-0.33	-0.720168073667965\\
-0.35	-0.698061402521676\\
-0.37	-0.672787390239213\\
-0.39	-0.643306757679645\\
-0.41	-0.60805524010955\\
-0.43	-0.564532725055726\\
-0.45	-0.508355530562068\\
-0.47	-0.430656995207905\\
-0.49	-0.307805746656662\\
};
\addlegendentry{ALM$^1$ ($\varepsilon = 7 \Delta x$)}

\addplot [color=red, dotted, line width=1.0pt, mark size=1.0pt, mark=diamond, mark options={solid, red}]
  table[row sep=crcr]{%
-0.01	-2.32203804415402\\
-0.03	-1.34088852180954\\
-0.05	-1.12355776032505\\
-0.07	-1.02078077327957\\
-0.09	-0.958750240725666\\
-0.11	-0.916021899585874\\
-0.13	-0.885210308812768\\
-0.15	-0.861080680513235\\
-0.17	-0.841400792252059\\
-0.19	-0.824217101810126\\
-0.21	-0.808507519378366\\
-0.23	-0.793385031897024\\
-0.25	-0.778268961883708\\
-0.27	-0.762776339422038\\
-0.29	-0.746391618406273\\
-0.31	-0.728711080461843\\
-0.33	-0.70923490693819\\
-0.35	-0.687379454851504\\
-0.37	-0.662387574098346\\
-0.39	-0.63325363317408\\
-0.41	-0.598529907018513\\
-0.43	-0.555903916073343\\
-0.45	-0.501226915587999\\
-0.47	-0.425699070422487\\
-0.49	-0.30547226563462\\
};
\addlegendentry{ALM$^2$ ($\varepsilon = 3.5 \Delta x$)}

\addplot [color=orange, line width=1.0pt, only marks, mark size=2.0pt, mark=x, mark options={solid, orange}]
  table[row sep=crcr]{%
-0.01	-2.33137366538355\\
-0.03	-1.34255029718676\\
-0.05	-1.12440476691898\\
-0.07	-1.02370826856117\\
-0.09	-0.963930519520399\\
-0.11	-0.92318671946863\\
-0.13	-0.892743553869514\\
-0.15	-0.868379356763039\\
-0.17	-0.847785364578692\\
-0.19	-0.829568722799362\\
-0.21	-0.812818864964467\\
-0.23	-0.796861818381205\\
-0.25	-0.781093220925434\\
-0.27	-0.765155335717444\\
-0.29	-0.748540527735733\\
-0.31	-0.730787328822625\\
-0.33	-0.711393338010204\\
-0.35	-0.689742978086389\\
-0.37	-0.665042055468121\\
-0.39	-0.63620980938574\\
-0.41	-0.601661970889903\\
-0.43	-0.558894215737203\\
-0.45	-0.503535769616043\\
-0.47	-0.426776465100793\\
-0.49	-0.305169582574807\\
};
\addlegendentry{ALM$^2$ ($\varepsilon = 7 \Delta x$)}
\legend{}
\end{axis}
\end{tikzpicture}%
%
%
\definecolor{mycolor1}{rgb}{0.00000,1.00000,1.00000}%
\begin{tikzpicture}

\begin{axis}[%
width=0.48\textwidth,
height=0.22\textwidth,
scale only axis,
xmin=-0.5,
xmax=0.5,
xtick={-0.5,-0.4,...,0.5},
xlabel style={font=\color{white!15!black}},
xlabel={$X$},
ymin=-0.5,
ymax=0.1,
ytick={-0.5,-0.4,...,0.1},
ylabel style={font=\color{white!15!black}},
ylabel={$(C_{l U_0}^c-C_{l U_0 LL}))/C_{l ref}$},
axis background/.style={fill=white},
axis x line*=bottom,
axis y line*=left,
xmajorgrids,
ymajorgrids,
tick label style={font=\scriptsize},
legend style={legend cell align=left, align=left, draw=white!15!black}
]
\addplot [color=blue, dashed, line width=1.0pt, mark size=1.5pt, mark=o, mark options={solid, blue}]
  table[row sep=crcr]{%
-0.49	-0.00511951086407269\\
-0.47	-0.00904992821664276\\
-0.45	-0.0119375803030507\\
-0.43	-0.0136997980585371\\
-0.41	-0.0144852616861347\\
-0.39	-0.0146329433733442\\
-0.37	-0.0144352936473343\\
-0.35	-0.0140979480384135\\
-0.33	-0.0137320571749972\\
-0.31	-0.0133852900656052\\
-0.29	-0.0130972825684181\\
-0.27	-0.0128886757164756\\
-0.25	-0.0128335930264511\\
-0.23	-0.0128897619569287\\
-0.21	-0.0132099188961121\\
-0.19	-0.0139918563947555\\
-0.17	-0.0153441888196755\\
-0.15	-0.0178639140763247\\
-0.13	-0.0221826965278206\\
-0.11	-0.0303297735293182\\
-0.09	-0.0436622924060927\\
-0.07	-0.0667463599485391\\
-0.05	-0.103701963966192\\
-0.03	-0.169266407144248\\
-0.01	-0.379012114048137\\
0.01	0.00895931333435179\\
0.03	-0.0413029045410085\\
0.05	-0.0522645033354976\\
0.07	-0.0513265000611422\\
0.09	-0.0466431517112189\\
0.11	-0.0407781340564371\\
0.13	-0.03603584203497\\
0.15	-0.0317887913522488\\
0.17	-0.0284348951844241\\
0.19	-0.0255615354333484\\
0.21	-0.0231695405873595\\
0.23	-0.021097118292731\\
0.25	-0.0192534021669391\\
0.27	-0.0176591546021326\\
0.29	-0.016204617491417\\
0.31	-0.0148954781945006\\
0.33	-0.0137052955828871\\
0.35	-0.0126039327270983\\
0.37	-0.0115533921569098\\
0.39	-0.0104729684576927\\
0.41	-0.00921518510125008\\
0.43	-0.00757518224927498\\
0.45	-0.00533043670420065\\
0.47	-0.00248880173303824\\
0.49	0.000421484026921693\\
};
\addlegendentry{ALM$^1$ ($\varepsilon = 3.5 \Delta x$)}

\addplot [color=cyan, line width=1.0pt, only marks, mark size=2.0pt, mark=+, mark options={solid, cyan}]
  table[row sep=crcr]{%
-0.49	-0.00432065385953121\\
-0.47	-0.00612179744194721\\
-0.45	-0.00728393897980631\\
-0.43	-0.00809658166880913\\
-0.41	-0.00865238240772479\\
-0.39	-0.009006896109917\\
-0.37	-0.0092155181261799\\
-0.35	-0.0093381780666587\\
-0.33	-0.00944397972892643\\
-0.31	-0.00960277330203362\\
-0.29	-0.00990206201885424\\
-0.27	-0.0104547016697956\\
-0.25	-0.0113874218349811\\
-0.23	-0.0127992160756364\\
-0.21	-0.0149711156704758\\
-0.19	-0.0180376568203998\\
-0.17	-0.022228505298876\\
-0.15	-0.0277733104371561\\
-0.13	-0.0348840965679536\\
-0.11	-0.0438050068308888\\
-0.09	-0.0548954570539785\\
-0.07	-0.0689692953418951\\
-0.05	-0.0882706920897836\\
-0.03	-0.121345554601361\\
-0.01	-0.24298329260498\\
0.01	0.00884159169728372\\
0.03	-0.0194368167019187\\
0.05	-0.0277806049251576\\
0.07	-0.0306788660201872\\
0.09	-0.0308167916410214\\
0.11	-0.0293894590488166\\
0.13	-0.0271305040175348\\
0.15	-0.024527924745039\\
0.17	-0.0218912117048604\\
0.19	-0.0193989767286385\\
0.21	-0.0171523150947834\\
0.23	-0.0151828188271898\\
0.25	-0.0134417942899948\\
0.27	-0.0119498312430682\\
0.29	-0.0106163711178032\\
0.31	-0.0093951213072814\\
0.33	-0.00825458899123507\\
0.35	-0.00714958498726164\\
0.37	-0.00603330293496696\\
0.39	-0.00485805756306196\\
0.41	-0.00360481366640388\\
0.43	-0.00226506591640819\\
0.45	-0.000860301620722553\\
0.47	0.000528524677016721\\
0.49	0.0015966144551267\\
};
\addlegendentry{ALM$^1$ ($\varepsilon = 7 \Delta x$)}

\addplot [color=red, dotted, line width=1.0pt, mark size=1.0pt, mark=diamond, mark options={solid, red}]
  table[row sep=crcr]{%
-0.49	-0.00578292377501015\\
-0.47	-0.0101697606078933\\
-0.45	-0.0135620700458225\\
-0.43	-0.0159014731580957\\
-0.41	-0.017312509008937\\
-0.39	-0.0181013806856948\\
-0.37	-0.0185405162585385\\
-0.35	-0.0188328620482362\\
-0.33	-0.0190989523686131\\
-0.31	-0.0194026056073841\\
-0.29	-0.0197951193488177\\
-0.27	-0.0203060181838538\\
-0.25	-0.0209910640828509\\
-0.23	-0.0217905139146836\\
-0.21	-0.0227806843812916\\
-0.19	-0.0239470668481867\\
-0.17	-0.0251317879193705\\
-0.15	-0.0263314980166359\\
-0.13	-0.0272524358203496\\
-0.11	-0.0282295897734775\\
-0.09	-0.028845825422156\\
-0.07	-0.0309149783844947\\
-0.05	-0.0352917180541361\\
-0.03	-0.0459823759326544\\
-0.01	-0.0807996729805503\\
0.01	-0.00519062679140797\\
0.03	-0.0233662517031937\\
0.05	-0.0275081138072578\\
0.07	-0.0279432916139749\\
0.09	-0.0270153550854083\\
0.11	-0.0249716700871233\\
0.13	-0.0232222460329912\\
0.15	-0.0213743166206811\\
0.17	-0.0198766809886847\\
0.19	-0.0185119174328105\\
0.21	-0.0173548527321265\\
0.23	-0.0163031105207372\\
0.25	-0.0152847237367018\\
0.27	-0.0144020005477931\\
0.29	-0.0135445982304967\\
0.31	-0.0127319503290756\\
0.33	-0.0119524299947265\\
0.35	-0.0111844842676773\\
0.37	-0.0103934901284141\\
0.39	-0.00950407053128434\\
0.41	-0.00838122129538774\\
0.43	-0.00684762824605203\\
0.45	-0.00471860044373104\\
0.47	-0.00203780894258162\\
0.49	0.000650256068398691\\
};
\addlegendentry{ALM$^2$ ($\varepsilon = 3.5 \Delta x$)}

\addplot [color=orange, line width=1.0pt, only marks, mark size=2.0pt, mark=x, mark options={solid, orange}]
  table[row sep=crcr]{%
-0.49	-0.00549492481708822\\
-0.47	-0.00795781518255401\\
-0.45	-0.00966234937860693\\
-0.43	-0.0109741486323826\\
-0.41	-0.0120021714770995\\
-0.39	-0.0127999709707501\\
-0.37	-0.013416087846482\\
-0.35	-0.0138923414835658\\
-0.33	-0.014264721057495\\
-0.31	-0.0145586177501393\\
-0.29	-0.0147979345797169\\
-0.27	-0.0150119424272344\\
-0.25	-0.0152150195208182\\
-0.23	-0.0153848976480485\\
-0.21	-0.0156335822892034\\
-0.19	-0.015947902289769\\
-0.17	-0.0163936164543644\\
-0.15	-0.0170611283400236\\
-0.13	-0.0180553590655747\\
-0.11	-0.0195179147378365\\
-0.09	-0.0216390214737126\\
-0.07	-0.0247446417581077\\
-0.05	-0.0295080186497043\\
-0.03	-0.0382404279808084\\
-0.01	-0.0702373260907151\\
0.01	-0.00119972564826317\\
0.03	-0.013357754851172\\
0.05	-0.0162136409410412\\
0.07	-0.0171888332371664\\
0.09	-0.017314482587085\\
0.11	-0.0169258201691974\\
0.13	-0.0162051559379724\\
0.15	-0.0152674222548827\\
0.17	-0.0142095723947763\\
0.19	-0.0131063058393771\\
0.21	-0.012018519905274\\
0.23	-0.0109832616836396\\
0.25	-0.0099997161499351\\
0.27	-0.0091076343725975\\
0.29	-0.00826850305834702\\
0.31	-0.0074648846641318\\
0.33	-0.00667786683648765\\
0.35	-0.00587523839098842\\
0.37	-0.00502314606133143\\
0.39	-0.0040907485585218\\
0.41	-0.00305941405214132\\
0.43	-0.00192582786129043\\
0.45	-0.000707452652113587\\
0.47	0.000512558043415628\\
0.49	0.00144918722149113\\
};
\addlegendentry{ALM$^2$ ($\varepsilon = 7 \Delta x$)}
\legend{}
\end{axis}
\end{tikzpicture}%
%
%
\definecolor{mycolor1}{rgb}{0.00000,1.00000,1.00000}%
\begin{tikzpicture}

\begin{axis}[%
width=0.24\textwidth,
height=0.22\textwidth,
scale only axis,
xmin=-0.5,
xmax=0,
xtick={-0.5,-0.4,...,0.0},
xlabel style={font=\color{white!15!black}},
xlabel={$Y$},
ymin=-0.5,
ymax=0.1,
ytick={-0.5,-0.4,...,0.1},
ylabel style={font=\color{white!15!black}},
ylabel={$(C_{l U_0}^c-C_{l U_0 LL}))/C_{l ref}$},
axis background/.style={fill=white},
axis x line*=bottom,
axis y line*=left,
xmajorgrids,
ymajorgrids,
tick label style={font=\scriptsize},
legend style={legend cell align=left, align=left, draw=white!15!black}
]
\addplot [color=blue, dashed, line width=1.0pt, mark size=1.5pt, mark=o, mark options={solid, blue}]
  table[row sep=crcr]{%
-0.01	-0.413419167189086\\
-0.03	-0.169769427931064\\
-0.05	-0.0984555105438389\\
-0.07	-0.0570384124837805\\
-0.09	-0.0301739953860319\\
-0.11	-0.0142781277336193\\
-0.13	-0.0050225850809921\\
-0.15	-0.000740689346863845\\
-0.17	0.001068033378084\\
-0.19	0.00136778698708887\\
-0.21	0.000974022790563794\\
-0.23	0.000114788359231362\\
-0.25	-0.00097750278117195\\
-0.27	-0.00211935373028897\\
-0.29	-0.00335060094786967\\
-0.31	-0.00459976536231408\\
-0.33	-0.00585684324478432\\
-0.35	-0.00709262788005628\\
-0.37	-0.00827253163274673\\
-0.39	-0.00930400064280361\\
-0.41	-0.0100134464581194\\
-0.43	-0.0101486571683081\\
-0.45	-0.00940666967423497\\
-0.47	-0.00763300018876303\\
-0.49	-0.00481125309468533\\
};
\addlegendentry{ALM$^1$ ($\varepsilon = 3.5 \Delta x$)}

\addplot [color=cyan, line width=1.0pt, only marks, mark size=2.0pt, mark=+, mark options={solid, cyan}]
  table[row sep=crcr]{%
-0.01	-0.26161033237633\\
-0.03	-0.120300625611082\\
-0.05	-0.0850727004129975\\
-0.07	-0.065166020966189\\
-0.09	-0.0506020888332518\\
-0.11	-0.0389331673063229\\
-0.13	-0.0293829476769128\\
-0.15	-0.0216642173631692\\
-0.17	-0.0156293425033204\\
-0.19	-0.0111099437602487\\
-0.21	-0.00789091656894229\\
-0.23	-0.0057527363881883\\
-0.25	-0.00455044181885527\\
-0.27	-0.00394561204146293\\
-0.29	-0.00382571411376251\\
-0.31	-0.00403904386877934\\
-0.33	-0.0044449908937034\\
-0.35	-0.00493895158889516\\
-0.37	-0.0054445304390513\\
-0.39	-0.00588554151169802\\
-0.41	-0.00619810661372693\\
-0.43	-0.00632261796118069\\
-0.45	-0.00619938126963449\\
-0.47	-0.00571951730919545\\
-0.49	-0.00451616329907522\\
};
\addlegendentry{ALM$^1$ ($\varepsilon = 7 \Delta x$)}

\addplot [color=red, dotted, line width=1.0pt, mark size=1.0pt, mark=diamond, mark options={solid, red}]
  table[row sep=crcr]{%
-0.01	-0.0809437726058645\\
-0.03	-0.0315023076464592\\
-0.05	-0.0208118430459059\\
-0.07	-0.0182840602534491\\
-0.09	-0.018053272067964\\
-0.11	-0.0187374845583405\\
-0.13	-0.0185269606107253\\
-0.15	-0.0181690178464457\\
-0.17	-0.017406891081987\\
-0.19	-0.0166730462676136\\
-0.21	-0.0159796067328183\\
-0.23	-0.0154874164677616\\
-0.25	-0.015209810851917\\
-0.27	-0.0150451042838666\\
-0.29	-0.0150542108299999\\
-0.31	-0.015173184237991\\
-0.33	-0.0153781576221848\\
-0.35	-0.0156208992578032\\
-0.37	-0.0158443465786877\\
-0.39	-0.0159386660160736\\
-0.41	-0.0157234397036369\\
-0.43	-0.0149514269425429\\
-0.45	-0.0133279962428601\\
-0.47	-0.0106774420940268\\
-0.49	-0.00684964432084113\\
};
\addlegendentry{ALM$^2$ ($\varepsilon = 3.5 \Delta x$)}

\addplot [color=orange, line width=1.0pt, only marks, mark size=2.0pt, mark=x, mark options={solid, orange}]
  table[row sep=crcr]{%
-0.01	-0.0716081513774381\\
-0.03	-0.029840532269435\\
-0.05	-0.019964836452082\\
-0.07	-0.0153565649721983\\
-0.09	-0.0128729932738439\\
-0.11	-0.0115726646764323\\
-0.13	-0.0109937155548707\\
-0.15	-0.0108703415975052\\
-0.17	-0.0110223187561094\\
-0.19	-0.0113214252790107\\
-0.21	-0.0116682611472274\\
-0.23	-0.0120106299839919\\
-0.25	-0.0123855518105252\\
-0.27	-0.012666107988742\\
-0.29	-0.0129053015007942\\
-0.31	-0.0130969358774546\\
-0.33	-0.0132197265504262\\
-0.35	-0.0132573760231979\\
-0.37	-0.0131898652092268\\
-0.39	-0.0129824898047632\\
-0.41	-0.0125913758326175\\
-0.43	-0.0119611272790366\\
-0.45	-0.0110191422150892\\
-0.47	-0.00960004741584828\\
-0.49	-0.00715232738061833\\
};
\addlegendentry{ALM$^2$ ($\varepsilon = 7 \Delta x$)}
\legend{}
\end{axis}
\end{tikzpicture}%
  \caption{Comparison of the circulation and lift coefficient results for the T-tail (legend in figure \ref{fig:res_resultswinglet}).}
  \label{fig:res_resultstail}
\end{figure}

The angle of attack of $\alpha_g=-9.189^{\circ}$ might be a little extreme for the vertical tail of an airplane in a steady flow, however, this case is very interesting from a validation perspective because the order of magnitude of the induced velocity is 1 near the intersection of the surfaces. In other words, $|u_z^c-U_0|/U_0 \approx 1$ ($u_z^c/U_0 \approx 0$ and $u_z^c/U_0 \approx 2$), as can be seen in figure \ref{fig:res_resultsutail}. For this case, the limitations of the method should become more evident, as seen in figure \ref{fig:res_resultstail}, since the small perturbation approximation is not valid.

For this case, the iterative version of the smearing correction~\citep{kleine2022non} with a low relaxation parameter was used for a few time steps after the smearing correction was turned on. It was needed because the velocity $u_z^c$ is close to zero at one point near the intersection, causing instabilities in the non-iterative correction in the first time steps. Nevertheless, the use of the iterative version did not affect the final results because the effects of this different method were washed away with the initial conditions. The presence of numerical instabilities in this case is related to a limitation of both the LL and the ALM with smearing correction: the methods do not allow local negative velocities in the streamwise direction. For example, a small change to the angle of attack to $\alpha_g=-10.5^{\circ}$ makes the methods not converge for this discretization, because of a negative velocity in one of the points. If the control-point discretization is reduced, then this problem is avoided. This pattern is further evidence of how the knowledge about the LL can be used to provide guidance and understanding on the use of the ALM with smearing correction.

As can be seen in figure \ref{fig:res_resultstail}, the results far from the intersection and the general trend can be captured well by both methods. Nevertheless, the differences near the intersection are even greater for this case, as expected. Again the vorticity magnitude correction improves the results and makes them less dependent on the value of the smearing parameter.

\begin{figure}
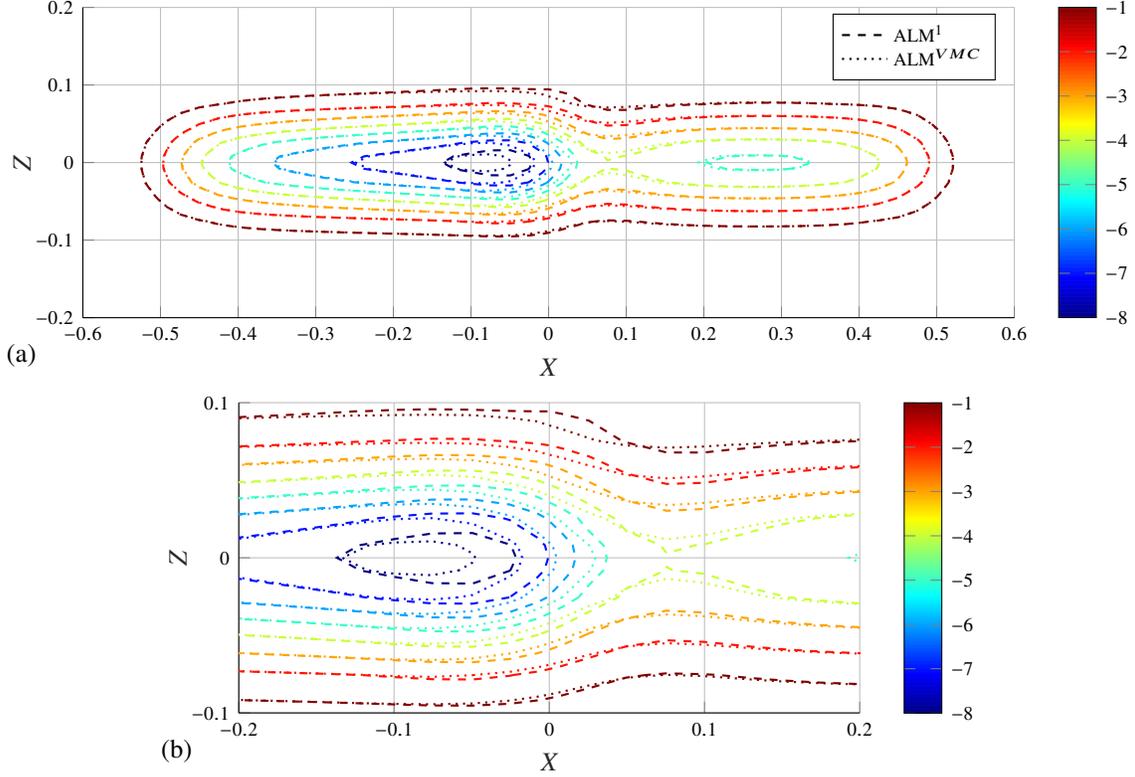

  \centering
  \input{contour_s2_htail_new}
  \input{contour_s2_htailzoom_new}
  \caption{Contour of the vorticity $\omega_X^s$ created in the numerical simulation by the different formulations of the ALM, for the T-tail configuration ($\varepsilon = 3.5 \Delta x$), in the plane $Y=0$. (a) Region including the entirety of the bound vortex of the horizontal tail. (b) Highlight of region near the intersection ($X=0$).}
  \label{fig:vort_htail}
\end{figure}

In figure \ref{fig:vort_htail} the vorticity in the global $X$ direction, $\omega_X^s$, is shown for the $X-Z$-plane ($Y=0$). The vorticity in the $X$ direction corresponds mostly to the bound vortex of the horizontal tail, so it is directly related to the circulation of figure \ref{fig:res_resultstail}. Far from the intersection, both formulations of the ALM agree well, as expected, as can be observed in figure~\ref{fig:vort_htail}(a). On the other hand, they differ near $X=0$, as shown in figure~\ref{fig:vort_htail}(b). For $X \approx -0.05$, for example, it can be noted that the magnitude of $\omega_X^s$ is greater for ALM$^1$. This result is different from the observed behavior of $\Gamma^c$ in figure~\ref{fig:res_resultstail}, where the magnitude is greater for ALM$^{VMC}$. However, there is no contradiction, this is the expected behavior because figure \ref{fig:vort_htail} displays the vorticity related to $\Gamma^s$, not $\Gamma^c$. For ALM$^1$, the missing velocity $u_z^m$ is positive in this region ($X \approx -0.05$), so we should expect $\Gamma^s>\Gamma^c$ in this region, according to equation \eqref{eq:vortx_corr}. Around $X \approx 0.1$ the opposite effect is observed, as expected, because $u_z^m$ is negative.

Interestingly, for positive values of $X$ very close to $0$, for example $X \approx 0.02$, the vorticity $\omega_X^s$ is greater for ALM$^1$ even though $u_z^m$ is negative. This can be explained by the smearing in the $x$ direction. The convolution of the force with the Gaussian function makes the vorticity spread in all directions, including in the $X$ direction. So, the vorticity at a point is a weighted combination of the smeared vorticity of the neighboring points. For $X \approx 0.02$, in particular, the effects of the points at $X<0$ are stronger than the effects of the points at $X>0$, because the magnitude of the circulation of the former is higher. Hence, the effect at $X \approx 0.02$ follows the effect at $X<0$, which is for ALM$^1$ to have higher values of $\omega_X^s$. It is the same reason why the minimum vorticity occurs near $X \approx 0.07$, and not near $X \approx 0.02$, as the circulation from figure \ref{fig:res_resultstail} would suggest.

For many applications, the differences observed between the ALM$^1$ and ALM$^{VMC}$ may be considered negligible. One argument that can be raised to disregard the vorticity magnitude correction is that the higher differences are obtained near the intersection of the surfaces, where the lifting line method might not represent well a real flow. In the intersection region, the assumption that the flow in each spanwise section can be treated independently may not be applicable. The flow in this region is highly three-dimensional, hence, the predictions of the LL may not agree with the actual forces. Also, in a real situation, the interaction between the boundary layers of the two surfaces may cause viscous effects to be relevant. These viscous effects of these interacting boundary layers are not modeled by the LL or the ALM. The order of magnitude of the viscous interference effects might be higher than the order of magnitude of the errors related to the magnitude of the vorticity.

On the other hand, besides the independence of the smearing parameter, two other important arguments can be made for its adoption. First, it makes the ALM more consistent with the ideal case. Without it not only a second-order error in the computed circulation and forces are present, but also a first-order error in the vorticity generated in the numerical simulation is caused by the uncorrected body forces. Second, the numerical cost of the correction is practically null. If the original smearing correction is already implemented, the implementation of the vorticity magnitude correction requires only the definition of one extra small vector that is obtained directly by element-wise product and division of other small vectors. Hence, these are also reasons to implement this correction even for cases where the difference is negligible, such as planar wings.

Additionally, an important consequence of the vorticity magnitude correction is that it not only corrects the results but also provides a better understanding on the theoretical aspects of the ALM. The improved agreement after the introduction of the correction confirms the theoretical explanation for the differences and solves the ambiguity in the definition of the forces. The resolution of such ambiguity might be even more important when developing smearing corrections for other actuator methods, such as the actuator surface method~\citep{shen2009actuator}.

Therefore, by using the vorticity magnitude correction, some known errors of the ALM are avoided and the method becomes even closer to the LL. Even for an induced velocity in the order of one, the ALM agrees very well with the LL. The suitability of the ALM for a specific geometry or flow case is still an open question that depends on the application. However, the results presented here and in previous studies~\citep{dag2020new,meyer2019vortex,martinez2019filtered,kleine2022non} point to a direction: for this formulation of the ALM (with the smearing correction and vorticity magnitude correction), the ALM is as suitable as a non-linear iterative lifting line method. Since the lifting line method is used for many fixed-wing aircraft applications, it is possible to conclude that the ALM is also adequate for these applications.

Hence, the advantages and limitations of the LL can also be extended to the ALM. This is an important piece of information, because it would be possible to use the known limitations of the lifting line method, such as its weakness near intersections and for low aspect ratio geometries, to guide applications of the ALM. In this sense, for configurations for which the LL has low accuracy, the ALM with vortex-based smearing correction would perform no better than the LL. Nevertheless, corrections that are applicable to LL can be used in conjunction with the smearing correction, for example, a viscous core LL can be easily implemented by using a finite core in the reference (ideal) vortices and a viscous decambering correction has already been applied on top of this correction~\citep{dag2017combined}. For wings with low aspect ratio, the ``line'' assumption may hamper both the ALM and the LL, however, the ideas presented in this work may contribute to the development of more accurate and general actuator surface methods, which may not suffer from this limitation.

On the other hand, configurations for which the LL is known to have good results can be studied with the ALM, with the added advantage of the latter being part of a full Navier-Stokes simulation, thus allowing more complex boundary conditions, more complex geometries and consideration of viscous effects.

This method allows more versatility compared to simpler models, and lowers simulation costs compared to simulations of wings using the wall boundary conditions or immersed boundary methods. For example, an airplane with propellers can be simulated using the ALM for the wings and the actuator disk method for the propellers, in the initial design stages. Or, the simulation of a wing in an open-jet wind tunnel (of any cross-section shape) can be used to derive wind tunnel corrections that take into account the shear layer displacement caused by the lift force, at a fraction of the cost of a simulation that includes the geometry as no-slip boundary conditions.

\section{Conclusions} \label{sec:conclusions}
In this work, two configurations typical of fixed-wing aircraft were simulated using the actuator line method for the first time: a wing with winglets and a T-tail. Even for such complex configurations, the ALM with vortex-based smearing correction was shown to agree well with the non-linear lifting line method.

However, for these non-planar configurations, differences between the ALM with the original smearing correction and the LL were observed near the intersection of the wings. These errors were shown to be related to differences in the magnitude of the vorticity generated in the numerical simulation and the ideal vorticity, caused by an ambiguity in the velocity that defines the body forces.

A vorticity magnitude correction is proposed, which improves the agreement of the calculated forces, induced velocities and circulation. This improvement in the results confirms the theoretical development and resolves the ambiguity in the definition of the forces. This second-order correction is very easily implemented, does not increase the computational time and does not add complexity to the code. For its low computational cost, it is reasonable to implement the vorticity magnitude correction even for straight planar wings, where the differences are negligible.

The concepts and theory developed in this work and in previous studies~\citep{forsythe2015coupled,dag2020new,meyer2019vortex,martinez2019filtered,kleine2022non} on the vorticity generated by the smeared body forces could be applied to improve the accuracy of other actuator-based method, such as the actuator surface~\citep{shen2009actuator} and actuator sector~\citep{storey2015actuator} methods.

The results presented here, together with previous promising results of~\citep{dag2020new,meyer2019vortex,martinez2019filtered,kleine2022non}, indicate that the ALM with vortex-based smearing correction and vorticity magnitude correction can be used for simulating the aerodynamics of airplanes. It provides lift calculations as accurate as the lifting line method, with the added advantage of being integrated to a Navier-Stokes solver, allowing more complex geometries and the consideration of viscous effects. It can become a relevant tool for middle and low-fidelity simulations of fixed-wing aircraft and other applications beyond rotating blades.

\appendix

\section*{Appendix}

\section{Order of magnitude of the correction} \label{app:errorvelocity}

In this section, to estimate the order of the correction, we assume that the vorticity magnitude correction is not applied and that the corrected lift force is used as body force. If the missing velocity can be considered small ($u_y^m,u_z^m << u_y^s,u_z^s$), the linearization of the corrected relative velocity gives
\begin{equation}
    u_r^c = \sqrt{(u_y^{s}+u_y^{m})^2 + (u_z^{s}+u_z^{m})^2} \approx u_r^{s}  +\frac{u_y^{m} u_y^{s}}{u_r^{s}} +\frac{u_z^{m} u_z^{s}}{u_r^{s}} .
\end{equation}
From equation~\eqref{eq:vortx_corr}, the circulation $\Gamma^{s}$ generated in the CFD simulation if the corrected lift force ($F_l^s=F_l^c=\rho u_{r}^c \Gamma^c$) is imposed as body force is 
\begin{equation}
    \label{eq:Gammaserror}
    \Gamma^{s} = \frac{F_l^c}{\rho u_r^s} = \frac{u_r^c}{u_r^s} \Gamma^{c} \approx \left(1+\frac{u_y^m u_y^s}{u_r^{s\,2}}+\frac{u_z^m u_z^s}{u_r^{s\,2}} \right) \Gamma^{c} .
\end{equation}
The normalized difference in the corrected circulation and the circulation generated in the numerical simulation is denoted as $\beta^s$
\begin{equation}
    \beta^s = \frac{\Gamma^{s}-\Gamma^{c}}{\Gamma^{c}} \approx \frac{u_y^m u_y^s}{u_r^{s\,2}}+\frac{u_z^m u_z^s}{u_r^{s\,2}} ,
\end{equation}
which shows that the difference in the corrected circulation and the circulation generated in the numerical simulation is of first order with respect to the velocity ratio $u^m /u_r^s$. As a consequence, the magnitude of the vorticity created in the numerical simulation also has a first order difference to the vorticity created by a vortex with a Gaussian distribution of vorticity with total circulation $\Gamma^{c}$ (which is the vorticity considered by the smearing correction without vorticity magnitude correction).

Nevertheless, here we are interested in the difference of the corrected circulation, $\Gamma^{c}$, and a goal corrected circulation $\Gamma^{ll}$ that is equivalent to the lifting line method
\begin{equation}
    \beta^{ll} = \frac{\Gamma^{c}-\Gamma^{ll}}{\Gamma^{c}}
\end{equation}

The goal of the vortex-based smearing correction is to calculate a velocity the that simulates the effect of ideal vortices, hence, the value of $\beta^{ll}$, ideally, should be zero. We denote this velocity, $\mathbf{u}^{ll}$, that can be calculated using the influence matrices $\mathsfbi{A_y}^{vi}$ and $\mathsfbi{A_z}^{vi}$, from the lifting line method:
\begin{equation}
    \mathsfbi{u_y}^{ll} = \mathsfbi{U_y} + \mathsfbi{u_y}^{vi,ll} = \mathsfbi{U_y} + \mathsfbi{A_y}^{vi} \boldsymbol{\Gamma}^{ll}
\end{equation}
\begin{equation}
    \mathsfbi{u_z}^{ll} = \mathsfbi{U_z} + \mathsfbi{u_z}^{vi,ll} = \mathsfbi{U_z} + \mathsfbi{A_z}^{vi} \boldsymbol{\Gamma}^{ll}
\end{equation}
while the corrected velocity of the ALM, in the linearized method, is calculated as
\begin{equation}
    \mathsfbi{u_y}^{c} = \mathsfbi{u_y}^{s} + \mathsfbi{u_y}^{m} = \mathsfbi{u_y}^{s} + \mathsfbi{u_y}^{vi,c} - \mathsfbi{u_y}^{v,c} = \mathsfbi{u_y}^{s} + \mathsfbi{A_y}^{vi} \boldsymbol{\Gamma}^{c} - \mathsfbi{A_y}^{v} \boldsymbol{\Gamma}^{c}
\end{equation}
\begin{equation}
    \mathsfbi{u_z}^{c} = \mathsfbi{u_z}^{s} + \mathsfbi{u_z}^{m} = \mathsfbi{u_z}^{s} + \mathsfbi{u_z}^{vi,c} - \mathsfbi{u_z}^{v,c} = \mathsfbi{u_z}^{s} + \mathsfbi{A_z}^{vi} \boldsymbol{\Gamma}^{c} - \mathsfbi{A_z}^{v} \boldsymbol{\Gamma}^{c}
\end{equation}
where we assumed that the matrix of influence $\mathsfbi{A}^{vi}$ is the same for the ALM and the LL. This may not be the case, since the vortex sheet in each method is created in a different way (see, for example, figure \ref{fig:tail_VSheet} and further discussion at~\citep{kleine2022non}). However, the errors caused by differences in $\mathsfbi{A}^{vi}$ are not of interest here and are, therefore, neglected.

The velocity in the CFD simulation, as a linear approximation, can be interpreted as the undisturbed velocity summed to the effects of the vortices with Gaussian core:
\begin{equation}
    \mathsfbi{u_y}^{s} \approx \mathsfbi{U_y} + \mathsfbi{u_y}^{v,s} = \mathsfbi{U_y} + \mathsfbi{A_y}^{v} \boldsymbol{\Gamma}^s
\end{equation}
\begin{equation}
    \mathsfbi{u_z}^{s} \approx \mathsfbi{U_z} + \mathsfbi{u_z}^{v,s} = \mathsfbi{U_z} + \mathsfbi{A_z}^{v} \boldsymbol{\Gamma}^s
\end{equation}
The corrected velocity can then be written as
\begin{equation}
    \mathsfbi{u_y}^{c} = \mathsfbi{U_y} + \mathsfbi{A_y}^{v} \boldsymbol{\Gamma}^s + \mathsfbi{A_y}^{vi} \boldsymbol{\Gamma}^{c} - \mathsfbi{A_y}^{v} \boldsymbol{\Gamma}^{c} = \mathsfbi{u_y}^{ll} + \mathsfbi{A_y}^{vi} ( \boldsymbol{\Gamma}^{c} - \boldsymbol{\Gamma}^{ll}) +  \mathsfbi{A_y}^{v} \left( \boldsymbol{\Gamma}^s - \boldsymbol{\Gamma}^{c} \right)
\end{equation}
\begin{equation}
    \mathsfbi{u_z}^{c} = \mathsfbi{U_z} + \mathsfbi{A_z}^{v} \boldsymbol{\Gamma}^s + \mathsfbi{A_z}^{vi} \boldsymbol{\Gamma}^{c} - \mathsfbi{A_z}^{v} \boldsymbol{\Gamma}^{c} = \mathsfbi{u_z}^{ll} + \mathsfbi{A_z}^{vi} ( \boldsymbol{\Gamma}^{c} - \boldsymbol{\Gamma}^{ll}) +  \mathsfbi{A_z}^{v} \left( \boldsymbol{\Gamma}^s - \boldsymbol{\Gamma}^{c} \right)
\end{equation}
and the errors in the component of the velocity, $\mathsfbi{u_y}^{c} - \mathsfbi{u_y}^{ll}$ and $\mathsfbi{u_z}^{c} - \mathsfbi{u_z}^{ll}$ are
\begin{equation}
  \label{eq:deltauy}
  \mathsfbi{u_y}^{c} - \mathsfbi{u_y}^{ll} = \mathsfbi{A_y}^{vi} ( \boldsymbol{\Gamma}^{c} - \boldsymbol{\Gamma}^{ll}) +  \mathsfbi{A_y}^{v} \left( \boldsymbol{\Gamma}^s - \boldsymbol{\Gamma}^{c} \right) = \mathsfbi{A_y}^{vi} (\boldsymbol{\beta}^{ll} \circ \boldsymbol{\Gamma}^{c}) +  \mathsfbi{A_y}^{v} (\boldsymbol{\beta}^{s} \circ \boldsymbol{\Gamma}^{c}) 
\end{equation}
\begin{equation}
  \label{eq:deltauz}
  \mathsfbi{u_z}^{c} - \mathsfbi{u_z}^{ll} = \mathsfbi{A_z}^{vi} ( \boldsymbol{\Gamma}^{c} - \boldsymbol{\Gamma}^{ll}) +  \mathsfbi{A_z}^{v} \left( \boldsymbol{\Gamma}^s - \boldsymbol{\Gamma}^{c} \right) = \mathsfbi{A_z}^{vi} (\boldsymbol{\beta}^{ll} \circ \boldsymbol{\Gamma}^{c}) +  \mathsfbi{A_z}^{v} (\boldsymbol{\beta}^{s} \circ \boldsymbol{\Gamma}^{c})
\end{equation}
where $\circ$ denotes the element-wise product and $\boldsymbol{\beta}$ is the vector formed by the values of $\beta$. Hence, each element of $\boldsymbol{\beta}$ is the difference of the circulation at a control point, normalized by the local value of $\Gamma^{c}$.

From the linearization of the equations used to calculate the circulation (see~\citep{kleine2022non}),
\begin{equation}
  \label{eq:deltaGamma}
  \boldsymbol{\Gamma}^{c} - \boldsymbol{\Gamma}^{ll} = \mathsfbi{b_y} \circ (\mathsfbi{u_y}^{c} - \mathsfbi{u_y}^{ll}) + \mathsfbi{b_z} \circ (\mathsfbi{u_z}^{c} - \mathsfbi{u_z}^{ll}) .
\end{equation}

According to equations \eqref{eq:deltauy} and \eqref{eq:deltauz}, a difference in the circulation in one region of the wing affects the velocity difference in the whole wing. However, in order to estimate the order of magnitude of the error, we assume that the induced velocity is dominated by a few terms around the control point and the values of $\beta^{ll}$ and $\beta^s$ are approximately constant around this region. With these simplifying assumptions, each control point can be treated independently. Then, $\mathsfbi{A_y}^{vi} (\boldsymbol{\beta}^{ll} \circ \boldsymbol{\Gamma}^{c}) \approx \boldsymbol{\beta}^{ll} \circ \mathsfbi{u}^{vi}_y$, and analogously for the other terms. Equations \eqref{eq:deltauy} and \eqref{eq:deltauz} for each control point become
\begin{equation}
  \label{eq:deltauylocal}
  u_y^{c} - u_y^{ll} \approx \beta^{ll} u^{vi}_y +  \beta^{s} u^{v}_y 
\end{equation}
\begin{equation}
  \label{eq:deltauzlocal}
  u_z^{c} - u_z^{ll} \approx \beta^{ll} u^{vi}_z +  \beta^{s} u^{v}_z .
\end{equation}
Applying to equation \eqref{eq:deltaGamma}
\begin{equation}
  \Gamma^{c} - \Gamma^{ll} \approx b_y (\beta^{ll} u^{vi}_y +  \beta^{s} u^{v}_y ) + b_z (\beta^{ll} u^{vi}_z +  \beta^{s} u^{v}_z) ,
\end{equation}
so
\begin{equation}
  \beta^{ll} \approx \left(b_y^* \frac{u^{v}_y}{u^{c}} + b_z^* \frac{u^{v}_z}{u^{c}} \right) \left(1+b_y^* \frac{u^{vi}_y}{u^{c}} + b_z^* \frac{u^{vi}_z}{u^{c}} \right) \beta^{s} ,
\end{equation}
where the superscript $b^*$ indicates that $b$ is normalized by the local values of $u_r^c$ and $\Gamma^c$ ($b^*=b u_r^c/\Gamma^c$), which assumes the values $b_y^* \approx (1/C_l) \partial C_l / \partial \alpha$ and $b_z^* \approx 1$ for the cases studied in this work~\citep{kleine2022non}.

Therefore, while the error in the circulation created in the numerical simulation, $\beta^{s}$, is of first order relative to the velocity ratio, the error in the corrected circulation relative to the LL, $\beta^{ll}$, is of second order. From equations \eqref{eq:deltauylocal} and \eqref{eq:deltauzlocal}, it is possible to see that the error in the induced velocities are of second order as well. As a consequence, a second order error is also observed in the corrected forces.

On the other hand, if the vorticity magnitude correction is applied, $\Gamma^{s} = \Gamma^c$ and $\beta^{s}=0$, which imply $\beta^{ll} \approx 0$. Hence these errors are null, within the limits of the current approximations. It should be noted, however, that the method is still of second-order accuracy, since the linear method is of second-order accuracy. Nevertheless, by using the vorticity magnitude correction, we are able to eliminate the main term of the second-order error relative to the LL.

\section*{Acknowledgements}
The computations were performed on resources provided by the Swedish National Infrastructure for Computing (SNIC) at the High Performance Computing Center North (HPC2N). This work was conducted within StandUp for Wind. V.G.K. thanks KTH Engineering Mechanics for partially funding this work.

\bibliography{references}

\listofchanges

\end{document}